\documentclass[prd,showkeys,floatfix,twocolumn,amsmath,amssymb,floatfix]{revtex4}
\usepackage{graphicx}
\usepackage{epstopdf}
\usepackage{multirow}
\usepackage{subfigure}
\usepackage{color}
\usepackage[colorlinks,citecolor=blue,anchorcolor=red,menucolor=red,linkcolor=red,filecolor=red,runcolor=red,urlcolor=blue,frenchlinks=red]{hyperref}
\allowdisplaybreaks[3]

\begin{document}
\title{$P$-wave charmed baryons of the $SU(3)$ flavor $\mathbf{6}_F$}
%

\author{Hui-Min Yang$^1$}
\author{Hua-Xing Chen$^2$}
\email{hxchen@seu.edu.cn}

\affiliation{
$^1$School of Physics, Beihang University, Beijing 100191, China \\
$^2$School of Physics, Southeast University, Nanjing 210094, China
}

\begin{abstract}
We use QCD sum rules to study mass spectra of $P$-wave charmed baryons of the $SU(3)$ flavor $\mathbf{6}_F$. We also use light-cone sum rules to study their $S$- and $D$-wave decays into ground-state charmed baryons together with light pseudoscalar and vector mesons. We work within the framework of heavy quark effective theory, and we also consider the mixing effect. Our results can explain many excited charmed baryons as a whole, including the $\Sigma_c(2800)^0$, $\Xi_c(2923)^0$, $\Xi_c(2939)^0$, $\Xi_{c}(2965)^{0}$, $\Omega_c(3000)^0$, $\Omega_c(3050)^0$, $\Omega_c(3066)^0$,  $\Omega_c(3090)^0$, and $\Omega_c(3119)^0$. Their masses, mass splittings within the same multiplets, and decay properties are extracted for future experimental searches.
\end{abstract}
\maketitle
\pagenumbering{arabic}
%
%
%
\section{Introduction}\label{sec:intro}
%

The singly heavy baryon system is an ideal platform to study the fine structure of hadron spectra~\cite{Chen:2016spr,Copley:1979wj,Karliner:2008sv,pdg}, where light quarks and gluons circle around the nearly static heavy quark, and the whole system behaves as the QCD analogue of the hydrogen~\cite{Korner:1994nh,Manohar:2000dt,Bianco:2003vb,Klempt:2009pi}. In the past years important experimental progresses have been made in the field of excited singly charmed baryons, {\it e.g.}, the $\Lambda_c(2595)$, $\Lambda_c(2625)$, $\Xi_c(2790)$, and $\Xi_c(2815)$ can be well interpreted as the $P$-wave charmed baryons completing two $SU(3)$ flavor $\mathbf{\bar 3}_F$ multiplets of $J^P=1/2^-$ and $3/2^-$~\cite{Frabetti:1993hg,Albrecht:1993pt,Edwards:1994ar,Alexander:1999ud}. The $\Sigma_c(2800)$, $\Xi_c(2930)$, and $\Xi_c(2970)$ are also $P$-wave charmed baryon candidates of the $SU(3)$ flavor $\mathbf{6}_F$, whose experimental parameters are~\cite{Mizuk:2004yu,Chistov:2006zj,Aubert:2007dt,Aubert:2008ax,Yelton:2016fqw,Kato:2016hca,Aaij:2017vbw,Belle:2017jrt,Belle:2018yob}:
\begin{eqnarray}
\nonumber              \Sigma_{c}(2800)^{++}:M&=&2801 ^{+4}_{-6} \mbox{ MeV} \, ,
\\                         \Gamma&=&75{^{+18}_{-13}}{^{+12}_{-11}} \mbox{ MeV} \, ,
\\ \nonumber              \Sigma_{c}(2800)^{+}:M&=&2792 ^{+14}_{-~5} \mbox{ MeV} \, ,
\\                         \Gamma&=&62{^{+37}_{-23}}{^{+52}_{-38}} \mbox{ MeV} \, ,
\\ \nonumber              \Sigma_{c}(2800)^{0}:M&=&2806 ^{+5}_{-7} \mbox{ MeV} \, ,
\\                         \Gamma&=&72{^{+22}_{-15}} \mbox{ MeV} \, ,
\\ \nonumber        \Xi_{c}(2930)^{+}:M&=&2942.3 \pm 4.4 \pm 1.5 \mbox{ MeV} \, ,
\\                         \Gamma&=&14.8 \pm 8.8 \pm 2.5 \mbox{ MeV} \, ,
\\ \nonumber        \Xi_{c}(2930)^{0}:M&=&2929.7{^{+2.8}_{-5.0}}\mbox{ MeV} \, ,
\\                         \Gamma&=&26 \pm 8 \mbox{ MeV} \, ,
\\ \nonumber        \Xi_{c}(2970)^{+}:M&=&2966.34{^{+0.17}_{-1.00}}\mbox{ MeV} \, ,
\\                                 \Gamma&=&20.9^{+2.4}_{-3.5} \mbox{ MeV} \, ,
\\ \nonumber        \Xi_{c}(2970)^{0}:M&=&2970.9 ^{+0.4}_{-0.6}\mbox{ MeV} \, ,
\\                                 \Gamma&=&28.1^{+3.4}_{-4.0} \mbox{ MeV} \, .
\end{eqnarray}
Besides, in the past five years the LHCb Collaboration discovered as many as eight excited charmed baryons:
\begin{itemize}

\item In 2017 the LHCb Collaboration observed the $\Omega_c(3000)^0$, $\Omega_c(3050)^0$, $\Omega_c(3066)^0$, $\Omega_c(3090)^0$, and $\Omega_c(3119)^0$ in the $\Xi_c^+ K^-$ invariant mass spectrum with a sample of $pp$ collision data, whose experimental parameters are~\cite{Aaij:2017nav}:
    \begin{eqnarray}
    \nonumber \Omega_c(3000)^{0}   &:& M = 3000.4 \pm 0.2 \pm 0.1^{+0.3}_{-0.5} ~{\rm MeV} \, ,
    \\                 && \Gamma = 4.5 \pm 0.6 \pm 0.3 ~{\rm MeV} \, ,
    \\ \nonumber \Omega_c(3050)^{0}&:& M = 3050.2 \pm 0.1 \pm 0.1^{+0.3}_{-0.5} ~{\rm MeV} \, ,
    \\                 && \Gamma = 0.8 \pm 0.2 \pm 0.1 ~{\rm MeV} \, ,
    \\ \nonumber \Omega_c(3066)^{0}&:& M = 3065.6 \pm 0.1 \pm 0.3^{+0.3}_{-0.5} ~{\rm MeV} \, ,
    \\                 && \Gamma = 3.5 \pm 0.4 \pm 0.2 ~{\rm MeV} \, ,
    \\ \nonumber \Omega_c(3090)^{0}&:& M = 3090.2 \pm 0.3 \pm 0.5^{+0.3}_{-0.5} ~{\rm MeV} \, ,
    \\                 && \Gamma = 8.7 \pm 1.0 \pm 0.8 ~{\rm MeV} \, ,
    \\ \nonumber \Omega_c(3119)^{0}&:& M = 3119.1 \pm 0.3 \pm 0.9^{+0.3}_{-0.5} ~{\rm MeV} \, ,
    \\                 && \Gamma = 1.1 \pm 0.8 \pm 0.4 ~{\rm MeV} \, .
    \end{eqnarray}
    Some of them are confirmed in the latter Belle experiment~\cite{Yelton:2017qxg} and in the $\Omega_b^- \to \Xi_c^+ K^-\pi^-$ decay process by LHCb~\cite{LHCb:2021ptx}.

\item In 2020 the LHCb Collaboration observed the  $\Xi_{c}(2923)^0$, $\Xi_{c}(2939)^0$, and $\Xi_{c}(2965)^0$ in the $\Lambda^+_c K^-$ invariant mass spectrum, whose experimental parameters are~\cite{Aaij:2020yyt}:
    \begin{eqnarray}
    \nonumber              \Xi_{c}(2923)^{0}:M&=&2923.04 \pm 0.25 \pm 0.20 \pm 0.14 \mbox{ MeV} \, ,
    \\                         \Gamma&=&7.1 \pm 0.8 \pm 1.8 \mbox{ MeV} \, ,
    \\ \nonumber        \Xi_{c}(2939)^{0}:M&=&2938.55 \pm 0.21 \pm 0.17 \pm 0.14 \mbox{ MeV} \, ,
    \\                         \Gamma&=&10.2 \pm 0.8 \pm 1.1 \mbox{ MeV} \, ,
    \\ \nonumber        \Xi_{c}(2965)^{0}:M&=&2964.88 \pm 0.26 \pm 0.14 \pm 0.14 \mbox{ MeV} \, ,
    \\                                 \Gamma&=&14.1 \pm 0.9 \pm 1.3 \mbox{ MeV} \, .
    \end{eqnarray}

\end{itemize}

Based on the above experimental observations, many phenomenological methods and models were proposed to study excited singly charmed baryons, such as
various quark models~\cite{Capstick:1986bm,Chen:2007xf,Ebert:2007nw,Garcilazo:2007eh,Roberts:2007ni,Zhong:2007gp,Valcarce:2008dr,Ebert:2011kk,Ortega:2012cx,Yoshida:2015tia,Nagahiro:2016nsx,Wang:2017kfr,Ye:2017yvl,Lu:2020ivo,Chen:2021eyk},
various molecular explanations~\cite{GarciaRecio:2008dp,Liang:2014eba,Yu:2018yxl,Nieves:2019jhp,Huang:2017dwn,Montana:2017kjw,Debastiani:2017ewu},
the hyperfine interaction~\cite{Copley:1979wj,Karliner:2008sv},
the chiral perturbation theory~\cite{Lu:2014ina,Cheng:2015naa},
the Lattice QCD~\cite{Bowler:1996ws,Burch:2008qx,Brown:2014ena,Padmanath:2013bla,Padmanath:2017lng,Bahtiyar:2020uuj},
and QCD sum rules~\cite{Bagan:1991sg,Neubert:1991sp,Broadhurst:1991fc,Huang:1994zj,Dai:1996yw,Groote:1996em,Colangelo:1998ga,Huang:2000tn,Zhu:2000py,Lee:2000tb,Wang:2003zp,Duraes:2007te,Liu:2007fg,Zhang:2008pm,Aliev:2009jt,Wang:2010it,Zhou:2014ytp,Zhou:2015ywa,Wang:2017zjw,Aliev:2018ube,Xu:2020ofp}, etc.
Their productions and decay properties were studied in Refs.~\cite{Cheng:2006dk,Kim:2014qha,Xie:2015zga,Huang:2016ygf,Wang:2020gkn}, and we refer to the reviews~\cite{Korner:1994nh,Manohar:2000dt,Bianco:2003vb,Klempt:2009pi,Crede:2013kia,Cheng:2015rra,Chen:2016qju,Chen:2016spr} for their recent progress.

In this paper we shall systematically investigate $P$-wave charmed baryons of the $SU(3)$ flavor $\mathbf{6}_F$. In Refs.~\cite{Chen:2015kpa,Mao:2015gya} we have studied mass spectra of $P$-wave bottom baryons using the method of QCD sum rules~\cite{Shifman:1978bx,Reinders:1984sr}, and in the present study we shall replace the $bottom$ quark by the $charm$ quark, and reanalyse those results. In Ref.~\cite{Yang:2020zrh} we have studied decay properties of $P$-wave bottom baryons using the method of light-cone sum rules~\cite{Balitsky:1989ry,Braun:1988qv,Chernyak:1990ag,Ball:1998je,Ball:2006wn}, and in the present study we shall apply the same method to study $P$-wave charmed baryons of the $SU(3)$ flavor $\mathbf{6}_F$. We shall study their $S$- and $D$-wave decays into ground-state charmed baryons together with pseudoscalar mesons $\pi/K$ and vector mesons $\rho/K^*$. We shall work within the framework of the heavy quark effective theory (HQET)~\cite{Grinstein:1990mj,Eichten:1989zv,Falk:1990yz}, and we shall also consider the mixing effect between two different HQET multiplets.

This paper is organized as follows. In Sec.~\ref{sec:sumrule} we briefly introduce our notations, and use the method of QCD sum rule to study mass spectra of $P$-wave charmed baryons of the $SU(3)$ flavor $\mathbf{6}_F$. The obtained results are further used in Sec.~\ref{sec:decay} to study their $S$- and $D$-wave decays into ground-state charmed baryons together with light pseudoscalar and vector mesons. The mixing effect between different HQET multiplets is investigated in Sec.~\ref{sec:mixing}, and the obtained results are summarized in Sec.~\ref{sec:summary}, where we conclude this paper.

%
\section{Mass spectra through QCD sum rules}
\label{sec:sumrule}
%

\begin{figure*}[hbtp]
\begin{center}
\scalebox{0.63}{\includegraphics{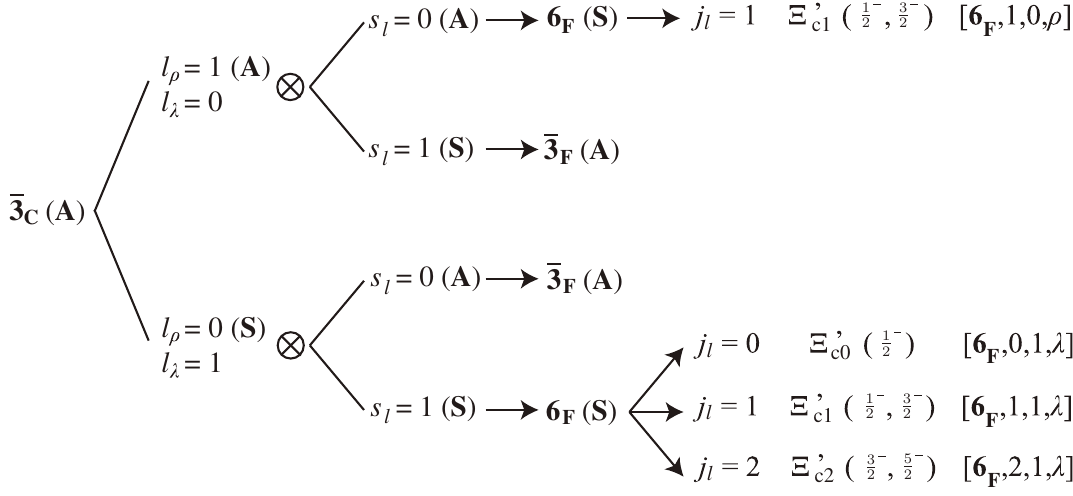}}
\end{center}
\caption{$P$-wave charmed baryons belonging to the $SU(3)$ flavor $\mathbf{6}_F$ representation.}
\label{fig:pwave}
\end{figure*}

In this section we follow Ref.~\cite{Chen:2007xf} and classify $P$-wave charmed baryons. A singly charmed baryon consists of one $charm$ quark and two light $up/down/strange$ quarks, and its internal symmetries are:
\begin{itemize}

\item The color structure of the two light quarks is antisymmetric ($\mathbf{\bar 3}_C$).

\item The flavor structure of the two light quarks is either symmetric ($\mathbf{6}_F$) or antisymmetric ($\mathbf{\bar 3}_F$).

\item The spin structure of the two light quarks is either symmetric ($s_l \equiv s_{qq} = 1$) or antisymmetric ($s_l = 0$).

\item The orbital structure of the two light quarks is either symmetric or antisymmetric. We call the former $\lambda$-type with $l_\rho = 0$ and $l_\lambda = 1$, and the latter $\rho$-type with $l_\rho = 1$ and $l_\lambda = 0$. Here $l_\rho$ denotes the orbital angular momentum between the two light quarks, and $l_\lambda$ denotes the orbital angular momentum between the charm quark and the two-light-quark system.

\end{itemize}
According to the Pauli principle, the total symmetry of the two light quarks is antisymmetric, so that we can categorize $P$-wave charmed baryons into eight multiplets. Four of them belong to the $SU(3)$ flavor $\mathbf{6}_F$ representation, as shown in Fig.~\ref{fig:pwave}. We denote them as $[F({\rm flavor}), j_l, s_l, \rho/\lambda]$, where $j_l = l_\lambda \otimes l_\rho \otimes s_l$ is the total angular momentum of the light components. There are one or two charmed baryons contained in each multiplet, with the total angular momenta $j = j_l \otimes s_b = |j_l \pm 1/2|$.

\begin{table*}[hbtp]
\begin{center}
\renewcommand{\arraystretch}{1.5}
\caption{Parameters of $P$-wave charmed baryons belonging to the $SU(3)$ flavor $\mathbf{6}_F$ representation, extracted from their mass sum rules. In the last column we list decay constant, satisfying $f_{\Sigma^{++}_c} = f_{\Sigma^0_c} = \sqrt2 f_{\Sigma^+_c}$ and $f_{\Xi^{\prime+}_c} = f_{\Xi^{\prime0}_c}$.}
\begin{tabular}{c | c | c | c | c | c c | c | c}
\hline\hline
\multirow{2}{*}{Multiplet} & \multirow{2}{*}{~~B~~} & $\omega_c$ & ~~~Working region~~~ & ~~~~~~~$\overline{\Lambda}$~~~~~~~ & ~~~Baryon~~~ & ~~~~Mass~~~~~ & ~Difference~ & Decay constant
\\                                              &  & (GeV)      & (GeV)                & (GeV)                              & ($j^P$)       & (GeV)      & (MeV)        & (GeV$^{4}$)
\\ \hline\hline
\multirow{6}{*}{$[\mathbf{6}_F, 1, 0, \rho]$}
& \multirow{2}{*}{$\Sigma_c$} & \multirow{2}{*}{1.74} & \multirow{2}{*}{$0.27< T < 0.32$} & \multirow{2}{*}{$1.25 \pm 0.11$} & $\Sigma_c(1/2^-)$ & $2.77 \pm 0.14$ & \multirow{2}{*}{$15 \pm 6$} & $0.067 \pm 0.017~(\Sigma^-_c(1/2^-))$
\\ \cline{6-7}\cline{9-9}
& & & & & $\Sigma_c(3/2^-)$ & $2.79 \pm 0.14$ & &$0.031 \pm 0.008~(\Sigma^-_c(3/2^-))$
\\ \cline{2-9}
& \multirow{2}{*}{$\Xi^\prime_c$} & \multirow{2}{*}{1.87} & \multirow{2}{*}{$0.26< T < 0.34$} & \multirow{2}{*}{$1.36 \pm 0.10$} & $\Xi^\prime_c(1/2^-)$ & $2.88 \pm 0.14$ & \multirow{2}{*}{$13 \pm 5$} & $0.059 \pm 0.014~(\Xi^{\prime-}_c(1/2^-))$
\\ \cline{6-7}\cline{9-9}
& & & & & $\Xi^\prime_c(3/2^-)$ & $2.89 \pm 0.14$ & &$0.028 \pm 0.007~(\Xi^{\prime-}_c(3/2^-))$
\\ \cline{2-9}
& \multirow{2}{*}{$\Omega_c$} & \multirow{2}{*}{2.00} & \multirow{2}{*}{$0.26< T < 0.35$} & \multirow{2}{*}{$1.48 \pm 0.09$} & $\Omega_c(1/2^-)$ & $2.99 \pm 0.15$ & \multirow{2}{*}{$12 \pm 5$} & $0.105 \pm 0.023~(\Omega^-_c(1/2^-))$
\\ \cline{6-7}\cline{9-9}
& & & & & $\Omega_c(3/2^-)$ & $3.00 \pm 0.15$ & &$0.049 \pm 0.011~(\Omega^-_c(3/2^-))$
\\ \hline
\multirow{3}{*}{$[\mathbf{6}_F, 0, 1, \lambda]$} & $\Sigma_c$ & $1.35$ & $T=0.27$ & $1.10 \pm 0.04$ & $\Sigma_c(1/2^-)$ & $2.83 \pm 0.05$ & -- & $0.045 \pm 0.008~(\Sigma^-_c(1/2^-))$
\\ \cline{2-9}
                                                 & $\Xi^\prime_c$ & $1.57$ & $0.27< T < 0.29$ & $1.22 \pm 0.08$ & $\Xi^\prime_c(1/2^-)$ & $2.90 \pm 0.13$ & -- & $0.041 \pm 0.009~(\Xi^{\prime-}_c(1/2^-))$
\\ \cline{2-9}
                                                 & $\Omega_c$ & 1.78 & $0.27< T < 0.31$ & $1.37 \pm 0.09$ & $\Omega_c(1/2^-)$ & $3.03 \pm 0.18$ & -- & $0.081 \pm 0.020~(\Omega^-_c(1/2^-))$
\\ \hline
\multirow{6}{*}{$[\mathbf{6}_F, 1, 1, \lambda]$}
& \multirow{2}{*}{$\Sigma_c$} & \multirow{2}{*}{1.72} & \multirow{2}{*}{$T=0.33$} & \multirow{2}{*}{$1.03 \pm 0.12$} & $\Sigma_c(1/2^-)$ & $2.73 \pm 0.17$ & \multirow{2}{*}{$41 \pm 16$} & $0.045 \pm 0.011~(\Sigma^-_c(1/2^-))$
\\ \cline{6-7}\cline{9-9}
& & & & & $\Sigma_c(3/2^-)$ & $2.77 \pm 0.17$ & &$0.021 \pm 0.005~(\Sigma^-_c(3/2^-))$
\\ \cline{2-9}
& \multirow{2}{*}{$\Xi^\prime_c$} & \multirow{2}{*}{1.72} & \multirow{2}{*}{$T=0.34$} & \multirow{2}{*}{$1.14 \pm 0.09$} & $\Xi^\prime_c(1/2^-)$ & $2.91 \pm 0.12$ & \multirow{2}{*}{$38 \pm 14$} & $0.041 \pm 0.008~(\Xi^{\prime-}_c(1/2^-))$
\\ \cline{6-7}\cline{9-9}
& & & & & $\Xi^\prime_c(3/2^-)$ & $2.95 \pm 0.12$ & &$0.019 \pm 0.004~(\Xi^{\prime-}_c(3/2^-))$
\\ \cline{2-9}
& \multirow{2}{*}{$\Omega_c$} & \multirow{2}{*}{1.72} & \multirow{2}{*}{$T=0.35$} & \multirow{2}{*}{$1.22 \pm 0.07$} & $\Omega_c(1/2^-)$ & $3.04 \pm 0.10$ & \multirow{2}{*}{$36\pm 13$} & $0.069 \pm 0.011~(\Omega^-_c(1/2^-))$
\\ \cline{6-7}\cline{9-9}
& & & & & $\Omega_c(3/2^-)$ & $3.07 \pm 0.09$ & &$0.032 \pm 0.005~(\Omega^-_c(3/2^-))$
\\ \hline
\multirow{6}{*}{$[\mathbf{6}_F, 2, 1, \lambda]$}
& \multirow{2}{*}{$\Sigma_c$} & \multirow{2}{*}{1.50} & \multirow{2}{*}{$0.28< T < 0.29$} & \multirow{2}{*}{$1.09 \pm 0.09$} & $\Sigma_c(3/2^-)$ & $2.78 \pm 0.13$ & \multirow{2}{*}{$86 \pm 36$} & $0.055 \pm 0.013~(\Sigma^-_c(3/2^-))$
\\ \cline{6-7}\cline{9-9}
& & & & & $\Sigma_c(5/2^-)$ & $2.87 \pm 0.11$ & &$0.033 \pm 0.008 ~(\Sigma^-_c(5/2^-))$
\\ \cline{2-9}
 & \multirow{2}{*}{$\Xi^\prime_c$} & \multirow{2}{*}{1.72} & \multirow{2}{*}{$0.27< T < 0.32$} & \multirow{2}{*}{$1.24 \pm 0.12$} & $\Xi^\prime_c(3/2^-)$ & $2.96 \pm 0.20$ & \multirow{2}{*}{$66 \pm 27$} & $0.057 \pm0.016 ~(\Xi^{\prime-}_c(3/2^-))$
\\ \cline{6-7}\cline{9-9}
& & & & & $\Xi^\prime_c(5/2^-)$ & $3.02 \pm 0.18$ & &$0.034 \pm 0.009 ~(\Xi^{\prime-}_c(5/2^-))$
\\ \cline{2-9}
& \multirow{2}{*}{$\Omega_c$} & \multirow{2}{*}{1.85} & \multirow{2}{*}{$0.26< T < 0.33$} & \multirow{2}{*}{$1.35 \pm 0.11$} & $\Omega_c(3/2^-)$ & $3.08\pm 0.19$ & \multirow{2}{*}{$59 \pm 24$} & $0.103 \pm 0.026~(\Omega^-_c(3/2^-))$
\\ \cline{6-7}\cline{9-9}
& & & & & $\Omega_c(5/2^-)$ & $3.14 \pm 0.18$ & &$0.062 \pm 0.016~(\Omega^-_c(5/2^-))$
\\ \hline \hline
\end{tabular}
\label{tab:pwaveparameter}
\end{center}
\end{table*}

We have systematically studied mass spectra of $P$-wave bottom baryons in Refs.~\cite{Chen:2015kpa,Mao:2015gya}. In the present study we just need to replace the $bottom$ quark by the $charm$ quark, and reanalyse those results. The newly obtained results for charmed baryons are summarized in Table~\ref{tab:pwaveparameter}. In the calculation we have used the following QCD parameters at the renormalization scale 1~GeV~\cite{pdg,Yang:1993bp,Hwang:1994vp,Ovchinnikov:1988gk,Jamin:2002ev,Ioffe:2002be,Gimenez:2005nt,Colangelo:1998ga}:
\begin{eqnarray}
\nonumber && \langle \bar qq \rangle = - (0.24 \pm 0.01 \mbox{ GeV})^3 \, ,
\\ \nonumber && \langle \bar ss \rangle = (0.8\pm 0.1)\times \langle\bar qq \rangle \, ,
\\ && \langle g_s \bar q \sigma G q \rangle = M_0^2 \times \langle \bar qq \rangle\, ,
\label{eq:condensates}
\\ \nonumber && \langle g_s \bar s \sigma G s \rangle = M_0^2 \times \langle \bar ss \rangle\, ,
\\ \nonumber && M_0^2= 0.8 \mbox{ GeV}^2\, ,
\\ \nonumber && \langle g_s^2GG\rangle =(0.48\pm 0.14) \mbox{ GeV}^4\, .
\end{eqnarray}
Besides, we have used the PDG value $m_c = 1.275 \pm 0.025$ GeV~\cite{pdg} for the $charm$ quark mass in the $\overline{\rm MS}$ scheme.

To better understand $P$-wave charmed baryons, we shall further investigate their decay properties in the next section. The parameters given in Table~\ref{tab:pwaveparameter} will be used as inputs. To better describe $P$-wave charmed baryons, we select the following mass values when calculating their decay widths:
\begin{itemize}

\item We shall find it possible to interpret the $\Omega_c(3000)^0$ as a $P$-wave $\Omega_c$ baryon belonging to the $[\mathbf{6}_F, 1, 0, \rho]$ doublet. Accordingly, we choose:
\begin{eqnarray}
\nonumber M_{[\Sigma_c(1/2^-), 1, 0, \rho]} &=& 2.77~{\rm GeV} \, ,
\\ \nonumber M_{[\Sigma_c(3/2^-), 1, 0, \rho]} &=& 2.79~{\rm GeV} \, ,
\\ M_{[\Xi_c^{\prime}(1/2^-), 1, 0, \rho]} &=& 2.88~{\rm GeV} \, ,
\\ \nonumber M_{[\Xi_c^{\prime}(3/2^-), 1, 0, \rho]} &=& 2.89~{\rm GeV} \, ,
\\ \nonumber M_{[\Omega_c(1/2^-), 1, 0, \rho]} &=& 3000.4~{\rm MeV}~\mbox{\cite{Aaij:2017nav}} \, ,
\\ \nonumber M_{[\Omega_c(3/2^-), 1, 0, \rho]} &=& 3000.4~{\rm MeV}~\mbox{\cite{Aaij:2017nav}} \, .
\end{eqnarray}

\item For the $[\mathbf{6}_F, 0, 1, \lambda]$ singlet, we choose:
\begin{eqnarray}
\nonumber M_{[\Sigma_c(1/2^-), 0, 1, \lambda]} &=& 2.83~{\rm GeV}\, ,
\\ M_{[\Xi_c^{\prime}(1/2^-), 0, 1, \lambda]} &=& 2.90~{\rm GeV}\, ,
\\ \nonumber M_{[\Omega_c(1/2^-), 0, 1, \lambda]} &=& 3.03~{\rm GeV} \, .
\end{eqnarray}

\item We shall find it reasonable to interpret the $\Xi_c(2923)^0$, $\Xi_c(2939)^0$, $\Omega_c(3050)^0$, and $\Omega_c(3066)^0$ as $P$-wave charmed baryons belonging to the $[\mathbf{6}_F, 1, 1, \lambda]$ doublet. We shall also find it possible to explain the $\Sigma_c(2800)^0$ as the combination of such charmed baryons. Accordingly, we choose:
\begin{eqnarray}
\nonumber && M_{[\Sigma_c(1/2^-), 1, 1, \lambda]} \sim 2800~{\rm MeV}~\mbox{\cite{Zyla:2020zbs}} \, ,
\\ \nonumber && M_{[\Sigma_c(3/2^-), 1, 1, \lambda]} \sim 2800~{\rm MeV}~\mbox{\cite{Zyla:2020zbs}} \, ,
\\ && M_{[\Xi_c^{\prime}(1/2^-), 1, 1, \lambda]} = 2923.04~{\rm MeV}~\mbox{\cite{Aaij:2020yyt}} \, ,
\\ \nonumber && M_{[\Xi_c^{\prime}(3/2^-), 1, 1, \lambda]} = 2938.55~{\rm MeV}~\mbox{\cite{Aaij:2020yyt}} \, ,
\\ \nonumber && M_{[\Omega_c(1/2^-), 1, 1, \lambda]} = 3050.2~{\rm MeV}~\mbox{\cite{Aaij:2017nav}} \, ,
\\ \nonumber && M_{[\Omega_c(3/2^-), 1, 1, \lambda]} = 3065.6~{\rm MeV}~\mbox{\cite{Aaij:2017nav}} \, .
\end{eqnarray}
The mass difference between the $\Omega_c(3050)^0$ and $\Omega_c(3066)^0$ baryons is slightly smaller than that between the $[\Omega_c(1/2^-), 1, 1, \lambda]$ and $[\Omega_c(3/2^-), 1, 1, \lambda]$ baryons, as given in Table~\ref{tab:pwaveparameter}. This is because the HQET is an effective theory, which works quite well for bottom baryons~\cite{Yang:2020zrh}, but not so perfect for charmed baryons~\cite{Yang:2020zjl}. Accordingly, we shall investigate the mixing effect between different HQET multiplets in Sec.~\ref{sec:mixing}, especially, between the $[\mathbf{6}_F, 1, 1, \lambda]$ and $[\mathbf{6}_F, 2, 1, \lambda]$ multiplets.

\item We shall find it reasonable to interpret the $\Xi_{c}(2965)^{0}$, $\Omega_c(3090)^0$, and $\Omega_c(3119)^0$ as $P$-wave charmed baryons belonging to the $[\mathbf{6}_F, 2, 1, \lambda]$ doublet. We shall also find it possible to explain the $\Sigma_c(2800)^0$ as the combination of such charmed baryons. Accordingly, we choose:
\begin{eqnarray}
\nonumber && M_{[\Sigma_c(3/2^-), 2, 1, \lambda]} \sim 2800~{\rm MeV}~\mbox{\cite{Zyla:2020zbs}} \, ,
\\ \nonumber && M_{[\Sigma_c(5/2^-), 2, 1, \lambda]} \sim 2800~{\rm MeV}~\mbox{\cite{Zyla:2020zbs}} \, ,
\\ && M_{[\Xi_c^{\prime}(3/2^-), 2, 1, \lambda]} = 2964.88~{\rm MeV}~\mbox{\cite{Aaij:2020yyt}}\, ,
\\ \nonumber && M_{[\Xi_c^{\prime}(5/2^-), 2, 1, \lambda]} -  M_{[\Xi_c^{\prime}(3/2^-), 2, 1, \lambda]} = 56~{\rm MeV} \, ,
\\ \nonumber && M_{[\Omega_c(3/2^-), 2, 1, \lambda]} = 3090.2~{\rm MeV}~\mbox{\cite{Aaij:2017nav}}  \, ,
\\ \nonumber && M_{[\Omega_c(5/2^-), 2, 1, \lambda]} = 3119.1~{\rm MeV}~\mbox{\cite{Aaij:2017nav}}  \, .
\end{eqnarray}

\end{itemize}
Note that the above interpretations are just possible explanations, and there exist many other possible explanations for the $\Sigma_c(2800)^0$, $\Xi_c(2923)^0$, $\Xi_c(2939)^0$, $\Xi_{c}(2965)^{0}$, $\Omega_c(3000)^0$, $\Omega_c(3050)^0$, $\Omega_c(3066)^0$,  $\Omega_c(3090)^0$, and $\Omega_c(3119)^0$.

We shall use the following mass values for ground-state charmed baryons~\cite{pdg}:
\begin{eqnarray}
   \nonumber        \Lambda_{c}(1/2^+)  ~:~ m&=&2286.46 \mbox{ MeV} \, ,
\\ \nonumber        \Xi_{c}(1/2^+)  ~:~ m&=&2469.34 \mbox{ MeV} \, ,
\\ \nonumber        \Sigma_{c}(1/2^+)    ~:~ m&=&2453.54 \mbox{ MeV} \, ,
\\         \Sigma_{c}^{*}(3/2^+)    ~:~ m&=&2518.1 \mbox{ MeV} \, , \,
\\ \nonumber        \Xi_{c}^{\prime}(1/2^+)  ~:~ m&=&2576.8 \mbox{ MeV} \, ,
\\ \nonumber        \Xi_{b}^{*}(3/2^+)  ~:~ m&=&2645.9 \mbox{ MeV} \, , \,
\\ \nonumber        \Omega_b(1/2^+)  ~:~ m&=&2695.2 \mbox{ MeV}\, ,
\\ \nonumber        \Omega_b^*(3/2^+)  ~:~ m&=&2765.9 \mbox{ MeV} \, .
\end{eqnarray}
We shall use the following parameters for light pseudoscalar and vector mesons~\cite{pdg}:
\begin{eqnarray}
\nonumber \pi(0^-) &:& m= 138.04 {\rm~MeV} \, ,
\\ \nonumber K(0^-) &:& m= 495.65 {\rm~MeV} \, ,
\\ \nonumber \rho(1^-) &:& m= 775.21 {\rm~MeV} \, ,\,
\\                               && \Gamma= 148.2 {\rm~MeV} \, ,
\\ \nonumber  && {g}_{\rho \pi \pi} = 5.94 \, ,
\\  \nonumber         K^*(1^-) &:& m= 893.57 {\rm~MeV} \, ,\,
\\   \nonumber                 && \Gamma = 49.1 {\rm~MeV} \, ,\,
\\ \nonumber && {g}_{K^* K \pi} = 3.20 \, ,
\end{eqnarray}
which are calculated through:
\begin{eqnarray}
\nonumber \mathcal{L}_{\rho \pi \pi} &=& {g}_{\rho \pi \pi} \times \left( \rho_\mu^0 \pi^+ \partial^\mu \pi^- - \rho_\mu^0 \pi^- \partial^\mu \pi^+ \right) + \cdots \, ,
\\ \nonumber \mathcal{L}_{K^* K \pi} &=& {g}_{K^* K \pi} K^{*+}_\mu \times \left( K^- \partial^\mu \pi^0 - \partial^\mu K^- \pi^0 \right) + \cdots \, .
\\
\end{eqnarray}

\section{Decay properties through light-cone sum rules}
\label{sec:decay}

We have systematically studied decay properties of $P$-wave bottom baryons of the $SU(3)$ flavor $\mathbf{6}_F$ in Ref.~\cite{Yang:2020zrh} using the method of light-cone sum rules within HQET. In this paper we apply the same method to study $P$-wave charmed baryons of the $SU(3)$ flavor $\mathbf{6}_F$. We shall study their $S$- and $D$-wave decays into ground-state charmed baryons together with pseudoscalar mesons $\pi/K$ and vector mesons $\rho/K^*$, including:
\begin{widetext}
\begin{eqnarray}
 &(a1)& {\bf \Gamma\Big[} \Sigma_c[1/2^-] \rightarrow \Lambda_c + \pi {\Big ]}
=  {\bf \Gamma\Big[} \Sigma_c^{0}[1/2^-] \rightarrow \Lambda_c^+ +\pi^- {\Big ]} \, ,
\label{eq:couple1}
\\ &(a2)& {\bf \Gamma\Big[} \Sigma_c[1/2^-] \rightarrow \Sigma_c + \pi {\Big ]}
= 2\times {\bf \Gamma\Big[} \Sigma_c^{0}[1/2^-] \rightarrow \Sigma_c^+\pi^- {\Big ]} \, ,
\\ &(a3)& {\bf \Gamma\Big[} \Sigma_c[1/2^-] \rightarrow \Sigma_c^{*} + \pi {\Big ]}
= 2 \times {\bf \Gamma\Big[} \Sigma_c^{0}[1/2^-] \rightarrow \Sigma_c^{*+} +\pi^- {\Big ]} \, ,
\\ &(a4)& {\bf\Gamma\Big[} \Sigma_c[1/2^-] \rightarrow \Lambda_c + \rho \rightarrow\Lambda_c+\pi+\pi{\Big ]}
= {\bf \Gamma\Big[} \Sigma_c^{0}[1/2^-] \rightarrow \Lambda_c^{+} +\pi^0+ \pi^- {\Big ]} \, ,
\\ &(a5)& { \bf\Gamma\Big[}\Sigma_c[1/2^-] \rightarrow \Sigma_c + \rho\rightarrow\Sigma_c+\pi+\pi{\Big ]}
= 2 \times { \bf\Gamma\Big[}\Sigma_c^{0}[1/2^-] \rightarrow \Sigma_c^{+} +\pi^0+ \pi^-{\Big ]} \, ,
\\ &(a6)&{\bf \Gamma\Big[}\Sigma_c[1/2^-] \rightarrow \Sigma_c^{*} + \rho\rightarrow\Sigma_c^{*}+\pi+\pi{\Big ]}
= 2 \times { \bf\Gamma\Big[}\Sigma_c^{0}[1/2^-] \rightarrow \Sigma_c^{*+} +\pi^0+ \pi^-{\Big ]} \, ,
\\ &(b1)& {\bf \Gamma\Big[}\Sigma_c[3/2^-] \rightarrow \Lambda_c + \pi{\Big ]}
= {\bf \Gamma\Big[}\Sigma_c^{0}[3/2^-] \rightarrow \Lambda_c^{+} +\pi^-{\Big ]} \, ,
\\ &(b2)&{\bf \Gamma\Big[}\Sigma_c[3/2^-] \rightarrow \Sigma_c + \pi{\Big ]}
= 2 \times {\bf \Gamma\Big[}\Sigma_c^{0}[3/2^-] \rightarrow \Sigma_c^{+} +\pi^-{\Big ]} \, ,
\\ &(b3)&{\bf \Gamma\Big[}\Sigma_c[3/2^-] \rightarrow \Sigma_c^{*} + \pi{\Big ]}
= 2 \times {\bf \Gamma\Big[}\Sigma_c^{0}[3/2^-] \rightarrow \Sigma_c^{*+} +\pi^-{\Big ]} \, ,
\\ &(b4)&{\bf \Gamma\Big[} \Sigma_c[3/2^-] \rightarrow \Lambda_c + \rho \rightarrow\Lambda_c+\pi+\pi{\Big ]}
= { \bf\Gamma\Big[} \Sigma_c^{0}[3/2^-] \rightarrow \Lambda_c^{+} +\pi^0+ \pi^- {\Big ]} \, ,
\\ &(b5)& { \bf\Gamma\Big[}\Sigma_c[3/2^-] \rightarrow \Sigma_c + \rho\rightarrow\Sigma_c+\pi+\pi{\Big ]}
= 2 \times { \bf\Gamma\Big[}\Sigma_c^{0}[3/2^-] \rightarrow \Sigma_c^{+} +\pi^0+ \pi^-{\Big ]} \, ,
\\&(b6)& { \bf\Gamma\Big[}\Sigma_c[3/2^-] \rightarrow \Sigma_c^{*} + \rho\rightarrow\Sigma_c^{*}+\pi+\pi {\Big ]}
= 2 \times {\bf \Gamma\Big[}\Sigma_c^{0}[3/2^-] \rightarrow \Sigma_c^{*+} + \pi^0+\pi^- {\Big ]} \, ,
\\ &(c1)& {\bf \Gamma\Big[}\Sigma_c[5/2^-] \rightarrow \Lambda_c + \pi{\Big ]}
= {\bf \Gamma\Big[}\Sigma_c^{0}[5/2^-] \rightarrow \Lambda_c^{+} +\pi^-{\Big ]} \, ,
\\ &(c2)&{\bf \Gamma\Big[}\Sigma_c[5/2^-] \rightarrow \Sigma_c + \pi{\Big ]}
= 2 \times {\bf \Gamma\Big[}\Sigma_c^{0}[5/2^-] \rightarrow \Sigma_c^{+} +\pi^-{\Big ]} \, ,
\\ &(c3)&{\bf \Gamma\Big[}\Sigma_c[5/2^-] \rightarrow \Sigma_c^{*} + \pi{\Big ]}
= 2 \times {\bf \Gamma\Big[}\Sigma_c^{0}[5/2^-] \rightarrow \Sigma_c^{*+} +\pi^-{\Big ]} \, ,
\\ &(c4)& { \bf\Gamma\Big[}\Sigma_c^{}[5/2^-] \rightarrow \Sigma_c^{*}+\rho\rightarrow\Sigma_c^{*}+\pi\pi {\Big]}
= 2\times{\bf \Gamma\Big[}\Sigma_c^{0}[5/2^-]\rightarrow\Sigma_c^{*+}+\pi^0+\pi^-{\Big]} \, ,
\\ &(d1)& {\bf \Gamma\Big[}\Xi_c^{\prime}[1/2^-] \rightarrow \Xi_c + \pi{\Big ]}
= {3 \over 2} \times {\bf \Gamma\Big[}\Xi_c^{\prime 0}[1/2^-] \rightarrow \Xi_c^+ + \pi^-{\Big ]} \, ,
\\ &(d2)& {\bf \Gamma\Big[}\Xi_c^{\prime}[1/2^-] \rightarrow \Xi_c^{\prime} + \pi{\Big ]}
= {3 \over 2} \times {\bf \Gamma\Big[}\Xi_c^{\prime 0}[1/2^-] \rightarrow \Xi_c^{\prime+} + \pi^-{\Big]} \, ,
\\ &(d3)& {\bf \Gamma\Big[}\Xi_c^{\prime}[1/2^-] \rightarrow \Lambda_c + \bar{K}{\Big ]}
=  {\bf \Gamma\Big[}\Xi_c^{\prime 0}[1/2^-] \rightarrow \Lambda_c^{+} + K^-{\Big ]} \, ,
\\ &(d4)& {\bf \Gamma\Big[}\Xi_c^{\prime}[1/2^-] \rightarrow \Sigma_c + \bar{K}{\Big ]}
= 3 \times {\bf \Gamma\Big[}\Xi_c^{\prime 0}[1/2^-] \rightarrow \Sigma_c^{+} + K^-{\Big ]} \, ,
\\ &(d5)& {\bf \Gamma\Big[}\Xi_c^{\prime}[1/2^-] \rightarrow \Xi_c^{*} + \pi{\Big ]}
= {3 \over 2} \times {\bf \Gamma\Big[}\Xi_c^{\prime 0}[1/2^-] \rightarrow \Xi_c^{*+} + \pi^-{\Big ]} \, ,
\\ &(d6)& {\bf \Gamma\Big[} \Xi_c^{\prime}[1/2^-] \rightarrow \Sigma_c^{*} + \bar{K} {\Big ]}
= 3 \times {\bf \Gamma\Big[} \Xi_c^{\prime 0}[1/2^-] \rightarrow \Sigma_c^{*+} + K^- {\Big ]} \, ,
\\ &(d7)& { \bf\Gamma\Big[}\Xi_c^{\prime}[1/2^-] \rightarrow \Xi_c + \rho\rightarrow\Xi_c+\pi+\pi{\Big ]}
= {3\over2} \times { \bf\Gamma\Big[}\Xi_c^{\prime 0}[1/2^-] \rightarrow \Xi_c^{+} + \pi^0+\pi^-{\Big ]} \, ,
\\ &(d8)& { \bf\Gamma\Big[} \Xi_c^{\prime}[1/2^-] \rightarrow \Lambda_c + \bar{K}^*\rightarrow\Lambda_c+\bar{K}+\pi {\Big ]}
={3\over2}\times {\bf \Gamma\Big[} \Xi_c^{\prime 0}[1/2^-] \rightarrow \Lambda_c^{+} + \bar{K}^0+\pi^- {\Big ]} \, ,
\\ &(d9)& { \bf\Gamma\Big[}\Xi_c^{\prime}[1/2^-] \rightarrow \Xi_c^{\prime} + \rho\rightarrow\Xi_c^{\prime}+\pi+\pi{\Big ]}
= {3\over2} \times  { \bf\Gamma\Big[}\Xi_c^{\prime 0}[1/2^-] \rightarrow \Xi_c^{\prime+} + \pi^0+\pi^-{\Big ]} \, ,
\\ &(d10)& {\bf \Gamma\Big[}\Xi_c^{\prime}[1/2^-] \rightarrow \Sigma_c + \bar{K}^*\rightarrow\Sigma_c+\bar{K}+\pi{\Big ]}
= {9\over2}\times { \bf\Gamma\Big[}\Xi_c^{\prime 0}[1/2^-] \rightarrow \Sigma_c^{+} + \bar{K}^0+\pi^-{\Big ]} \, ,
\\&(d11)&{ \bf\Gamma\Big[}\Xi_c^{\prime}[1/2^-] \rightarrow \Xi_c^{*} + \rho\rightarrow\Xi_c^{*}+\pi+\pi{\Big ]}
= {3\over2} \times  {\bf \Gamma\Big[}\Xi_c^{\prime 0}[1/2^-] \rightarrow \Xi_c^{*+} + \pi^0+\pi^-{\Big ]} \, ,
\\&(d12)&{ \bf\Gamma\Big[}\Xi_c^{\prime}[1/2^-] \rightarrow \Sigma_c^{*} + \bar{K}^*\rightarrow\Sigma_c^{*}+\bar{K}+\pi{\Big ]}
= {9\over2} \times  {\bf \Gamma\Big[}\Xi_c^{\prime 0}[1/2^-] \rightarrow \Sigma_c^{*+} + \bar{K}^0+\pi^-{\Big ]} \, ,
\\ &(e1)& {\bf \Gamma\Big[}\Xi_c^{\prime}[3/2^-] \rightarrow \Xi_c + \pi {\Big ]}
= {3 \over 2} \times  {\bf \Gamma\Big[}\Xi_c^{\prime 0}[3/2^-] \rightarrow \Xi_c^{+} + \pi^- {\Big ]} \, ,
\\ &(e2)& {\bf \Gamma\Big[}\Xi_c^{\prime}[3/2^-] \rightarrow \Lambda_c + \bar{K} {\Big ]}
= {\bf \Gamma\Big[}\Xi_c^{\prime 0}[3/2^-] \rightarrow \Lambda_c^{+} + K^- {\Big ]} \, ,
\\&(e3)&{\bf \Gamma\Big[}\Xi_c^{\prime}[3/2^-] \rightarrow \Xi_c^{\prime} + \pi {\Big ]}
= {3 \over 2} \times  {\bf \Gamma\Big[}\Xi_c^{\prime 0}[1/2^-] \rightarrow \Xi_c^{\prime +} + \pi^- {\Big ]} \, ,
\\&(e4)&{\bf \Gamma\Big[}\Xi_c^{\prime}[3/2^-] \rightarrow \Sigma_c + \bar{K} {\Big ]}
= 3 \times  {\bf \Gamma\Big[}\Xi_c^{\prime 0}[3/2^-] \rightarrow \Sigma_c^{+} + K^- {\Big ]} \, ,
\\ &(e5)& {\bf \Gamma\Big[}\Xi_c^{\prime}[3/2^-] \rightarrow \Xi_c^{*} + \pi {\Big ]}
= {3 \over 2} \times {\bf \Gamma\Big[}\Xi_c^{\prime0}[3/2^-] \rightarrow \Xi_c^{*+} + \pi^- {\Big ]} \, ,
\\ &(e6)& {\bf \Gamma\Big[}\Xi_c^{\prime}[3/2^-] \rightarrow \Sigma_c^{*} + \bar{K} {\Big ]}
= 3 \times {\bf \Gamma\Big[}\Xi_c^{\prime0}[3/2^-] \rightarrow \Sigma_c^{*+} + K^- {\Big ]} \, ,
\\ &(e7)& { \bf\Gamma\Big[}\Xi_c^{\prime}[3/2^-] \rightarrow \Xi_c + \rho\rightarrow\Xi_c+\pi+\pi{\Big ]}
= {3\over2} \times {\bf \Gamma\Big[}\Xi_c^{\prime 0}[3/2^-] \rightarrow \Xi_c^{+} + \pi^0+\pi^-{\Big ]} \, ,
\\ &(e8)& {\bf \Gamma\Big[} \Xi_c^{\prime}[3/2^-] \rightarrow \Lambda_c + \bar{K}^*\rightarrow\Lambda_c+\bar{K}+\pi {\Big ]}
={3\over2}\times { \bf\Gamma\Big[} \Xi_c^{\prime 0}[3/2^-] \rightarrow \Lambda_c^{+} + \bar{K}^0+\pi^- {\Big ]} \, ,
\\ &(e9)& {\bf \Gamma\Big[}\Xi_c^{\prime}[3/2^-] \rightarrow \Xi_c^{\prime} + \rho\rightarrow\Xi_c^{\prime}+\pi+\pi{\Big ]}
= {3\over2} \times  { \bf\Gamma\Big[}\Xi_c^{\prime 0}[3/2^-] \rightarrow \Xi_c^{\prime+} + \pi^0+\pi^-{\Big ]} \, ,
\\ &(e10)& {\bf \Gamma\Big[}\Xi_c^{\prime}[3/2^-] \rightarrow \Sigma_c + \bar{K}^*\rightarrow\Sigma_c^{\prime}+\bar{K}+\pi{\Big ]}
= {9\over2} \times  { \bf\Gamma\Big[}\Xi_c^{\prime 0}[3/2^-] \rightarrow \Sigma_c^{\prime+} + \bar{K}^0+\pi^-{\Big ]} \, ,
\\ &(e11)& {\bf \Gamma\Big[}\Xi_c^{\prime}[3/2^-] \rightarrow \Xi_c^{*} + \rho\rightarrow\Xi_c^{*}+\pi+\pi {\Big ]}
= {3\over2} \times {\bf \Gamma\Big[}\Xi_c^{\prime0}[3/2^-] \rightarrow \Xi_c^{*+} + \pi^0+\pi^- {\Big ]} \, ,
\\ &(e12)& {\bf \Gamma\Big[}\Xi_c^{\prime}[3/2^-] \rightarrow \Sigma_c^{*} + \bar{K}^* \rightarrow \Sigma_c^* + \bar{K} + \pi{\Big ]}
 = {9\over2} \times {\bf \Gamma\Big[}\Xi_c^{\prime0}[3/2^-] \rightarrow \Sigma_c^{*+} + \bar{K}^0+\pi^- {\Big ]} \, ,
\\ &(f1)& {\bf \Gamma\Big[}\Xi_c^{\prime}[5/2^-] \rightarrow \Xi_c + \pi {\Big ]}
= {3 \over 2} \times  {\bf \Gamma\Big[}\Xi_c^{\prime 0}[5/2^-] \rightarrow \Xi_c^{+} + \pi^- {\Big ]} \, ,
\\ &(f2)& {\bf \Gamma\Big[}\Xi_c^{\prime}[5/2^-] \rightarrow \Lambda_c + \bar{K} {\Big ]}
= {\bf \Gamma\Big[}\Xi_c^{\prime 0}[5/2^-] \rightarrow \Lambda_c^{+} + K^- {\Big ]} \, ,
\\&(f3)&{\bf \Gamma\Big[}\Xi_c^{\prime}[5/2^-] \rightarrow \Xi_c^{\prime} + \pi {\Big ]}
= {3 \over 2} \times  {\bf \Gamma\Big[}\Xi_c^{\prime 0}[1/2^-] \rightarrow \Xi_c^{\prime +} + \pi^- {\Big ]} \, ,
\\&(f4)&{\bf \Gamma\Big[}\Xi_c^{\prime}[5/2^-] \rightarrow \Sigma_c +\bar{K} {\Big ]}
= 3 \times  {\bf \Gamma\Big[}\Xi_c^{\prime 0}[5/2^-] \rightarrow \Sigma_c^{+} + K^- {\Big ]} \, ,
\\ &(f5)& {\bf \Gamma\Big[}\Xi_c^{\prime}[5/2^-] \rightarrow \Xi_c^{*} + \pi {\Big ]}
= {3 \over 2} \times {\bf \Gamma\Big[}\Xi_c^{\prime0}[5/2^-] \rightarrow \Xi_c^{*+} + \pi^- {\Big ]} \, ,
\\ &(f6)& {\bf \Gamma\Big[}\Xi_c^{\prime}[5/2^-] \rightarrow \Sigma_c^{*} + \bar{K} {\Big ]}
= 3 \times {\bf \Gamma\Big[}\Xi_c^{\prime0}[5/2^-] \rightarrow \Sigma_c^{*+} + K^- {\Big ]} \, ,
\\ &(f7)& {\bf \Gamma\Big[} \Xi_c^{\prime}[5/2^-]\rightarrow\Sigma_c^{*}+\bar{K}^*\rightarrow\Sigma_c^{*}+\bar{K}+\pi{\Big]}
={9\over2}\times{\bf\Gamma\Big[}\Xi_c^{\prime0}[5/2^-]\rightarrow\Sigma_c^{*+}+\bar{K}^0+\pi^-{\Big]}\, ,
\\&(f8)&  {\bf \Gamma\Big[}\Xi_c^{\prime}[5/2^-]\rightarrow\Xi_c^{*}+\rho\rightarrow\Xi_c^{*}+\pi+\pi{\Big]}
={3\over2}\times{\bf\Gamma\Big[}\Xi_c^{\prime0}[5/2^-]\rightarrow\Xi_c^{*+}+\pi^0+\pi^-{\Big]} \, ,
\\ &(g1)&{\bf \Gamma\Big[}\Omega_c[1/2^-] \rightarrow \Xi_c + \bar{K} {\Big ]}
= 2 \times {\bf \Gamma\Big[}\Omega_c^{0}[1/2^-] \rightarrow \Xi_c^{+} + K^- {\Big ]} \, ,
\\ &(g2)&{\bf \Gamma\Big[}\Omega_c[1/2^-] \rightarrow \Xi_c^{\prime} + \bar{K} {\Big ]}
= 2 \times {\bf \Gamma\Big[}\Omega_c^{0}[1/2^-] \rightarrow \Xi_c^{\prime+} + K^- {\Big ]} \, ,
\\ &(g3)&{\bf \Gamma\Big[}\Omega_c[1/2^-] \rightarrow \Xi_c^{*} + \bar{K} {\Big ]}
= 2 \times {\bf \Gamma\Big[}\Omega_c^{0}[1/2^-] \rightarrow \Xi_c^{*+} + K^- {\Big ]} \, ,
\\ &(g4)& {\bf \Gamma\Big[}\Omega_c[1/2^-] \rightarrow \Xi_c + \bar{K}^*\rightarrow\Xi_c+\bar{K}+\pi{\Big ]}
= 3 \times {\bf \Gamma\Big[}\Omega_c^{0}[1/2^-] \rightarrow \Xi_c^{+} + \bar{K}^0+\pi^-{\Big ]} \, ,
\\ &(g5)& {\bf \Gamma\Big[}\Omega_c[1/2^-] \rightarrow \Xi_c^{\prime} + \bar{K}^*\rightarrow\Xi_c^{\prime}+\bar{K}+\pi{\Big ]}
= 3 \times {\bf \Gamma\Big[}\Omega_c^{0}[1/2^-] \rightarrow \Xi_c^{\prime+} + \bar{K}^0+\pi^-{\Big ]} \, ,
\\ &(g6)&{\bf \Gamma\Big[}\Omega_c[1/2^-] \rightarrow \Xi_c^{*} + \bar{K}^*\rightarrow\Xi_c^{*}+\bar{K}+\pi{\Big ]}
= 3 \times {\bf \Gamma\Big[}\Omega_c^{0}[1/2^-] \rightarrow \Xi_c^{*+} + \bar{K}^0+\pi^-{\Big ]} \, ,
\\ &(h1)&{\bf \Gamma\Big[} \Omega_c[3/2^-] \rightarrow \Xi_c + \bar{K} {\Big ]}
= 2 \times {\bf \Gamma\Big[} \Omega_c^{0}[3/2^-] \rightarrow \Xi_c^{+} +K^- {\Big ]} \, ,
\\ &(h2)& {\bf \Gamma\Big[}\Omega_c[3/2^-] \rightarrow \Xi_c^{\prime} + \bar{K} {\Big ]}
= 2 \times {\bf \Gamma\Big[}\Omega_c^{0}[3/2^-] \rightarrow \Xi_c^{\prime+} + K^- {\Big ]} \, ,
\\ &(h3)& {\bf \Gamma\Big[}\Omega_c[3/2^-] \rightarrow \Xi_c^{*} + \bar{K} {\Big ]}
= 2 \times {\bf \Gamma\Big[}\Omega_c^{0}[3/2^-] \rightarrow \Xi_c^{*+} + K^- {\Big ]} \, ,
\\  &(h4)& {\bf \Gamma\Big[}\Omega_c[3/2^-] \rightarrow \Xi_c + \bar{K}^*\rightarrow\Xi_c +\bar{K}+\pi {\Big ]} = 3 \times { \bf\Gamma\Big[}\Omega_c^{0}[3/2^-] \rightarrow \Xi_c^+ + \bar{K}^0+\pi^- {\Big ]} \, ,
\\  &(h5)& {\bf \Gamma\Big[}\Omega_c[3/2^-] \rightarrow \Xi_c^{\prime} + \bar{K}^*\rightarrow\Xi_c^{\prime}+\bar{K}+\pi {\Big ]} = 3 \times {\bf \Gamma\Big[}\Omega_c^{0}[3/2^-] \rightarrow \Xi_c^{\prime+} + \bar{K}^0+\pi^- {\Big ]} \, ,
\\ &(h6)& { \bf\Gamma\Big[}\Omega_c[3/2^-] \rightarrow \Xi_c^{*} + \bar{K}^*\rightarrow\Xi_c^{*}+\bar{K}+\pi {\Big ]} = 3 \times {\bf \Gamma\Big[}\Omega_c^{0}[3/2^-] \rightarrow \Xi_c^{*+} + \bar{K}^0+\pi^- {\Big ]} \, ,
\\ &(i1)&{\bf \Gamma\Big[} \Omega_c[5/2^-] \rightarrow \Xi_c + \bar{K} {\Big ]}
= 2 \times {\bf \Gamma\Big[} \Omega_c^{0}[5/2^-] \rightarrow \Xi_c^{+} +K^- {\Big ]} \, ,
\\ &(i2)& {\bf \Gamma\Big[}\Omega_c[5/2^-] \rightarrow \Xi_c^{\prime} + \bar{K} {\Big ]}
= 2 \times {\bf \Gamma\Big[}\Omega_c^{0}[5/2^-] \rightarrow \Xi_c^{\prime+} + K^- {\Big ]} \, ,
\\ &(i3)& {\bf \Gamma\Big[}\Omega_c[5/2^-] \rightarrow \Xi_c^{*} + \bar{K} {\Big ]}
= 2 \times {\bf \Gamma\Big[}\Omega_c^{0}[5/2^-] \rightarrow \Xi_c^{*+} + K^- {\Big ]} \, ,
\\&(i4)& {\bf\Gamma\Big[}\Omega_c[5/2^-]\rightarrow\Xi_c^{*}+\bar{K}^*\rightarrow\Xi_c^{*}+\bar{K}+\pi{\Big]}
=3\times{\bf\Gamma\Big[}\Omega_c^0[5/2^-]\rightarrow\Xi_c^{*+}+\bar{K}^0+\pi^-{\Big]} \, .
\label{eq:couple28}
\end{eqnarray}
\end{widetext}
In the above expressions isospin factors are explicitly shown at right hand sides. Lagrangians of these decay channels are:
\begin{eqnarray}
&&\mathcal{L}^S_{X_c({1/2}^-) \rightarrow Y_c({1/2}^+) P}
\\ \nonumber&& ~~~~~~~~~~~~\, = g {\bar X_c}(1/2^-) Y_c(1/2^+) P \, ,
\\ &&\mathcal{L}^S_{X_c({3/2}^-) \rightarrow Y_c({3/2}^+) P}
\\ \nonumber&& ~~~~~~~~~~~~\, = g {\bar X_{c\mu}}(3/2^-)Y_c^{\mu}(3/2^+) P \, ,
\\ &&\mathcal{L}^S_{X_c({1/2}^-) \rightarrow Y_c({1/2}^+) V}
\\ \nonumber&& ~~~~~~~~~~~~\, = g {\bar X_c}(1/2^-) \gamma_\mu \gamma_5 Y_c(1/2^+) V^\mu \, ,
\\ &&\mathcal{L}^S_{X_c({1/2}^-) \rightarrow Y_c({3/2}^+) V}
\\ \nonumber&& ~~~~~~~~~~~~\, = g {\bar X_{c}}(1/2^-) Y_{c}^{\mu}(3/2^+) V_\mu \, ,
\\ &&\mathcal{L}^S_{X_c({3/2}^-) \rightarrow Y_c({1/2}^+) V}
\\ \nonumber&& ~~~~~~~~~~~~\, = g {\bar X_{c}^{\mu}}(3/2^-) Y_{c}(1/2^+) V_\mu \, ,
\\ &&\mathcal{L}^S_{X_c({3/2}^-) \rightarrow Y_c({3/2}^+) V}
\\ \nonumber&& ~~~~~~~~~~~~\, = g {\bar X_c}^{\nu}(3/2^-) \gamma_\mu \gamma_5 Y_{c\nu}(3/2^+) V^\mu \, ,
\\&& \mathcal{L}^S_{X_c({5/2}^-) \rightarrow Y_c({3/2}^+) V}
\\ \nonumber&& ~~~~~~~~~~~~\, = g {\bar X_{c}^{\mu\nu}}(5/2^-) Y_{c\mu}(3/2^+) V_\nu
\\ \nonumber &&~~~~~~~~~~~~\, + g {\bar X_{c}^{\nu\mu}}(5/2^-) Y_{c\mu}(3/2^+) V_\nu \, ,
\\ && \mathcal{L}^D_{X_c({1/2}^-) \rightarrow Y_c({3/2}^+) P}
\\ \nonumber && ~~~~~~~~~~~\, = g {\bar X_c}(1/2^-) \gamma_\mu \gamma_5 Y_{c\nu}(3/2^+) \partial^{\mu} \partial^{\nu}P \, ,
\\ && \mathcal{L}^D_{X_c({3/2}^-) \rightarrow Y_c({1/2}^+) P}
\\ \nonumber && ~~~~~~~~~~~\, = g {\bar X_{c\mu}}(3/2^-) \gamma_\nu \gamma_5 Y_{c}(1/2^+) \partial^{\mu} \partial^{\nu}P \, ,
\\ && \mathcal{L}^D_{X_c({3/2}^-) \rightarrow Y_c({3/2}^+) P}
\\ \nonumber && ~~~~~~~~~~~\, = g {\bar X_{c\mu}}(3/2^-) Y_{c\nu}(3/2^+) \partial^{\mu} \partial^{\nu}P \, ,
\\ && \mathcal{L}^D_{X_c({5/2}^-) \rightarrow Y_c({1/2}^+) P}
\\ \nonumber && ~~~~~~~~~~~\, = g {\bar X_{c\mu\nu}}(5/2^-) Y_{c}(1/2^+) \partial^{\mu} \partial^{\nu}P \, ,
\\ && \mathcal{L}^D_{X_c({5/2}^-) \rightarrow Y_c({3/2}^+) P}
\\ \nonumber && ~~~~~~~~~~~\, = g {\bar X_{c\mu\nu}}(5/2^-) \gamma_\rho \gamma_5 Y_{c}^{\mu}(3/2^+) \partial^{\nu} \partial^{\rho}P
\\ \nonumber && ~~~~~~~~~~~\, + g {\bar X_{c\mu\nu}}(5/2^-) \gamma_\rho \gamma_5 Y_{c}^{\nu}(3/2^+) \partial^{\mu} \partial^{\rho}P \, .
\end{eqnarray}
In the above expressions, the superscripts $S$ and $D$ denote $S$- and $D$-wave decays, respectively; $X_c^{(\mu\nu)}$, $Y_c^{(\mu)}$, $P$, and $V^\mu$ denote $P$-wave charmed baryons, ground-state charmed baryons, light pseudoscalar mesons, and light vector mesons, respectively.

We shall use $\Omega_c^0({3/2}^-)$ belonging to $[\mathbf{6}_F, 2, 1, \lambda]$ as an example, and study its $D$-wave decay into $\Xi_c^+(1/2^+)$ and $K^-(0^-)$ in Sec.~\ref{sec:example}. Then we shall apply the same method to systematically  investigate the four charmed baryon multiplets $[\mathbf{6}_F, 1, 0, \rho]$, $[\mathbf{6}_F, 0, 1, \lambda]$, $[\mathbf{6}_F, 1, 1, \lambda]$, and $[\mathbf{6}_F, 2, 1, \lambda]$, separately in the following subsections.

\subsection{$\Omega_c^0({3/2}^-)$ of $[\mathbf{6}_F, 2, 1, \lambda]$ decaying into  $\Xi_c^+ K^-$}
\label{sec:example}

In this subsection we use $\Omega_c^0({3/2}^-)$ belonging to $[\mathbf{6}_F, 2, 1, \lambda]$ as an example, and study its $D$-wave decay into $\Xi_c^+(1/2^+)$ and $K^-(0^-)$.

We consider the three-point correlation function,
\begin{eqnarray}
&& \Pi^\alpha(\omega, \omega^\prime)
\\ \nonumber &=& \int d^4 x~e^{-i k \cdot x}~\langle 0 | J^\alpha_{3/2,-,\Omega_c^0,2,1,\lambda}(0) \bar J_{\Xi_c^{+}}(x) | K^-(q) \rangle
\\ \nonumber &=& {1+v\!\!\!\slash\over2} G^\alpha_{\Omega_c^0[{3\over2}^-] \rightarrow   \Xi_c^{+}K^-} (\omega, \omega^\prime) \, ,
\end{eqnarray}
where $k^\prime = k + q$, $\omega = v \cdot k$, and $\omega^\prime = v \cdot k^\prime$. The currents inside this expression are~\cite{Liu:2007fg,Chen:2015kpa}:
\begin{eqnarray}
&& J^\alpha_{3/2,-,\Omega_c^0,2,1,\lambda}
\\ \nonumber &=& i \epsilon_{abc} \Big ( [\mathcal{D}_t^{\mu} s^{aT}] C \gamma_t^\nu s^b + s^{aT} C \gamma_t^\nu [\mathcal{D}_t^{\mu} s^b] \Big )
\\ \nonumber && ~~ \times \Big ( g_t^{\alpha\mu} \gamma_t^{\nu} \gamma_5 + g_t^{\alpha\nu} \gamma_t^{\mu} \gamma_5 - {2 \over 3} g_t^{\mu\nu} \gamma_t^{\alpha} \gamma_5 \Big ) h_v^c \, ,
\\ && J_{\Xi_c^{+}} = \epsilon_{abc} [u^{aT} C\gamma_{5} s^{b}] h_{v}^{c} \, ,
\end{eqnarray}
where $a \cdots c$ are color indices, $C$ is the charge-conjugation operator, $\mathcal{D}_t^{\mu}=\mathcal{D}^{\mu}-v\cdot \mathcal{D} v^{\mu}$, $\gamma_t^{\nu}=\gamma^{\nu}-v\!\!\!\slash v^{\nu}$, and $g_t^{\alpha\mu}=g^{\alpha\mu}-v^{\alpha} v^{\mu}$. These two currents couple to $\Omega_c^0({3/2}^-)$ and $\Xi_c^+(1/2^+)$, respectively.

\begin{widetext}
At the hadron level we can write $G^\alpha_{\Omega_c^0[{3\over2}^-] \rightarrow \Xi_c^{+}K^-}$ as:
\begin{eqnarray}
G^\alpha_{\Omega_c^0[{3\over2}^-] \rightarrow \Xi_c^{+}K^-} (\omega, \omega^\prime) &=& g_{\Omega_c^0[{3\over2}^-] \rightarrow \Xi_c^{+}K^-} \times { f_{\Omega_c^0[{3\over2}^-]} f_{\Xi_c^{+}} \over (\bar \Lambda_{\Omega_c^0[{3\over2}^-]} - \omega^\prime) (\bar \Lambda_{\Xi_c^{+}} - \omega)} \times \gamma \cdot q~\gamma_5~q^\alpha + \cdots \, , \label{G0C}
\end{eqnarray}
where $\cdots$ contains other possible amplitudes.

At the quark-gluon level we can calculate $G^\alpha_{\Omega_c^0[{3\over2}^-] \rightarrow \Xi_c^{+}K^-}$ using the method of operator product expansion (OPE):
\begin{eqnarray}
\label{eq:sumrule}
&& G^\alpha_{\Omega_c^0[{3\over2}^-] \rightarrow \Xi_c^{+}K^-} (\omega, \omega^\prime)
\\ \nonumber &=& \int_0^\infty dt \int_0^1 du e^{i (1-u) \omega^\prime t} e^{i u \omega t} \times 8 \times \Big (
\frac{f_K m_s u}{4\pi^2 t^2}\phi_{2;K}(u)+\frac{f_K m_s^2 u}{12(m_u+m_s)\pi^2 t^2}\phi_{3;K}^\sigma(u)
\\ \nonumber &&
+ \frac{f_K m_s^2 m_K^2 u}{48(m_u+m_s)\pi^2}\phi_{3;K}^\sigma(u)+\frac{f_K m_s u}{64\pi^2}\phi_{4;K}(u)+\frac{f_K u}{12}\langle \bar s s\rangle\phi_{2;K}(u)+\frac{f_K m_s m_K^2 u t^2}{288(m_u+m_s)}\langle s s\rangle\phi_{3;K}^\sigma(u)
\\ \nonumber &&
+ \frac{f_K u t^2}{192}\langle s s\rangle\phi_{4;K}(u)+\frac{f_K u t^2}{192}\langle g_s \bar s\sigma G s\rangle \phi_{2;K}(u)+\frac{f_K u t^4}{3072}\langle g_s \bar s \sigma G s\rangle\phi_{4;K}(u) \Big ) \times \gamma \cdot q~\gamma_5~q^\alpha
\\ \nonumber &-&
\int_0^\infty dt \int_0^1 du \int \mathcal{D} \underline{\alpha} e^{i \omega^{\prime} t(\alpha_2 + u \alpha_3)} e^{i \omega t(1 - \alpha_2 - u \alpha_3)} \times \Big (\frac{f_{3K} u}{2\pi^2 t^2}\Phi_{3;K}(\underline{\alpha})-\frac{f_{3K}}{2\pi^2 t^2}\Phi_{3;K}(\underline{\alpha})
\\ \nonumber &&
+\frac{i f_{3K} u^2 \alpha_3}{2\pi^2 t v \cdot q}\Phi_{3;K}(\underline{\alpha})+\frac{i f_{3K} u \alpha_2}{2\pi^2 t v \cdot q}\Phi_{3;K}(\underline{\alpha})-\frac{i f_{3K} u}{2\pi^2 t v \cdot q}\Phi_{3;K}(\underline{\alpha})\Big ) \times \gamma \cdot q~\gamma_5~q^\alpha + \cdots\, .
\end{eqnarray}

Then we perform the double-Borel transformation to both Eq.~(\ref{G0C}) at the hadron level and Eq.~(\ref{eq:sumrule}) at the quark-gluon level:
\begin{eqnarray}
&& g_{\Omega_c^0[{3\over2}^-] \rightarrow \Xi_c^{+}K^-} f_{\Omega_c^0[{3\over2}^-]} f_{\Xi_c^{+}} e^{- {\bar \Lambda_{\Omega_c^0[{3\over2}^-]} \over T_1}} e^{ - {\bar \Lambda_{\Xi_c^{+}} \over T_2}}
\label{eq:621lambda}
\\ \nonumber &=& 8 \times \Big ( -\frac{i f_k m_s u_0}{4\pi^2}T^3 f_2({\omega_c \over T})\phi_{2;K}(u_0)-\frac{i f_K m_K^2 u_0}{12(m_u+m_s)\pi^2}T^3 f_2({\omega_c \over T})\phi_{3;K}^\sigma(u_0)+\frac{i f_K m_s u_0}{64\pi^2}T f_0({\omega_c \over T})\phi_{4;K}(u_0)
\\ \nonumber &&
+\frac{i f_K u_0}{12}\langle \bar s s\rangle T f_0({\omega_c \over T})\phi_{2;K}(u_0)-\frac{i f_K m_s u_0}{288(m_u+m_s)}\langle \bar s s\rangle {1\over T}\phi_{3;K}^\sigma(u_0)-\frac{i f_K u_0}{192}\langle \bar s s\rangle {1\over T}\phi_{4;K}(u_0)
\\ \nonumber &&
-\frac{i f_K u_0}{192}\langle g_s \bar s\sigma G s\rangle {1\over T}\phi_{2;K}(u_0)+\frac{i f_K u_0}{3072}\langle g_s \bar s \sigma G s\rangle{1\over T^3}\phi_{4;K}(u_0) \Big )
\\ \nonumber &-&
\Big(-\frac{i f_{3K}}{2\pi^2}T^3f_2({\omega_c \over T}) \int_0^{1 \over 2} d\alpha_2 \int_{{1 \over 2}-\alpha_2}^{1-\alpha_2} d\alpha_3 ({u_0 \over \alpha_3} \Phi_{3;K}(\underline{\alpha})-{1 \over \alpha_3} \Phi_{3;K}(\underline{\alpha}))
\\ \nonumber &&
+\frac{i f_{3K}}{2\pi^2}T^3f_2({\omega_c\over T}) \int_0^{1 \over 2} d\alpha_2 \int_{{1 \over 2}-\alpha_2}^{1-\alpha_2} d\alpha_3 {1\over \alpha_3}{\partial\over\partial\alpha_3}(\alpha_3 u_0\Phi_{3;K}(\underline{\alpha})+\alpha_2\Phi_{3;K}(\underline{\alpha})-\Phi_{3;K}(\underline{\alpha}))\Big ) \, .
\end{eqnarray}
\end{widetext}
In the above expressions, $f_n(x) \equiv 1 - e^{-x} \sum_{k=0}^n {x^k \over k!}$; the parameters $\omega$ and $\omega^\prime$ are transformed to be $T_1$ and $T_2$, respectively; we choose the symmetric point $T_1 = T_2 = 2T$ so that $u_0 = {T_1 \over T_1 + T_2} = {1\over2}$; we choose $\omega_c = 1.55$~GeV to be the averaged threshold value of the $\Omega_c^0({3/2}^-)$ and $\Xi_c^{+}(1/2^+)$ mass sum rules; explicit forms of the light-cone distribution amplitudes contained in the above sum rule expressions can be found in Refs.~\cite{Ball:1998je,Ball:2006wn,Ball:2004rg,Ball:1998kk,Ball:1998sk,Ball:1998ff,Ball:2007rt,Ball:2007zt}, and more examples can be found in Appendix~\ref{sec:othersumrule}.

We extract the coupling constant from Eq.~(\ref{eq:621lambda}) to be:
\begin{eqnarray}
\nonumber g_{\Omega_c^0[{3\over2}^-] \rightarrow \Xi_c^{+}K^-} &=& 4.68~{^{+0.08}_{-0.18}}~{^{+1.82}_{-1.22}}~{^{+1.47}_{-1.11}}~{^{+2.02}_{-1.66}}~{\rm GeV}^{-2}
\\ &=& 4.68{^{+3.09}_{-2.35}}~{\rm GeV}^{-2} \, ,
\end{eqnarray}
where the uncertainties are due to the Borel mass, parameters of $\Xi_c^{+}(1/2^+)$, parameters of $\Omega_c^0({3/2}^-)$, and various QCD parameters given in Eqs.~(\ref{eq:condensates}), respectively.

Finally, we use the amplitude,
\begin{eqnarray}
&& \Gamma \left( \Omega_c^0({3/2}^-) \rightarrow \Xi_c^{+} + K^- \right)
\\ \nonumber &=& \frac{|\vec p_2|}{32\pi^2 m_0^2} \times g_{\Omega_c^0[{3\over2}^-] \rightarrow \Xi_c^{+}K^-}^2 \times p_{2,\mu}p_{2,\nu}p_{2,\rho}p_{2,\sigma}
\\ \nonumber && \times ~{\rm Tr}\Big[ \gamma^\nu\gamma_5 \left( p\!\!\!\slash_1 + m_1 \right) \gamma^\sigma \gamma_5
\\ \nonumber && \Big(g^{\rho\mu}-{\gamma^\rho\gamma^\mu\over3}-{p_{0}^{\rho}\gamma^\mu-p_{0}^{\mu}\gamma^\rho \over 3m_0} -{2p_{0}^{\rho}p_{0}^{\mu} \over 3m_0^2}  \Big) ( p\!\!\!\slash_0 + m_0 ) \Big],
\end{eqnarray}
to evaluate its partial decay width to be:
\begin{eqnarray}
\Gamma_{\Omega_c^-[{3\over2}^-] \rightarrow \Xi_c^{0}K^-}&=& 9.9{^{+17.4}_{-~7.4}}{\rm~MeV} \, .
\end{eqnarray}

In the following subsections we shall similarly investigate the four charmed baryon multiplets $[\mathbf{6}_F, 1, 0, \rho]$, $[\mathbf{6}_F, 0, 1, \lambda]$, $[\mathbf{6}_F, 1, 1, \lambda]$, and $[\mathbf{6}_F, 2, 1, \lambda]$. Some of their light-cone sum rule equations are given in Appendix~\ref{sec:othersumrule} as examples.

\subsection{The $[\mathbf{6}_F, 1, 0, \rho]$ doublet}

\begin{figure*}[hbt]
\begin{center}
\subfigure[]{
\scalebox{0.4}{\includegraphics{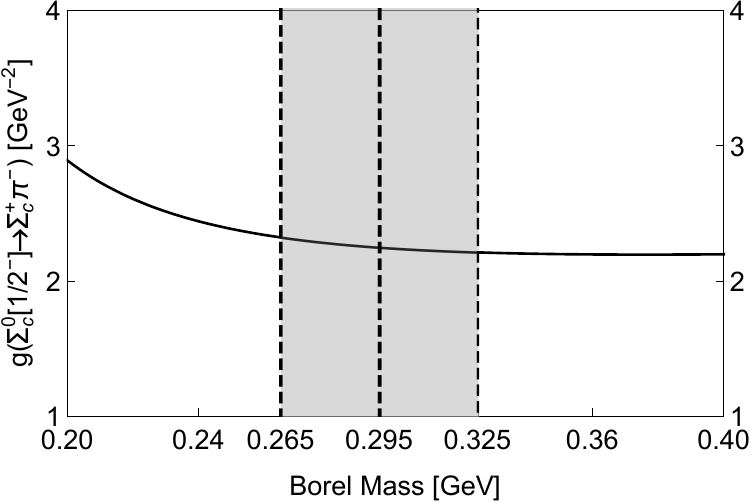}}}~~~~
\subfigure[]{
\scalebox{0.4}{\includegraphics{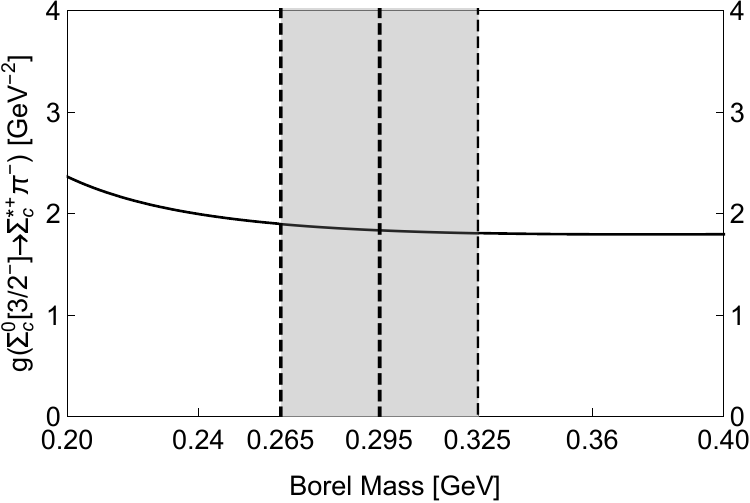}}}~~~~
\subfigure[]{
\scalebox{0.4}{\includegraphics{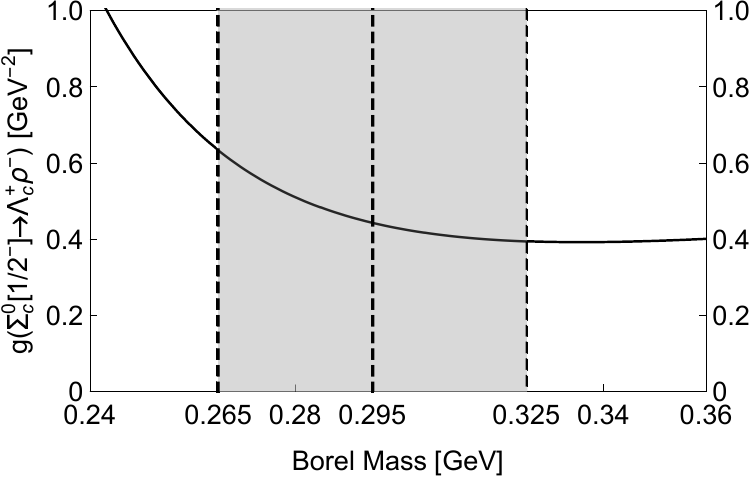}}}
\\
\subfigure[]{
\scalebox{0.4}{\includegraphics{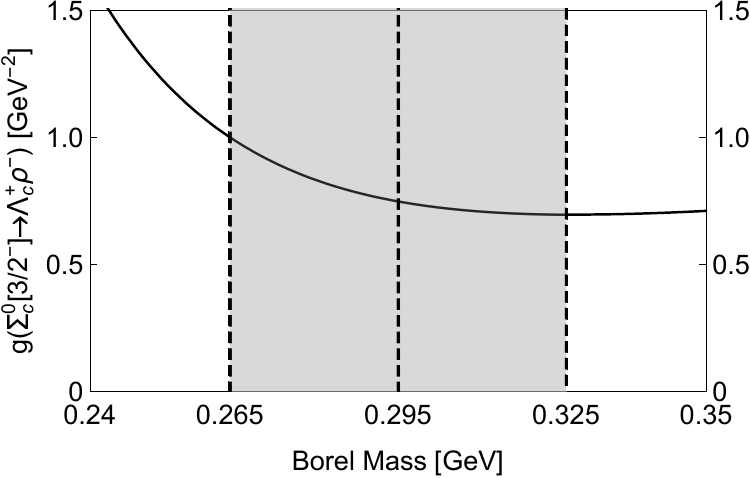}}}~~~~
\subfigure[]{
\scalebox{0.4}{\includegraphics{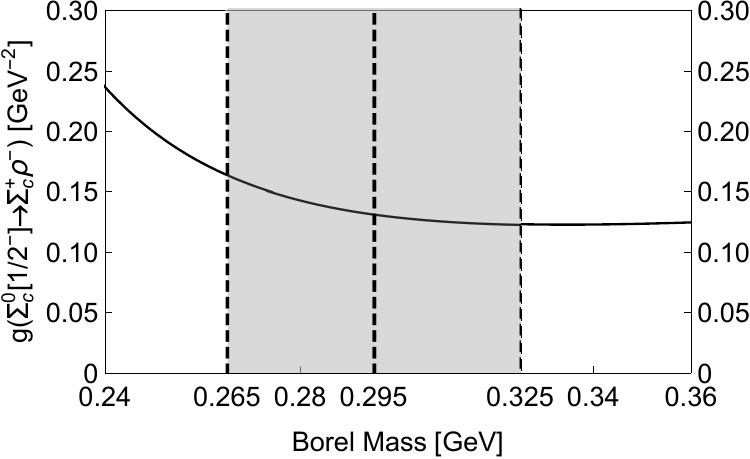}}}~~~~
\subfigure[]{
\scalebox{0.4}{\includegraphics{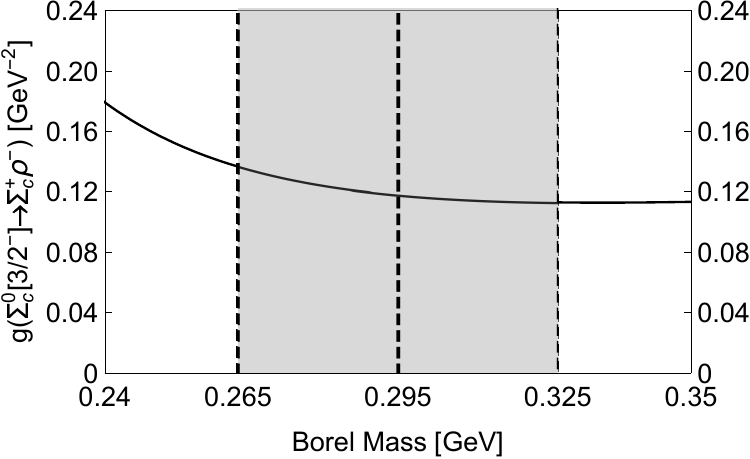}}}
\\
\subfigure[]{
\scalebox{0.4}{\includegraphics{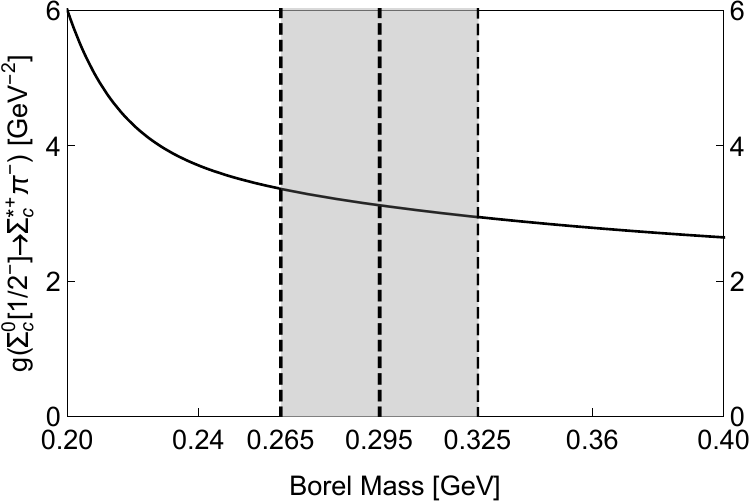}}}~~~~
\subfigure[]{
\scalebox{0.4}{\includegraphics{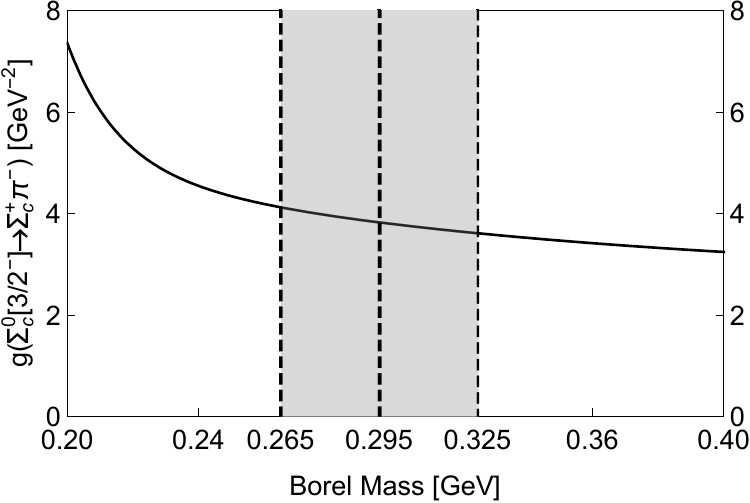}}}~~~~
\subfigure[]{
\scalebox{0.4}{\includegraphics{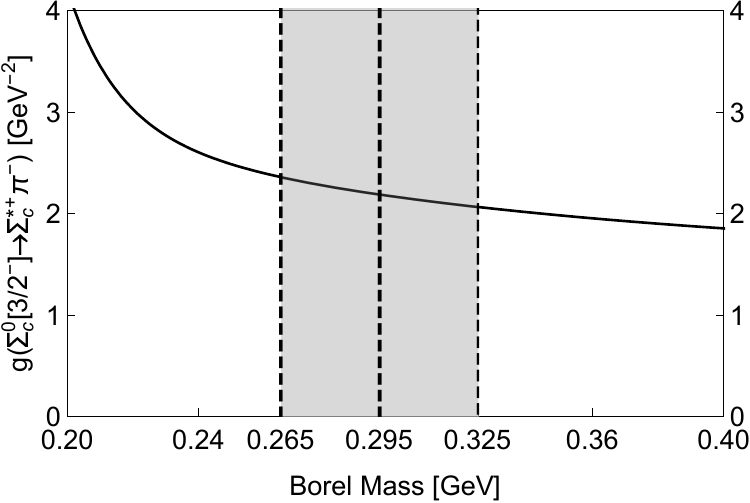}}}
\\
\subfigure[]{
\scalebox{0.4}{\includegraphics{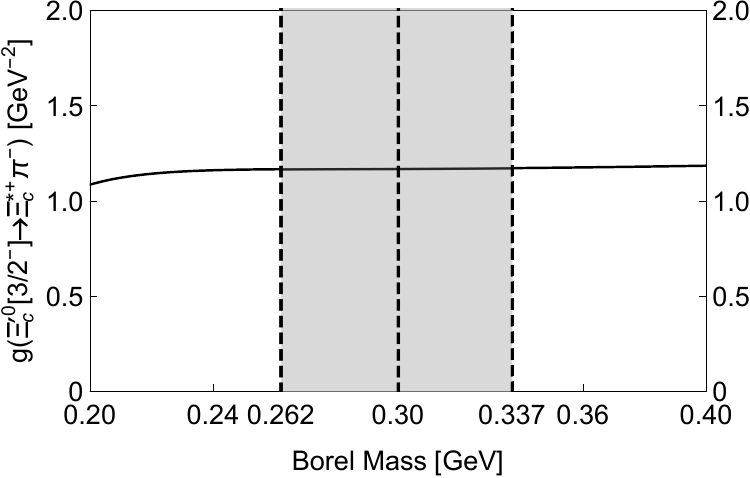}}}~~~~
\subfigure[]{
\scalebox{0.4}{\includegraphics{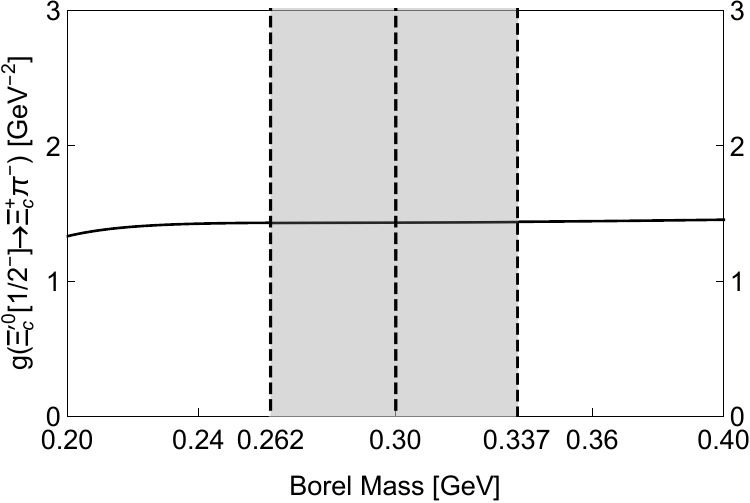}}}~~~~
\subfigure[]{
\scalebox{0.4}{\includegraphics{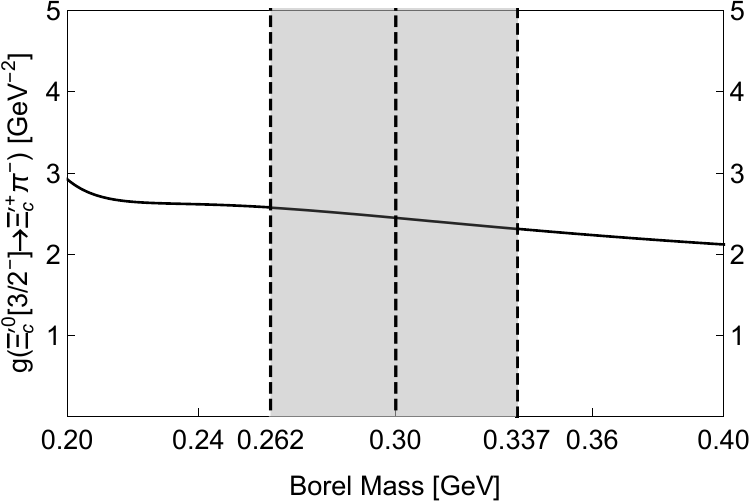}}}
\\
\subfigure[]{
\scalebox{0.3}{\includegraphics{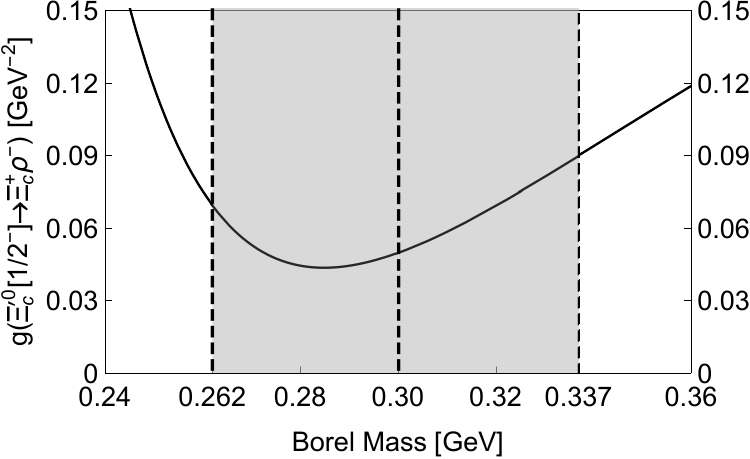}}}~~~~
\subfigure[]{
\scalebox{0.3}{\includegraphics{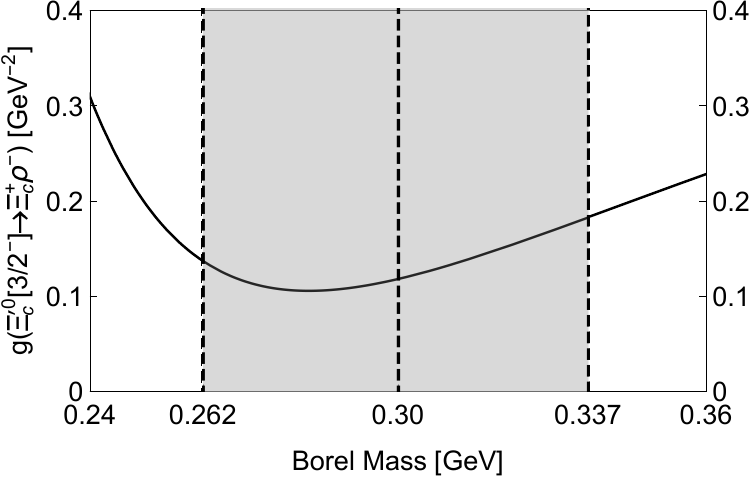}}}~~~~
\subfigure[]{
\scalebox{0.3}{\includegraphics{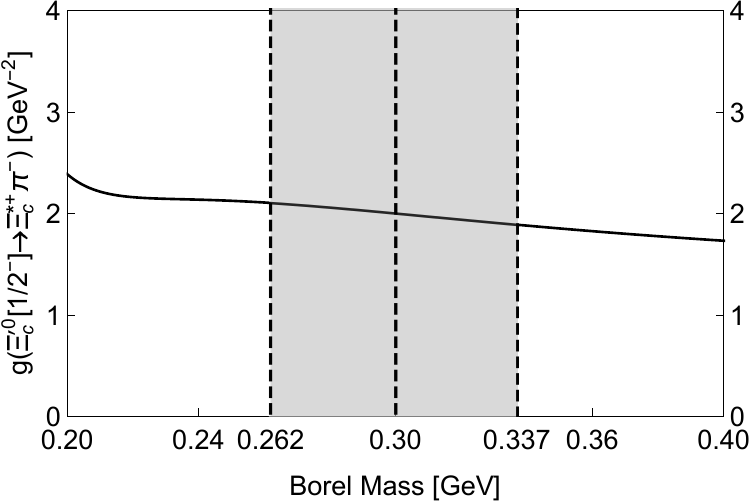}}}~~~~
\subfigure[]{
\scalebox{0.3}{\includegraphics{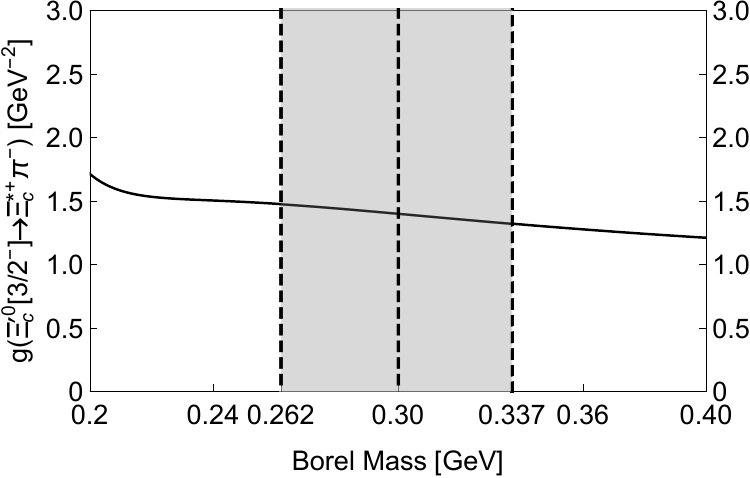}}}
\caption{Coupling constants as functions of the Borel mass $T$:
(a) $g^S_{\Sigma_c^0[{1\over2}^-] \rightarrow \Sigma_c^{+}\pi^-}$,
(b) $g^S_{\Sigma_c^0[{3\over2}^-] \rightarrow \Sigma_c^{*+}\pi^-}$,
(c) $g^S_{\Sigma_c^0[{1\over2}^-] \rightarrow \Lambda_c^{+}\rho^-}$,
(d) $g^S_{\Sigma_c^0[{3\over2}^-] \rightarrow \Lambda_c^{+}\rho^-}$,
(e) $g^S_{\Sigma_c^0[{1\over2}^-] \rightarrow \Sigma_c^{+}\rho^-}$,
(f) $g^S_{\Sigma_c^0[{3\over2}^-] \rightarrow \Sigma_c^{+}\rho^-}$,
(g) $g^D_{\Sigma_c^0[{1\over2}^-] \rightarrow \Sigma_c^{*+}\pi^-}$,
(h) $g^D_{\Sigma_c^0[{3\over2}^-] \rightarrow \Sigma_c^{+}\pi^-}$,
(i) $g^D_{\Sigma_c^0[{3\over2}^-] \rightarrow \Sigma_c^{*+}\pi^-}$,
(j) $g^S_{\Xi_c^{\prime0}[{3\over2}^-] \rightarrow \Xi_c^{*+}\pi^-}$,
(k) $g^S_{\Xi_c^{\prime0}[{1\over2}^-] \rightarrow \Xi_c^+ \pi^-}$,
(l) $g^S_{\Xi_c^{\prime0}[{3\over2}^-] \rightarrow \Xi_c^{\prime0} \pi^-}$,
(m) $g^S_{\Xi_c^{\prime0}[{1\over2}^-] \rightarrow \Xi_c^{0} \rho^-}$,
(n) $g^S_{\Xi_c^{\prime0}[{3\over2}^-] \rightarrow \Xi_c^{0} \rho^-}$,
(o) $g^D_{\Xi_c^{\prime0}[{1\over2}^-] \rightarrow \Xi_c^{*0} \pi^-}$,
and
(p) $g^D_{\Xi_c^{\prime0}[{3\over2}^-] \rightarrow \Xi_c^{*0} \pi^-}$.
The charmed baryon doublet $[\mathbf{6}_F, 1, 0, \rho]$ is investigated here.
\label{fig:610rho}}
\end{center}
\end{figure*}

The $[\mathbf{6}_F, 1, 0, \rho]$ doublet contains altogether six charmed baryons: $\Sigma_c({1\over2}^-/{3\over2}^-)$, $\Xi_c({1\over2}^-/{3\over2}^-)$, and $\Omega_c({1\over2}^-/{3\over2}^-)$. We study their $S$- and $D$-wave decays into ground-state charmed baryons together with light pseudoscalar and vector mesons. We derive the following non-zero coupling constants:
\begin{eqnarray}
\nonumber &(a2)& g^S_{\Sigma_c[{1\over2}^-]\to \Sigma_c[{1\over2}^+] \pi} = 2.25^{+1.41}_{-1.03}~{\rm GeV}^{-2} \, ,
\\
\nonumber &(b3)& g^S_{\Sigma_c[{3\over2}^-]\to \Sigma_c^{*}[{3\over2}^+] \pi} = 1.83^{+1.15}_{-0.84}~{\rm GeV}^{-2} \, ,
\\
\nonumber &(a4)& g^S_{\Sigma_c[{1\over2}^-]\to \Lambda_c[{1\over2}^+] \rho} =0.44~{\rm GeV}^{-2}\, ,
\\
\nonumber &(a5)&  g^S_{\Sigma_c[{1\over2}^-]\to \Sigma_c[{1\over2}^+] \rho} =0.13~{\rm GeV}^{-2}\, ,
\\
\nonumber &(a6)&  g^S_{\Sigma_c[{1\over2}^-]\to \Sigma_c^{*}[{3\over2}^+] \rho} =0.08~{\rm GeV}^{-2}\, ,
\\
\nonumber &(b4)&  g^S_{\Sigma_c[{3\over2}^-]\to \Lambda_c[{1\over2}^+] \rho} =0.75~{\rm GeV}^{-2}\, ,
\\
\nonumber &(b5)&  g^S_{\Sigma_c[{3\over2}^-]\to \Sigma_c[{1\over2}^+] \rho} =0.12~{\rm GeV}^{-2}\, ,
\\
\nonumber &(b6)&  g^S_{\Sigma_c[{3\over2}^-]\to \Sigma_c^{*}[{3\over2}^+] \rho} =0.11~{\rm GeV}^{-2}\, ,
\\
\nonumber &(a3)& g^D_{\Sigma_c[{1\over2}^-]\to \Sigma_c^{*}[{3\over2}^+] \pi} = 3.12^{+1.98}_{-1.48}~{\rm GeV}^{-2}\, ,
\\
\nonumber &(b2)& g^D_{\Sigma_c[{3\over2}^-]\to \Sigma_c[{1\over2}^+] \pi}= 3.82^{+2.36}_{-1.80}~{\rm GeV}^{-2} \, ,
\\
\nonumber &(b3)& g^D_{\Sigma_c[{3\over2}^-]\to \Sigma_c^{*}[{3\over2}^+] \pi} = 2.19^{+1.38}_{-1.03}~{\rm GeV}^{-2}\, ,
\\
\nonumber &(d2)& g^S_{\Xi_c^{\prime}[{1\over2}^-]\to \Xi_c^{\prime}[{1\over2}^+]\pi} = 1.43^{+0.85}_{-0.63}~{\rm GeV}^{-2}\, ,
\\
\nonumber &(d4)& g^S_{\Xi_c^{\prime}[{1\over2}^-]\to \Sigma_c[{1\over2}^+]\bar K} = 2.06~{\rm GeV}^{-2}\, ,
\\
\nonumber &(e5)& g^S_{\Xi_c^{\prime}[{3\over2}^-]\to \Xi_c^{*}[{3\over2}^+]\pi} = 1.17^{+0.69}_{-0.51}~{\rm GeV}^{-2}\, ,
\\
\nonumber &(e6)& g^S_{\Xi_c^{\prime}[{3\over2}^-]\to \Sigma_c^{*}[{3\over2}^+]\bar K} = 1.68~{\rm GeV}^{-2}\, ,
\\
\nonumber &(d7)& g^S_{\Xi_c^{\prime}[{1\over2}^-]\to \Xi_c[{1\over2}^+]\rho} = 0.05~{\rm GeV}^{-2}\, ,
\\
\nonumber &(d8)& g^S_{\Xi_c^{\prime}[{1\over2}^-]\to \Lambda_c[{1\over2}^+]\bar K^*} = 0.48~{\rm GeV}^{-2}\, ,
\\
\nonumber &(d9)& g^S_{\Xi_c^{\prime}[{1\over2}^-]\to \Xi_c^{\prime}[{1\over2}^+]\rho} = 0.003~{\rm GeV}^{-2}\, ,
\\
\nonumber &(d10)& g^S_{\Xi_c^{\prime}[{1\over2}^-]\to \Sigma_c[{1\over2}^+]\bar K^*} = 0.33~{\rm GeV}^{-2}\, ,
\\
\nonumber &(d11)& g^S_{\Xi_c^{\prime}[{1\over2}^-]\to \Xi_c^{*}[{3\over2}^+]\rho} = 0.002~{\rm GeV}^{-2}\, ,
\\
\nonumber &(d12)& g^S_{\Xi_c^{\prime}[{1\over2}^-]\to \Sigma_c^{*}[{3\over2}^+]\bar K^*} = 0.19~{\rm GeV}^{-2}\, ,
\\
\nonumber &(e7)& g^S_{\Xi_c^{\prime}[{3\over2}^-]\to \Xi_c[{1\over2}^+]\rho} = 0.12~{\rm GeV}^{-2}\, ,
\\
          &(e8)& g^S_{\Xi_c^{\prime}[{3\over2}^-]\to \Lambda_c[{1\over2}^+]\bar K^*} = 0.78~{\rm GeV}^{-2}\, ,
\\
\nonumber &(e9)& g^S_{\Xi_c^{\prime}[{3\over2}^-]\to \Xi_c^{\prime}[{1\over2}^+]\rho} = 0.002~{\rm GeV}^{-2}\, ,
\\
\nonumber &(e10)& g^S_{\Xi_c^{\prime}[{3\over2}^-]\to \Sigma_c[{1\over2}^+]\bar K^*} = 0.23~{\rm GeV}^{-2}\, ,
\\
\nonumber &(e11)& g^S_{\Xi_c^{\prime}[{3\over2}^-]\to \Xi_c^{*}[{3\over2}^+]\rho} = 0.003~{\rm GeV}^{-2}\, ,
\\
\nonumber &(e12)& g^S_{\Xi_c^{\prime}[{3\over2}^-]\to \Sigma_c^{*}[{3\over2}^+]\bar K^*} = 0.27~{\rm GeV}^{-2}\, ,
\\
\nonumber &(d5)& g^D_{\Xi_c^{\prime}[{1\over2}^-]\to \Xi_c^{*}[{3\over2}^+]\pi} = 2.00^{+1.21}_{-0.94}~{\rm GeV}^{-2}\, ,
\\
\nonumber &(d6)& g^D_{\Xi_c^{\prime}[{1\over2}^-]\to \Sigma_c^{*}[{3\over2}^+]\bar K} = 2.16~{\rm GeV}^{-2}\, ,
\\
\nonumber &(e3)& g^D_{\Xi_c^{\prime}[{3\over2}^-]\to \Xi_c^{\prime}[{1\over2}^+]\pi} = 2.45^{+1.47}_{-1.14}~{\rm GeV}^{-2}\, ,
\\
\nonumber &(e4)& g^D_{\Xi_c^{\prime}[{3\over2}^-]\to \Sigma_c[{1\over2}^+]\bar K} = 2.64~{\rm GeV}^{-2}\, ,
\\
\nonumber &(e5)& g^D_{\Xi_c^{\prime}[{3\over2}^-]\to \Xi_c^{*}[{3\over2}^+]\pi} = 1.40^{+0.84}_{-0.65}~{\rm GeV}^{-2}\, ,
\\
\nonumber &(e6)& g^D_{\Xi_c^{\prime}[{3\over2}^-]\to \Sigma_c^{*}[{3\over2}^+]\bar K} = 1.52~{\rm GeV}^{-2}\, ,
\\
\nonumber &(g2)& g^S_{\Omega_c[{1\over2}^-]\to \Xi_c^{\prime}[{1\over2}^+]\bar K}= 2.57~{\rm GeV}^{-2} \, ,
\\
\nonumber &(h3)& g^S_{\Omega_c[{3\over2}^-]\to \Xi_c^{*}[{3\over2}^+]\bar K}= 2.10~{\rm GeV}^{-2} \, ,
\\
\nonumber &(g4)& g^S_{\Omega_c[{1\over2}^-]\to \Xi_c[{1\over2}^+]\bar K^*}= 0.15~{\rm GeV}^{-2} \, ,
\\
\nonumber &(g5)& g^S_{\Omega_c[{1\over2}^-]\to \Xi_c^{\prime}[{1\over2}^+]\bar K^*}= 0.41~{\rm GeV}^{-2} \, ,
\\
\nonumber &(g6)& g^S_{\Omega_c[{1\over2}^-]\to \Xi_c^{*}[{3\over2}^+]\bar K^*}= 0.24~{\rm GeV}^{-2} \, ,
\\
\nonumber &(f4)& g^S_{\Omega_c[{3\over2}^-]\to \Xi_c[{1\over2}^+]\bar K^*}= 0.30~{\rm GeV}^{-2} \, ,
\\
\nonumber &(f5)& g^S_{\Omega_c[{3\over2}^-]\to \Xi_c^{\prime}[{1\over2}^+]\bar K^*}= 0.29~{\rm GeV}^{-2} \, ,
\\
\nonumber &(f6)& g^S_{\Omega_c[{3\over2}^-]\to \Xi_c^{*}[{3\over2}^+]\bar K^*}= 0.33~{\rm GeV}^{-2} \, ,
\\
\nonumber &(g3)& g^D_{\Omega_c[{1\over2}^-]\to \Xi_c^{*}[{3\over2}^+]\bar K}= 2.55~{\rm GeV}^{-2} \, ,
\\
\nonumber &(h2)& g^D_{\Omega_c[{3\over2}^-]\to \Xi_c^{\prime}[{1\over2}^+]\bar K}= 3.12~{\rm GeV}^{-2} \, ,
\\
\nonumber &(h3)& g^D_{\Omega_c[{3\over2}^-]\to \Xi_c^{*}[{3\over2}^+]\bar K}= 1.79~{\rm GeV}^{-2} \, .
\end{eqnarray}
Some of these coupling constants are shown in Fig.~\ref{fig:610rho} as functions of the Borel mass $T$. We further use these coupling constants to derive the following decay channels that are kinematically allowed:
\begin{eqnarray}
\nonumber  &(a2)& \Gamma^S_{\Sigma_c[{1\over2}^-]\to \Sigma_c[{1\over2}^+] \pi} =380^{+630}_{-270}~{\rm MeV} \, ,
\\
\nonumber  &(b3)& \Gamma^S_{\Sigma_c[{3\over2}^-]\to \Sigma_c^{*}[{3\over2}^+] \pi} =220^{+360}_{-150}~{\rm MeV} \, ,
\\
\nonumber  &(a4)& \Gamma^S_{\Sigma_c[{1\over2}^-]\to \Lambda_c[{1\over2}^+] \rho \to \Lambda_c[{1\over2}^+] \pi\pi} =0.06~{\rm MeV} \, ,
\\
\nonumber  &(a5)& \Gamma^S_{\Sigma_c[{1\over2}^-]\to \Sigma_c[{1\over2}^+] \rho \to\Sigma_c[{1\over2}^+] \pi\pi} =3\times 10^{-5}~{\rm MeV} \, ,
\\
\nonumber &(b4)& \Gamma^S_{\Sigma_c[{3\over2}^-]\to \Lambda_c[{1\over2}^+] \rho \to\Lambda_c[{1\over2}^+]\pi\pi} =0.08~{\rm MeV} \, ,
\\
\nonumber  &(b5)& \Gamma^S_{\Sigma_c[{3\over2}^-]\to \Sigma_c[{1\over2}^+] \rho \to \Sigma_c[{1\over2}^+]\pi\pi} =4\times 10^{-5}~{\rm MeV} \, ,
\\
\nonumber &(a3)& \Gamma^D_{\Sigma_c[{1\over2}^-]\to \Sigma_c^{*}[{3\over2}^+] \pi} = 0.82^{+1.37}_{-0.59}~{\rm MeV} \, ,
\\
\nonumber &(b2)& \Gamma^D_{\Sigma_c[{3\over2}^-]\to \Sigma_c[{1\over2}^+] \pi} = 3.1^{+4.6}_{-2.3}~{\rm MeV} \, ,
\\
\nonumber  &(b3)& \Gamma^D_{\Sigma_c[{3\over2}^-]\to \Sigma_c^{*}[{3\over2}^+] \pi} =0.21^{+0.34}_{-0.15}~{\rm MeV} \, ,
\\
          &(d2)& \Gamma^S_{\Xi_c^{\prime}[{1\over2}^-]\to\Xi_c^{\prime}[{1\over2}^+]\pi} = 110^{+170}_{-~80}~{\rm MeV} \, ,
\\
\nonumber &(e5)& \Gamma^S_{\Xi_c^{\prime}[{3\over2}^-]\to\Xi_c^{*}[{3\over2}^+]\pi} = 58^{+88}_{-39}~{\rm MeV} \, ,
\\
\nonumber &(d7)& \Gamma^S_{\Xi_c^{\prime}[{1\over2}^-]\to\Xi_c[{1\over2}^+]\rho\to\Xi_c[{1\over2}^+]\pi\pi} = 2\times 10^{-4}~{\rm MeV} \, ,
\\
\nonumber &(d9)& \Gamma^S_{\Xi_c^{\prime}[{1\over2}^-]\to\Xi_c^{\prime}[{1\over2}^+]\rho\to\Xi_c^{\prime}[{1\over2}^+]\pi\pi} = 5\times 10^{-9}~{\rm MeV} \, ,
\\
\nonumber &(e7)& \Gamma^S_{\Xi_c^{\prime}[{3\over2}^-]\to\Xi_c[{1\over2}^+]\rho\to\Xi_c[{1\over2}^+]\pi\pi} = 5\times 10^{-4}~{\rm MeV} \, ,
\\
\nonumber &(e9)& \Gamma^S_{\Xi_c^{\prime}[{3\over2}^-]\to\Xi_c^{\prime}[{1\over2}^+]\rho\to\Xi_c^{\prime}[{1\over2}^+]\pi\pi} = 2\times 10^{-9}~{\rm MeV} \, ,
\\
\nonumber &(d5)& \Gamma^D_{\Xi_c^{\prime}[{1\over2}^-]\to\Xi_c^{*}[{3\over2}^+]\pi} = 0.15^{+0.23}_{-0.11}~{\rm MeV} \, ,
\\
\nonumber &(e3)& \Gamma^D_{\Xi_c^{\prime}[{3\over2}^-]\to\Xi_c^{\prime}[{1\over2}^+]\pi} = 0.63^{+0.99}_{-0.45}~{\rm MeV} \, ,
\\
\nonumber &(e5)& \Gamma^D_{\Xi_c^{\prime}[{3\over2}^-]\to\Xi_c^{*}[{3\over2}^+]\pi} = 0.03~{\rm MeV} \, .
\end{eqnarray}
We summarize the above results in Table~\ref{tab:decay610rho}.

\subsection{The $[\mathbf{6}_F,0,1,\lambda]$ singlet}

\begin{figure*}[htb]
\begin{center}
\subfigure[]{
\scalebox{0.4}{\includegraphics{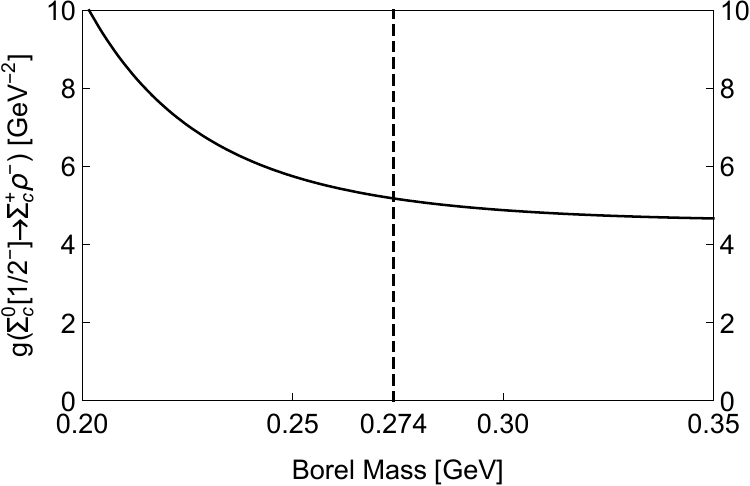}}}~~~~~
\subfigure[]{
\scalebox{0.4}{\includegraphics{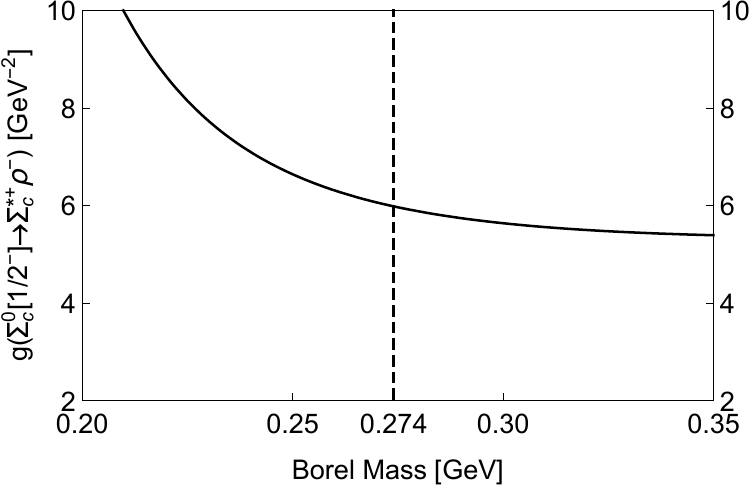}}}~~~~~
\subfigure[]{
\scalebox{0.4}{\includegraphics{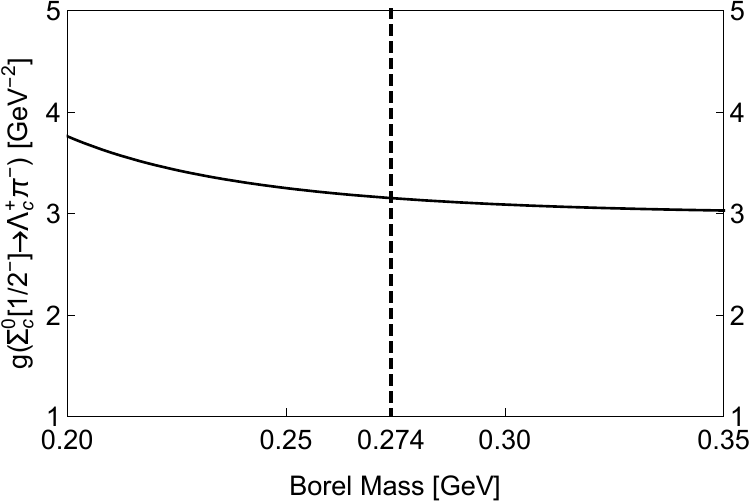}}}
\\
\subfigure[]{
\scalebox{0.5}{\includegraphics{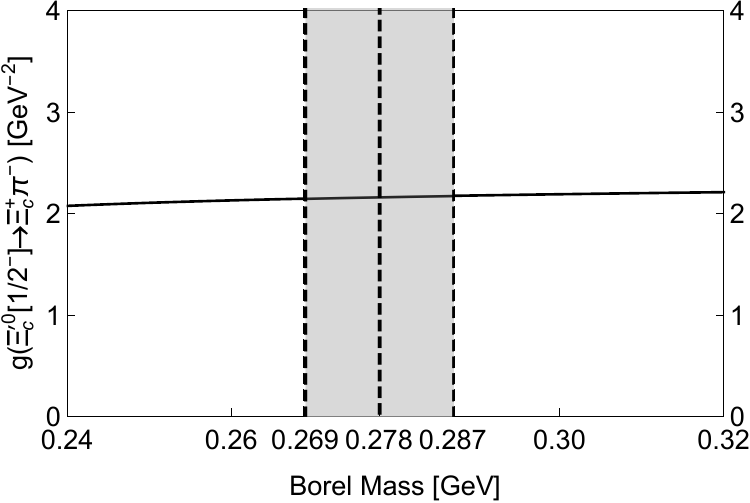}}}~~~~~
\subfigure[]{
\scalebox{0.5}{\includegraphics{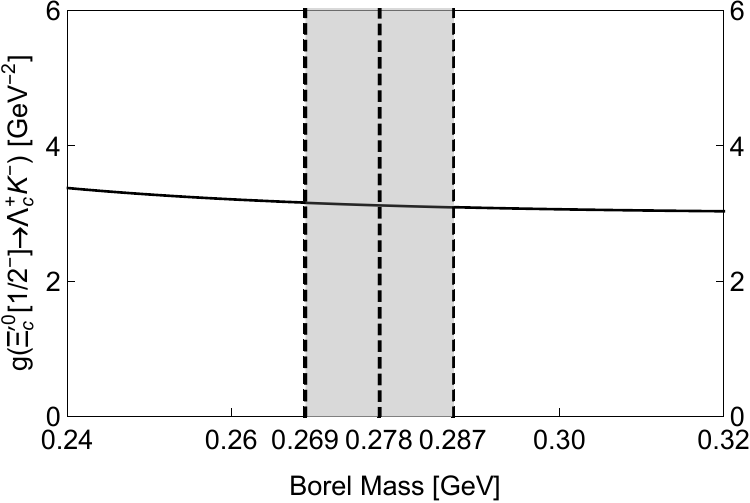}}}
\\
\subfigure[]{
\scalebox{0.5}{\includegraphics{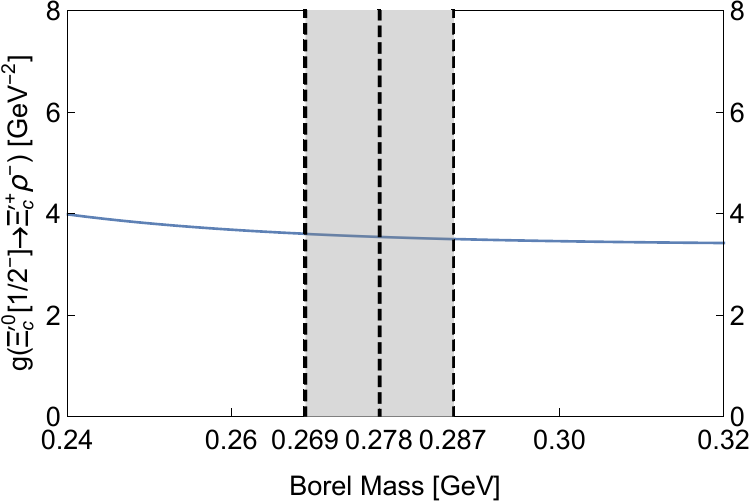}}}~~~~~
\subfigure[]{
\scalebox{0.5}{\includegraphics{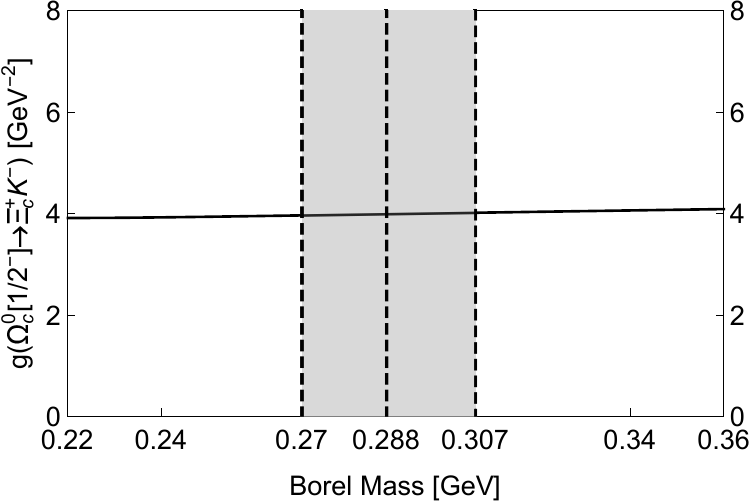}}}
\end{center}
\caption{Coupling constants as functions of the Borel mass $T$:
(a) $g^S_{\Sigma_c^0[{1\over2}^-] \rightarrow \Sigma_c^{+}\rho^-}$,
(b) $g^S_{\Sigma_c^0[{1\over2}^-] \rightarrow \Sigma_c^{*+}\rho^-}$,
(c) $g^S_{\Sigma_c^0[{1\over2}^-] \rightarrow \Lambda_c^{+}\pi^-}$,
(d) $g^S_{\Xi_c^{\prime0}[{1\over2}^-] \rightarrow \Xi_c^{+}\pi^-}$,
(e) $g_{\Xi_c^{\prime-}[{1\over2}^-] \rightarrow \Lambda_c^{0} K^-}$,
(f) $g_{\Xi_c^{\prime0}[{1\over2}^-] \rightarrow \Xi_c^{\prime0} \rho^-}$,
and
(g) $g_{\Omega_c^{0}[{1\over2}^-] \rightarrow \Xi_c^{0} K^-}$. The charmed baryon doublet $[\mathbf{6}_F, 0, 1, \lambda]$ is investigated here.
\label{fig:601lambda}}
\end{figure*}

The $[\mathbf{6}_F, 0, 1, \lambda]$ doublet contains altogether three charmed baryons: $\Sigma_c({1\over2}^-)$, $\Xi_c({1\over2}^-)$, and $\Omega_c({1\over2}^-)$. We study their $S$- and $D$-wave decays into ground-state charmed baryons together with light pseudoscalar and vector mesons. We derive the following non-zero coupling constants:
\begin{eqnarray}
\nonumber &(a1)&  g^S_{\Sigma_c[{1\over2}^-]\to \Lambda_c[{1\over2}^+] \pi} = 3.15^{+1.74}_{-1.35}~{\rm GeV}^{-2} \, ,
\\
\nonumber &(a5)& g^S_{\Sigma_c[{1\over2}^-]\to \Sigma_c[{1\over2}^+] \rho} = 5.18^{+2.59}_{-2.16}~{\rm GeV}^{-2} \, ,
\\
\nonumber &(a6)&  g^S_{\Sigma_c[{1\over2}^-]\to \Sigma_c^{*}[{3\over2}^+] \rho} = 5.98~{\rm GeV}^{-2} \, ,
\\
\nonumber &(d1)& g^S_{\Xi_c^{\prime}[{1\over2}^-]\to\Xi_c[{1\over2}^+]\pi} = 2.16^{+1.27}_{-0.96}~{\rm GeV}^{-2} \, ,
\\
\nonumber &(d3)& g^S_{\Xi_c^{\prime}[{1\over2}^-]\to\Lambda_c[{1\over2}^+]\bar K} = 3.12^{+1.85}_{-1.36}~{\rm GeV}^{-2} \, ,
\\
\nonumber &(d9)& g^S_{\Xi_c^{\prime}[{1\over2}^-]\to\Xi_c^{\prime}[{1\over2}^+]\rho} = 3.54~{\rm GeV}^{-2} \, ,
\\
          &(d10)& g^S_{\Xi_c^{\prime}[{1\over2}^-]\to\Sigma_c[{1\over2}^+]\bar K^*} = 4.13~{\rm GeV}^{-2} \, ,
\\
\nonumber &(d11)& g^S_{\Xi_c^{\prime}[{1\over2}^-]\to\Xi_c^{*}[{3\over2}^+]\rho} = 4.08~{\rm GeV}^{-2} \, ,
\\
\nonumber &(d12)& g^S_{\Xi_c^{\prime}[{1\over2}^-]\to\Sigma_c^{*}[{3\over2}^+]\bar K^*} = 4.83~{\rm GeV}^{-2}\, ,
\\
\nonumber &(g1)& g^D_{\Omega_c[{1\over2}^-]\to \Xi_c[{1\over2}^+]\bar K} = 3.98^{+2.40}_{-1.75}~{\rm GeV}^{-2} \, ,
\\
\nonumber &(g5)& g^D_{\Omega_c[{1\over2}^-]\to \Xi_c^{\prime}[{1\over2}^+]\bar K^*} = 5.21~{\rm GeV}^{-2} \, ,
\\
\nonumber &(g6)& g^D_{\Omega_c[{1\over2}^-]\to \Xi_c^{*}[{3\over2}^+]\bar K^*} = 6.06~{\rm GeV}^{-2} \, .
\end{eqnarray}
Some of these coupling constants are shown in Fig.~\ref{fig:601lambda} as functions of the Borel mass $T$. We further use these coupling constants to derive the following decay channels that are kinematically allowed:
\begin{eqnarray}
\nonumber &(a1)& \Gamma^S_{\Sigma_c[{1\over2}^-]\to \Lambda_c[{1\over2}^+] \pi} = 610{^{+860}_{-410}}~{\rm MeV} \, ,
\\
\nonumber &(a5)& \Gamma^S_{\Sigma_c[{1\over2}^-]\to \Sigma_c[{1\over2}^+] \rho \to\Sigma_c[{1\over2}^+]\pi\pi} = 1.1{^{+1.4}_{-0.7}}~{\rm MeV} \, ,
\\
\nonumber &(a6)& \Gamma^S_{\Sigma_c[{1\over2}^-]\to \Sigma_c^{*}[{3\over2}^+] \rho \to\Sigma_c^{*}[{3\over2}^+]\pi\pi} = 0.03~{\rm MeV} \, ,
\\
\nonumber &(d1)& \Gamma^S_{\Xi_c^{\prime}[{1\over2}^-]\to\Xi_c[{1\over2}^+]\pi} = 360^{+550}_{-250}~{\rm MeV} \, ,
\\
         &(d3)& \Gamma^S_{\Xi_c^{\prime}[{1\over2}^-]\to\Lambda_c[{1\over2}^+]\bar K} = 400^{+610}_{-270}~{\rm MeV} \, ,
\\
\nonumber &(d9)& \Gamma^S_{\Xi_c^{\prime}[{1\over2}^-]\to\Xi_c^{\prime}[{1\over2}^+]\rho\to\Xi_c^{\prime}[{1\over2}^+]\pi\pi} = 0.03~{\rm MeV} \, ,
\\
\nonumber &(a5)& \Gamma^S_{\Omega_c[{1\over2}^-]\to \Xi_c[{1\over2}^+]\bar K} = 980{^{+1530}_{-~670}}~{\rm MeV} \, .
\end{eqnarray}
We summarize the above results in Table~\ref{tab:decay601lambda}.

\subsection{The $[\mathbf{6}_F, 1, 1, \lambda]$ doublet}

\begin{figure*}[htb]
\begin{center}
\subfigure[]{
\scalebox{0.3}{\includegraphics{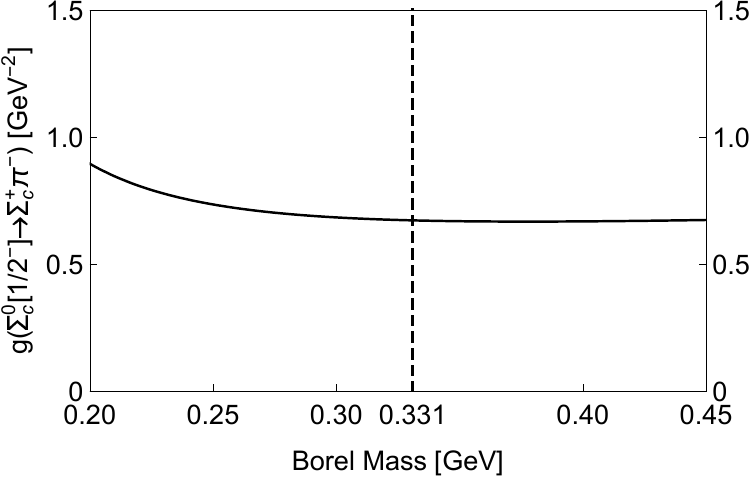}}}~~~~~
\subfigure[]{
\scalebox{0.3}{\includegraphics{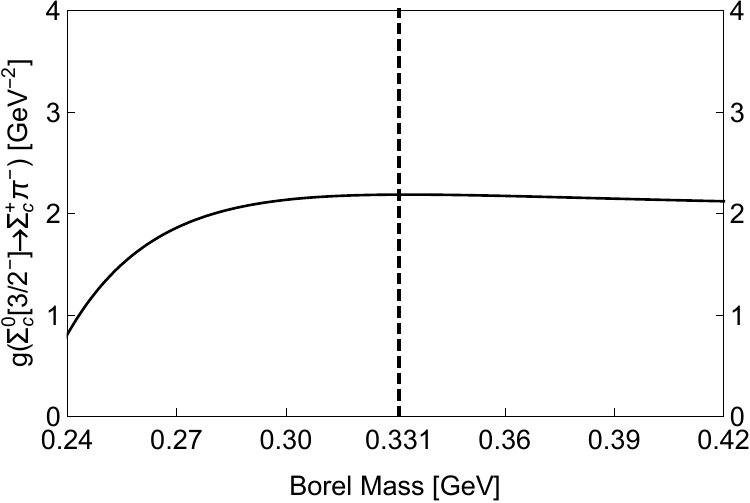}}}~~~~~
\subfigure[]{
\scalebox{0.3}{\includegraphics{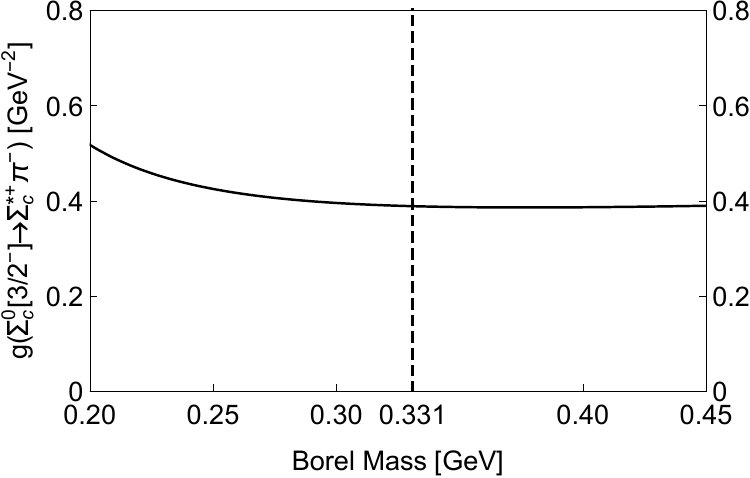}}}~~~~~
\subfigure[]{
\scalebox{0.3}{\includegraphics{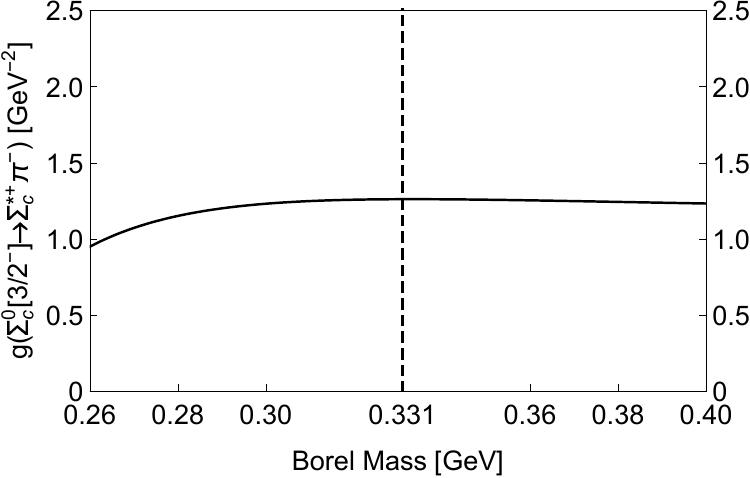}}}
\\
\subfigure[]{
\scalebox{0.3}{\includegraphics{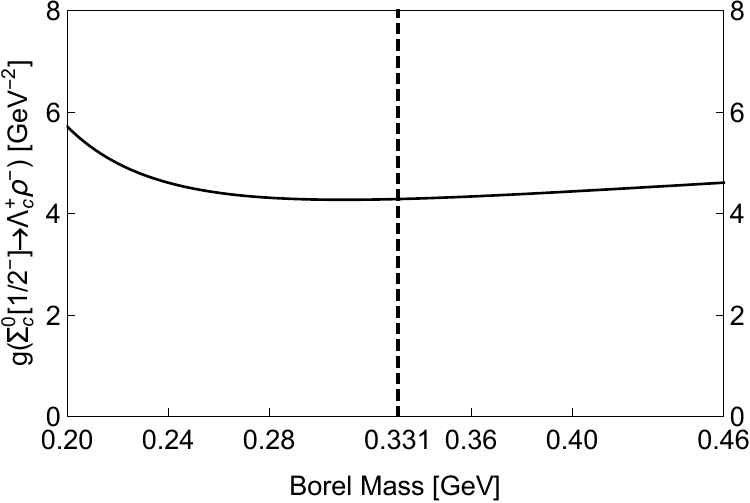}}}~~~~~
\subfigure[]{
\scalebox{0.3}{\includegraphics{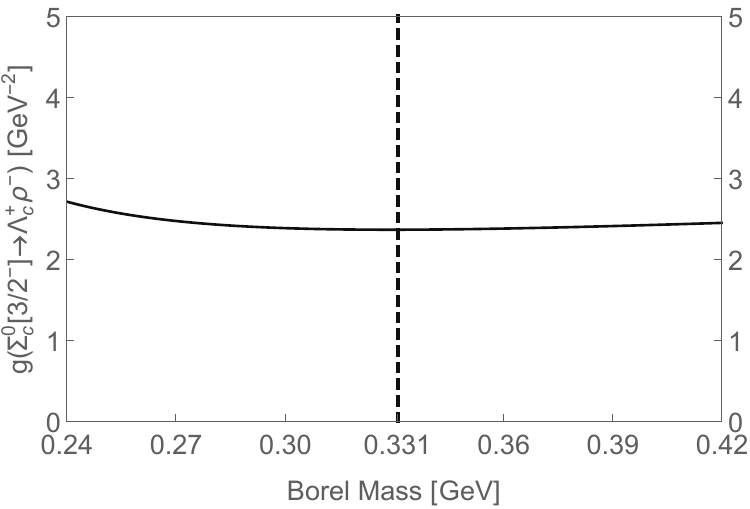}}}~~~~~
\subfigure[]{
\scalebox{0.3}{\includegraphics{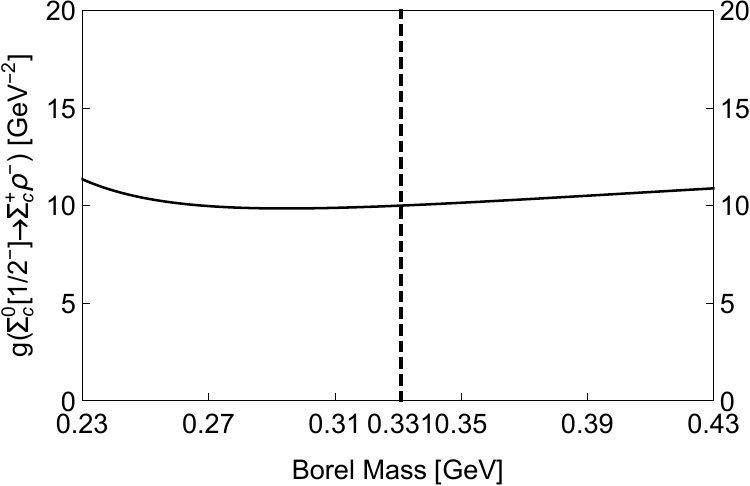}}}~~~~~
\subfigure[]{
\scalebox{0.3}{\includegraphics{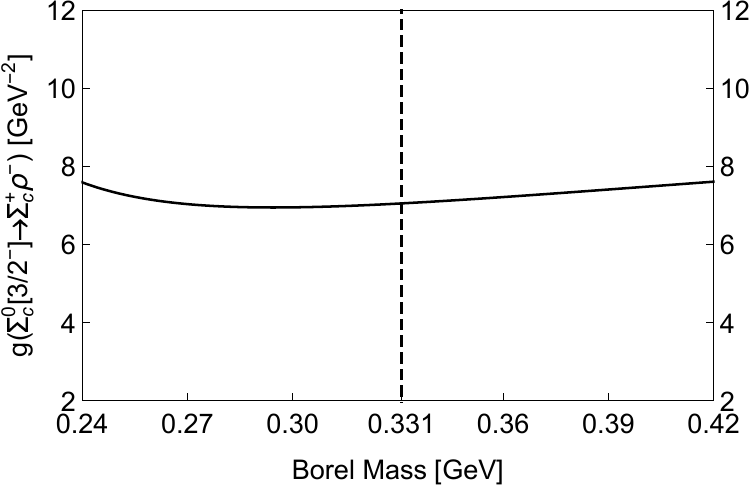}}}
\\
\subfigure[]{
\scalebox{0.3}{\includegraphics{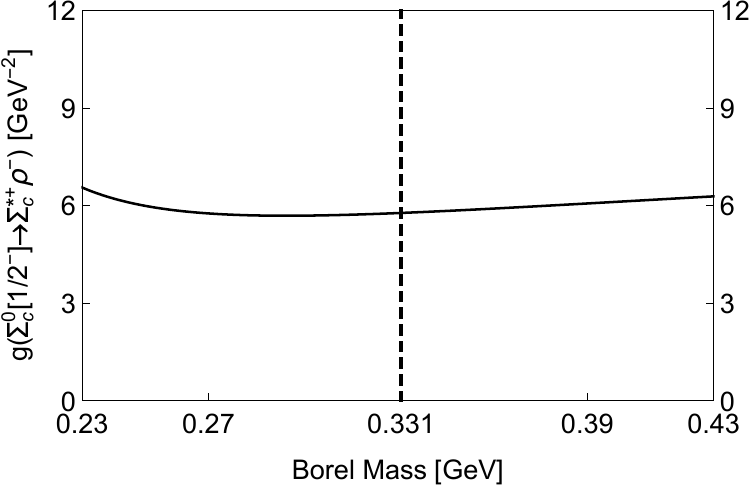}}}~~~~~
\subfigure[]{
\scalebox{0.3}{\includegraphics{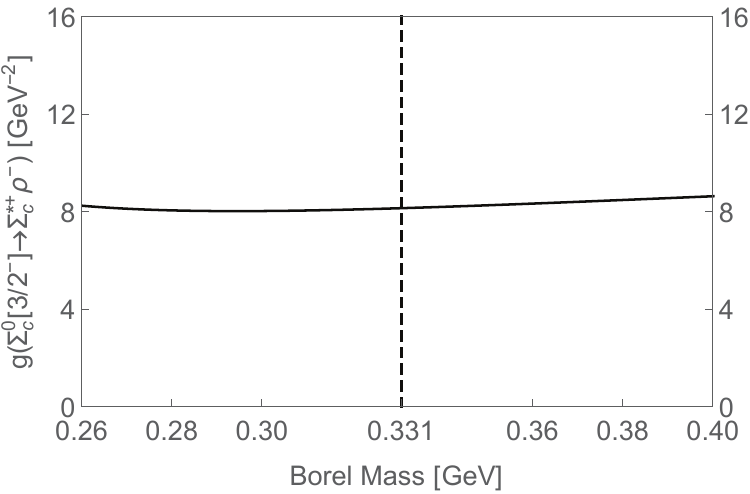}}}~~~~~
\subfigure[]{
\scalebox{0.3}{\includegraphics{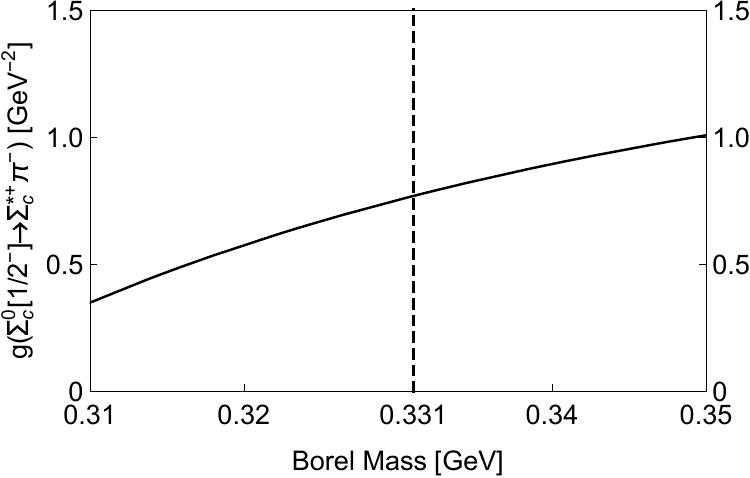}}}~~~~~
\subfigure[]{
\scalebox{0.3}{\includegraphics{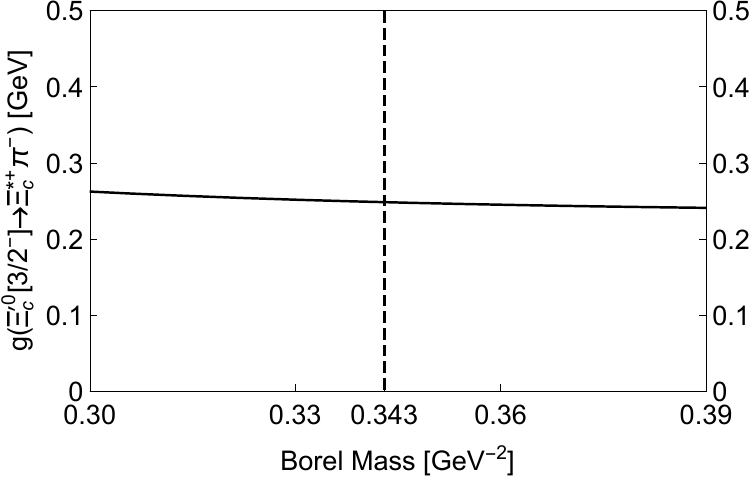}}}
\\
\subfigure[]{
\scalebox{0.3}{\includegraphics{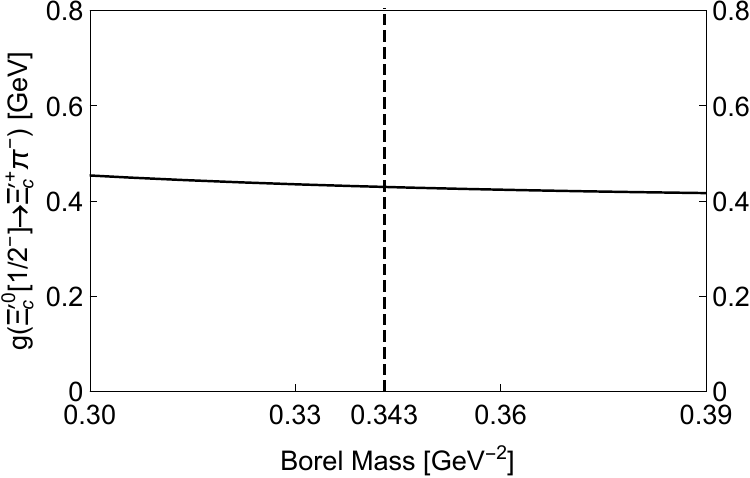}}}~~~~~
\subfigure[]{
\scalebox{0.3}{\includegraphics{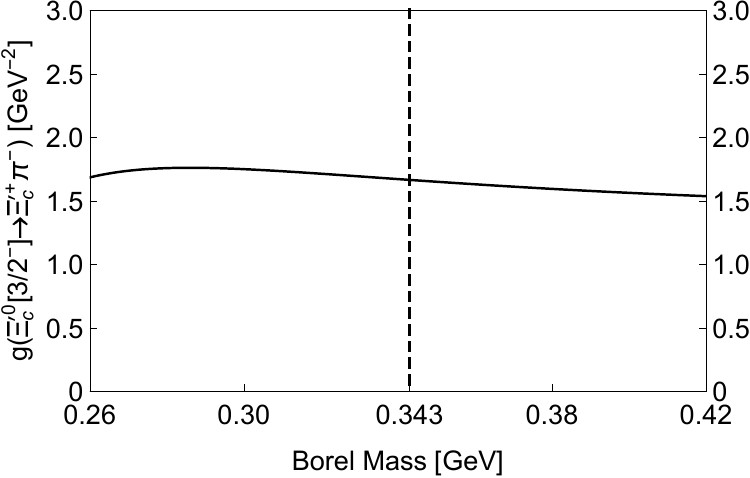}}}~~~~~
\subfigure[]{
\scalebox{0.3}{\includegraphics{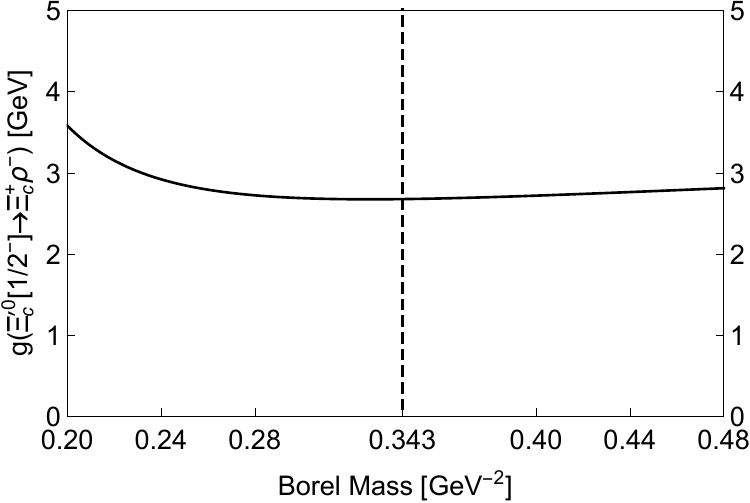}}}~~~~~
\subfigure[]{
\scalebox{0.3}{\includegraphics{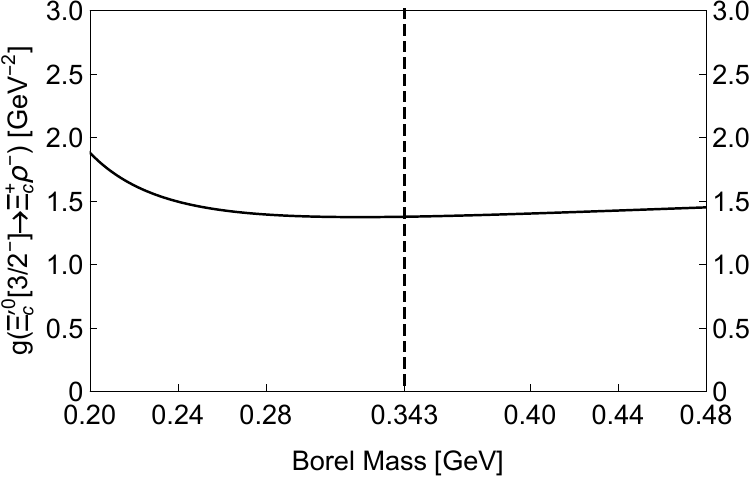}}}
\\
\subfigure[]{
\scalebox{0.4}{\includegraphics{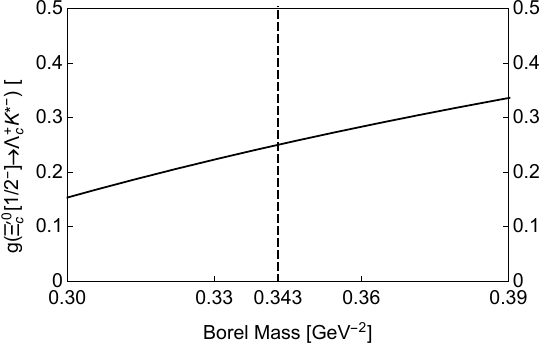}}}~~~~~
\subfigure[]{
\scalebox{0.3}{\includegraphics{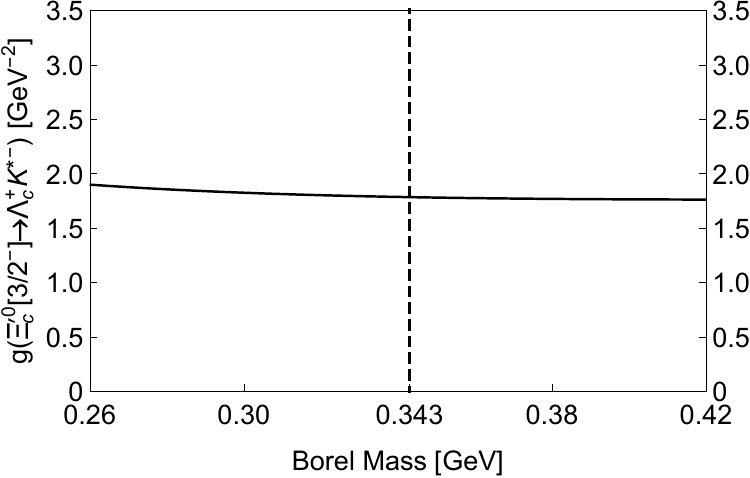}}}~~~~~
\subfigure[]{
\scalebox{0.3}{\includegraphics{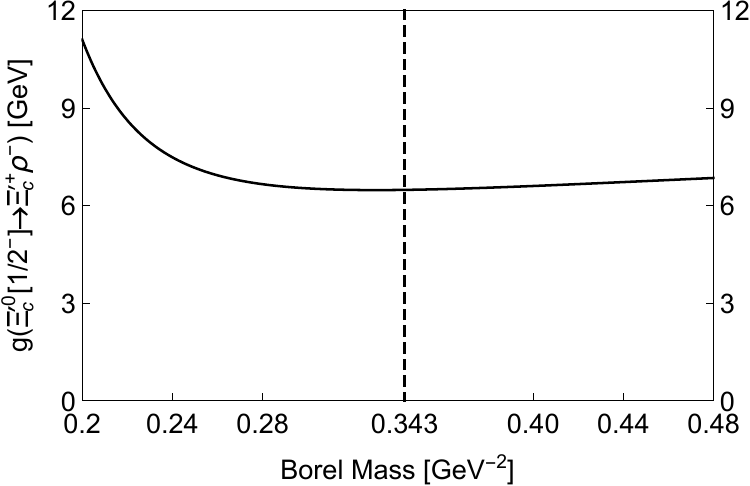}}}~~~~~
\subfigure[]{
\scalebox{0.3}{\includegraphics{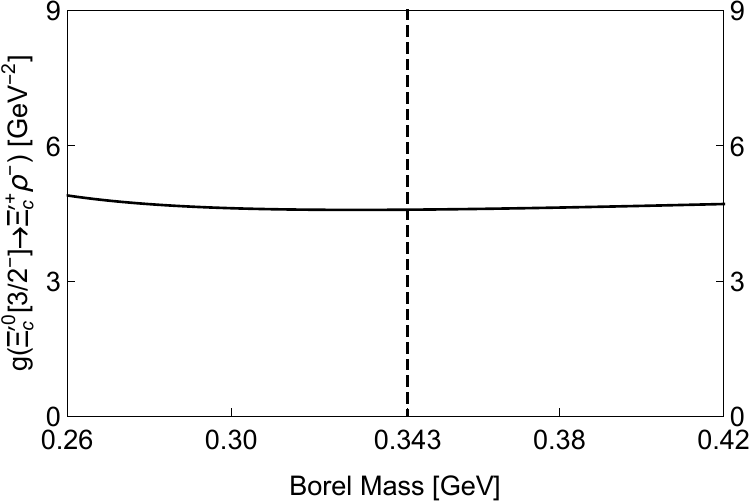}}}
\\
\subfigure[]{
\scalebox{0.3}{\includegraphics{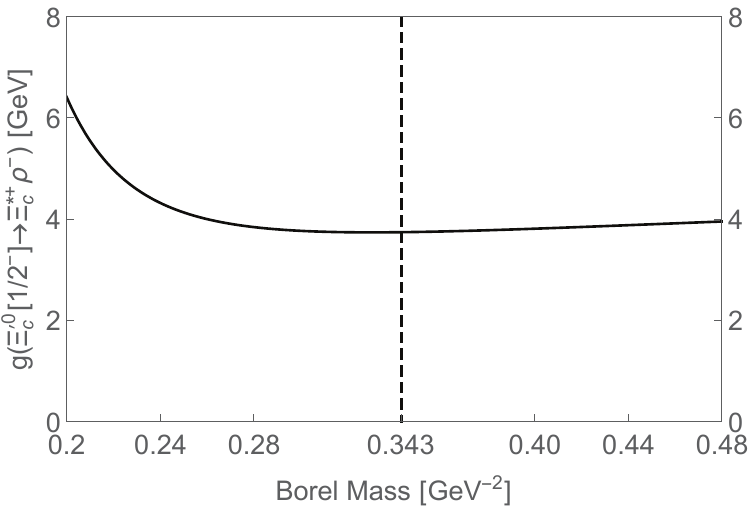}}}~~~~~
\subfigure[]{
\scalebox{0.3}{\includegraphics{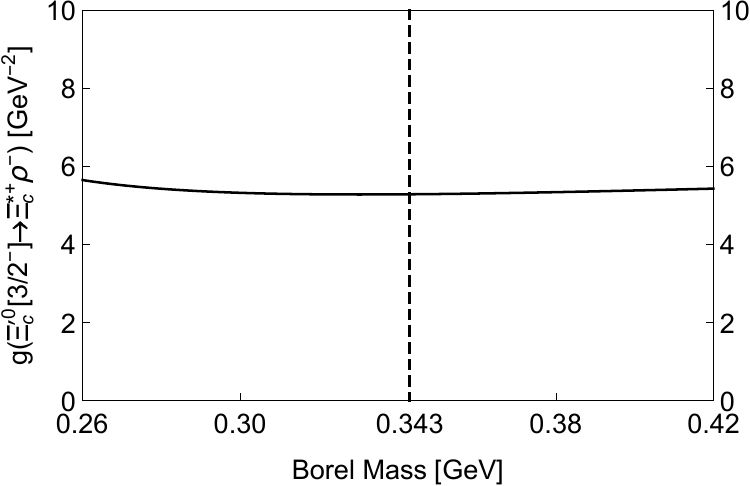}}}~~~~~
\subfigure[]{
\scalebox{0.3}{\includegraphics{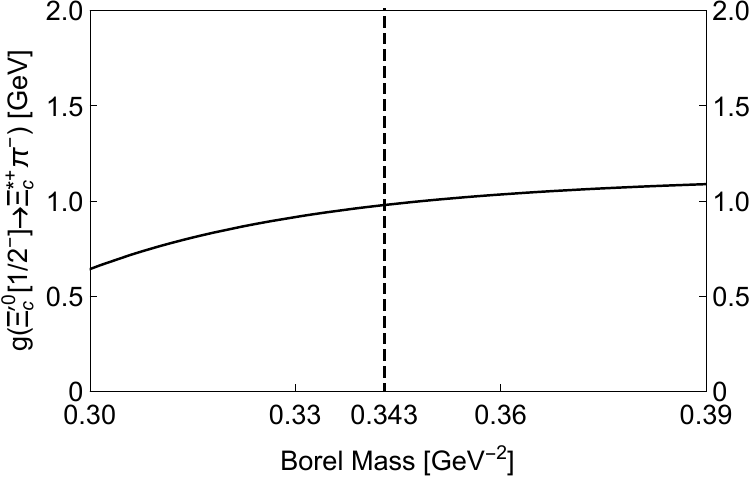}}}~~~~~
\subfigure[]{
\scalebox{0.3}{\includegraphics{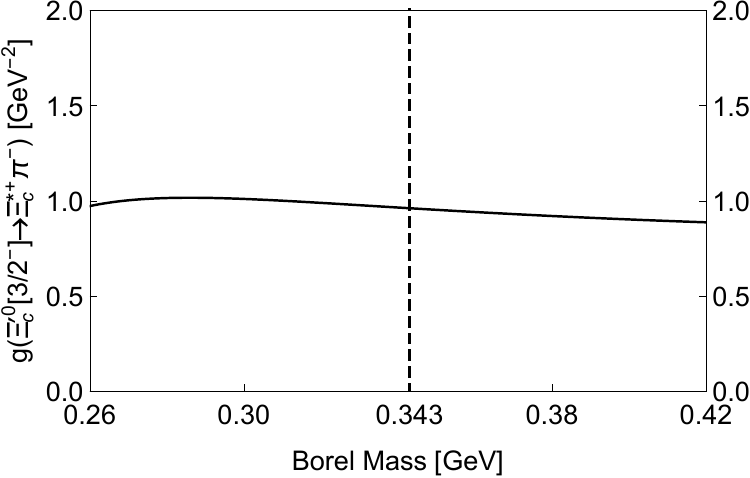}}}
\end{center}
\caption{The coupling constants as functions of the Borel mass $T$:
(a) $g^S_{\Sigma_c^0[{1\over2}^-] \rightarrow \Sigma_c^{+}\pi^-}$,
(b) $g^D_{\Sigma_c^0[{3\over2}^-] \rightarrow \Sigma_c^{+}\pi^-}$,
(c) $g^S_{\Sigma_c^0[{3\over2}^-] \rightarrow \Sigma_c^{*+}\pi^-}$,
(d) $g^D_{\Sigma_c^0[{3\over2}^-] \rightarrow \Sigma_c^{*+}\pi^-}$,
(e) $g^S_{\Sigma_c^0[{1\over2}^-] \rightarrow \Lambda_c^{+}\rho^-}$,
(f) $g^S_{\Sigma_c^0[{3\over2}^-] \rightarrow \Lambda_c^{+}\rho^-}$,
(g) $g^S_{\Sigma_c^0[{1\over2}^-] \rightarrow \Sigma_c^{+}\rho^-}$,
(h) $g^S_{\Sigma_c^0[{3\over2}^-] \rightarrow \Sigma_c^{+}\rho^-}$,
(i) $g^S_{\Sigma_c^0[{1\over2}^-] \rightarrow \Sigma_c^{*+}\rho^-}$,
(j) $g^S_{\Sigma_c^0[{3\over2}^-] \rightarrow \Sigma_c^{*+}\rho^-}$,
(k) $g^D_{\Sigma_c^0[{1\over2}^-] \rightarrow \Sigma_c^{*+}\pi^-}$,
(l) $g^S_{\Xi_c^{\prime0}[{3\over2}^-] \rightarrow \Xi_c^{*+} \pi^-}$,
(m) $g^S_{\Xi_c^{\prime0}[{1\over2}^-] \rightarrow \Xi_c^{\prime+} \pi^-}$,
(n) $g^D_{\Xi_c^{\prime0}[{3\over2}^-] \rightarrow \Xi_c^{\prime+} \pi^-}$,
(o) $g^S_{\Xi_c^{\prime0}[{1\over2}^-] \rightarrow \Xi_c^{+} \rho^-}$,
(p) $g^S_{\Xi_c^{\prime0}[{3\over2}^-] \rightarrow \Xi_c^{+} \rho^-}$,
(q) $g^S_{\Xi_c^{\prime0}[{1\over2}^-] \rightarrow \Lambda_c^{+} K^{*-}}$,
(r) $g^S_{\Xi_c^{\prime0}[{3\over2}^-] \rightarrow \Lambda_c^{+} K^{*-}}$,
(s) $g^S_{\Xi_c^{\prime0}[{1\over2}^-] \rightarrow \Xi_c^{\prime+} \rho^-}$,
(t) $g^S_{\Xi_c^{\prime0}[{3\over2}^-] \rightarrow \Xi_c^{\prime+} \rho^-}$,
(u) $g^S_{\Xi_c^{\prime0}[{1\over2}^-] \rightarrow \Xi_c^{*+} \rho^-}$,
(v) $g^S_{\Xi_c^{\prime0}[{3\over2}^-] \rightarrow \Xi_c^{*+} \rho^-}$,
(w) $g^D_{\Xi_c^{\prime0}[{1\over2}^-] \rightarrow \Xi_c^{*+} \pi^-}$,
and
(x) $g^D_{\Xi_c^{\prime0}[{3\over2}^-] \rightarrow \Xi_c^{*+} \pi^-}$.
The charmed baryon doublet $[\mathbf{6}_F, 1, 1, \lambda]$ is investigated here.
\label{fig:611lambda}}
\end{figure*}

The $[\mathbf{6}_F, 1, 1, \lambda]$ doublet contains altogether six charmed baryons: $\Sigma_c({1\over2}^-/{3\over2}^-)$, $\Xi_c({1\over2}^-/{3\over2}^-)$, and $\Omega_c({1\over2}^-/{3\over2}^-)$. We study their $S$- and $D$-wave decays into ground-state charmed baryons together with light pseudoscalar and vector mesons. We derive the following non-zero coupling constants:
\begin{eqnarray}
\nonumber &(a2)& g^S_{\Sigma_c[{1\over2}^-]\to \Sigma_c[{1\over2}^+] \pi} = 0.67^{+0.41}_{-0.33}~{\rm GeV}^{-2} \, ,
\\
\nonumber &(b3)& g^S_{\Sigma_c[{3\over2}^-]\to \Sigma_c^{*}[{3\over2}^+] \pi} = 0.39^{+0.24}_{-0.19}~{\rm GeV}^{-2} \, ,
\\
\nonumber &(a4)& g^S_{\Sigma_c[{1\over2}^-]\to \Lambda_c[{1\over2}^+] \rho} =4.28^{+5.32}_{-4.32}~{\rm GeV}^{-2}\, ,
\\
\nonumber &(a5)&  g^S_{\Sigma_c[{1\over2}^-]\to \Sigma_c[{1\over2}^+] \rho} =9.99^{+6.69}_{-5.68}~{\rm GeV}^{-2}\, ,
\\
\nonumber &(a6)&  g^S_{\Sigma_c[{1\over2}^-]\to \Sigma_c^{*}[{3\over2}^+] \rho} =5.77~{\rm GeV}^{-2}\, ,
\\
\nonumber &(b4)&  g^S_{\Sigma_c[{3\over2}^-]\to \Lambda_c[{1\over2}^+] \rho} =2.37^{+2.26}_{-2.10}~{\rm GeV}^{-2}\, ,
\\
\nonumber &(b5)&  g^S_{\Sigma_c[{3\over2}^-]\to \Sigma_c[{1\over2}^+] \rho} =7.05^{+4.74}_{-4.01}~{\rm GeV}^{-2}\, ,
\\
\nonumber &(b6)&  g^S_{\Sigma_c[{3\over2}^-]\to \Sigma_c^{*}[{3\over2}^+] \rho} =8.14^{+5.47}_{-4.63}~{\rm GeV}^{-2}\, ,
\\
\nonumber &(a3)& g^D_{\Sigma_c[{1\over2}^-]\to \Sigma_c^{*}[{3\over2}^+] \pi} = 0.76^{+0.99}_{-0.76}~{\rm GeV}^{-2}\, ,
\\
\nonumber &(b2)& g^D_{\Sigma_c[{3\over2}^-]\to \Sigma_c[{1\over2}^+] \pi}= 2.18^{+1.63}_{-1.40}~{\rm GeV}^{-2} \, ,
\\
\nonumber &(b3)& g^D_{\Sigma_c[{3\over2}^-]\to \Sigma_c^{*}[{3\over2}^+] \pi} = 1.26^{+0.94}_{-0.81}~{\rm GeV}^{-2}\, ,
\\
\nonumber &(d2)& g^S_{\Xi_c^{\prime}[{1\over2}^-]\to \Xi_c^{\prime}[{1\over2}^+]\pi} = 0.43^{+0.22}_{-0.19}~{\rm GeV}^{-2}\, ,
\\
\nonumber &(d4)& g^S_{\Xi_c^{\prime}[{1\over2}^-]\to \Sigma_c[{1\over2}^+]\bar K} = 0.36~{\rm GeV}^{-2}\, ,
\\
\nonumber &(e5)& g^S_{\Xi_c^{\prime}[{3\over2}^-]\to \Xi_c^{*}[{3\over2}^+]\pi} = 0.25^{+0.13}_{-0.11}~{\rm GeV}^{-2}\, ,
\\
\nonumber &(e6)& g^S_{\Xi_c^{\prime}[{3\over2}^-]\to \Sigma_c^{*}[{3\over2}^+]\bar K} = 0.14~{\rm GeV}^{-2}\, ,
\\
\nonumber &(d7)& g^S_{\Xi_c^{\prime}[{1\over2}^-]\to \Xi_c[{1\over2}^+]\rho} = 2.67^{+3.54}_{-2.33}~{\rm GeV}^{-2}\, ,
\\
\nonumber &(d8)& g^S_{\Xi_c^{\prime}[{1\over2}^-]\to \Lambda_c[{1\over2}^+]\bar K^*} = 0.25~{\rm GeV}^{-2}\, ,
\\
\nonumber &(d9)& g^S_{\Xi_c^{\prime}[{1\over2}^-]\to \Xi_c^{\prime}[{1\over2}^+]\rho} = 6.48^{+3.60}_{-3.51}~{\rm GeV}^{-2}\, ,
\\
\nonumber &(d10)& g^S_{\Xi_c^{\prime}[{1\over2}^-]\to \Sigma_c[{1\over2}^+]\bar K^*} = 4.32~{\rm GeV}^{-2}\, ,
\\
\nonumber &(d11)& g^S_{\Xi_c^{\prime}[{1\over2}^-]\to \Xi_c^{*}[{3\over2}^+]\rho} = 3.74~{\rm GeV}^{-2}\, ,
\\
\nonumber &(d12)& g^S_{\Xi_c^{\prime}[{1\over2}^-]\to \Sigma_c^{*}[{3\over2}^+]\bar K^*} = 2.50~{\rm GeV}^{-2}\, ,
\\
\nonumber &(e7)& g^S_{\Xi_c^{\prime}[{3\over2}^-]\to \Xi_c[{1\over2}^+]\rho} = 1.37^{+1.34}_{-1.29}~{\rm GeV}^{-2}\, ,
\\
          &(e8)& g^S_{\Xi_c^{\prime}[{3\over2}^-]\to \Lambda_c[{1\over2}^+]\bar K^*} = 1.78~{\rm GeV}^{-2}\, ,
\\
\nonumber &(e9)& g^S_{\Xi_c^{\prime}[{3\over2}^-]\to \Xi_c^{\prime}[{1\over2}^+]\rho} = 4.58^{+2.77}_{-2.48}~{\rm GeV}^{-2}\, ,
\\
\nonumber &(e10)& g^S_{\Xi_c^{\prime}[{3\over2}^-]\to \Sigma_c[{1\over2}^+]K^*} = 3.06~{\rm GeV}^{-2}\, ,
\\
\nonumber &(e11)& g^S_{\Xi_c^{\prime}[{3\over2}^-]\to \Xi_c^{*}[{3\over2}^+]\rho} = 5.29~{\rm GeV}^{-2}\, ,
\\
\nonumber &(e12)& g^S_{\Xi_c^{\prime}[{3\over2}^-]\to \Sigma_c^{*}[{3\over2}^+]\bar K^*} = 3.53~{\rm GeV}^{-2}\, ,
\\
\nonumber &(d5)& g^D_{\Xi_c^{\prime}[{1\over2}^-]\to \Xi_c^{*}[{3\over2}^+]\pi} = 0.98^{+0.68}_{-0.62}~{\rm GeV}^{-2}\, ,
\\
\nonumber &(d6)& g^D_{\Xi_c^{\prime}[{1\over2}^-]\to \Sigma_c^{*}[{3\over2}^+]\bar K} = 0.17~{\rm GeV}^{-2}\, ,
\\
\nonumber &(e3)& g^D_{\Xi_c^{\prime}[{3\over2}^-]\to \Xi_c^{\prime}[{1\over2}^+]\pi} = 1.67^{+1.01}_{-0.87}~{\rm GeV}^{-2}\, ,
\\
\nonumber &(e4)& g^D_{\Xi_c^{\prime}[{3\over2}^-]\to \Sigma_c[{1\over2}^+]\bar K} = 0.61~{\rm GeV}^{-2}\, ,
\\
\nonumber &(e5)& g^D_{\Xi_c^{\prime}[{3\over2}^-]\to \Xi_c^{*}[{3\over2}^+]\pi} = 0.96^{+0.58}_{-0.50}~{\rm GeV}^{-2}\, ,
\\
\nonumber &(e6)& g^D_{\Xi_c^{\prime}[{3\over2}^-]\to \Sigma_c^{*}[{3\over2}^+]\bar K} = 0.35~{\rm GeV}^{-2}\, ,
\\
\nonumber &(g2)& g^S_{\Omega_c[{1\over2}^-]\to \Xi_c^{\prime}[{1\over2}^+]\bar K}= 0.54~{\rm GeV}^{-2} \, ,
\\
\nonumber &(h3)& g^S_{\Omega_c[{1\over2}^-]\to \Xi_c^{*}[{3\over2}^+]\bar K}= 0.25~{\rm GeV}^{-2} \, ,
\\
\nonumber &(g4)& g^S_{\Omega_c[{1\over2}^-]\to \Xi_c[{1\over2}^+]\bar K^*}= 0.57~{\rm GeV}^{-2} \, ,
\\
\nonumber &(g5)& g^S_{\Omega_c[{1\over2}^-]\to \Xi_c^{\prime}[{1\over2}^+]\bar K^*}= 5.64~{\rm GeV}^{-2} \, ,
\\
\nonumber &(g6)& g^S_{\Omega_c[{1\over2}^-]\to \Xi_c^{*}[{3\over2}^+]\bar K^*}= 3.26~{\rm GeV}^{-2} \, ,
\\
\nonumber &(f4)& g^S_{\Omega_c[{3\over2}^-]\to \Xi_c^{*}[{3\over2}^+]\bar K^*}= 2.28~{\rm GeV}^{-2} \, ,
\\
\nonumber &(f5)& g^S_{\Omega_c[{3\over2}^-]\to \Xi_c^{\prime}[{1\over2}^+]\bar K^*}= 3.99~{\rm GeV}^{-2} \, ,
\\
\nonumber &(f6)& g^S_{\Omega_c[{3\over2}^-]\to \Xi_c^{*}[{3\over2}^+]\bar K^*}= 4.61~{\rm GeV}^{-2} \, ,
\\
\nonumber &(g3)& g^D_{\Omega_c[{1\over2}^-]\to \Xi_c^{*}[{3\over2}^+]\bar K}= 0.58~{\rm GeV}^{-2} \, ,
\\
\nonumber &(h2)& g^D_{\Omega_c[{3\over2}^-]\to \Xi_c^{\prime}[{1\over2}^+]\bar K}= 1.36~{\rm GeV}^{-2} \, ,
\\
\nonumber &(h3)& g^D_{\Omega_c[{3\over2}^-]\to \Xi_c^{*}[{3\over2}^+]\bar K}= 0.79~{\rm GeV}^{-2} \, .
\end{eqnarray}
Some of these coupling constants are shown in Fig.~\ref{fig:611lambda} as functions of the Borel mass $T$. We further use these coupling constants to derive the following decay channels that are kinematically allowed:
\begin{eqnarray}
\nonumber  &(a2)& \Gamma^S_{\Sigma_c[{1\over2}^-]\to \Sigma_c[{1\over2}^+] \pi} =37^{+60}_{-28}~{\rm MeV} \, ,
\\
\nonumber  &(b3)& \Gamma^S_{\Sigma_c[{3\over2}^-]\to \Sigma_c^{*}[{3\over2}^+] \pi} =10^{+16}_{-~8}~{\rm MeV} \, ,
\\
\nonumber  &(a4)& \Gamma^S_{\Sigma_c[{1\over2}^-]\to \Lambda_c[{1\over2}^+] \rho \to \Lambda_c[{1\over2}^+] \pi\pi} =9.2^{+37.0}_{-~9.2}~{\rm MeV} \, ,
\\
\nonumber  &(a5)& \Gamma^S_{\Sigma_c[{1\over2}^-]\to \Sigma_c[{1\over2}^+] \rho \to\Sigma_c[{1\over2}^+] \pi\pi} =1.2^{+2.1}_{-1.0}~{\rm MeV} \, ,
\\
\nonumber &(a6)& \Gamma^S_{\Sigma_c[{1\over2}^-]\to \Sigma_c^{*}[{3\over2}^+]\rho\to\Sigma_c^{*}[{3\over2}^+]\pi\pi} =1 \times 10^{-4}~{\rm MeV} \, ,
\\
\nonumber &(b4)& \Gamma^S_{\Sigma_c[{3\over2}^-]\to \Lambda_c[{1\over2}^+] \rho \to\Lambda_c[{1\over2}^+]\pi\pi} = 0.92^{+2.58}_{-0.91}~{\rm MeV} \, ,
\\
\nonumber  &(b5)& \Gamma^S_{\Sigma_c[{3\over2}^-]\to \Sigma_c[{1\over2}^+] \rho \to \Sigma_c[{1\over2}^+]\pi\pi} =0.20^{+0.36}_{-0.16}~{\rm MeV} \, ,
\\
\nonumber &(b6)& \Gamma^S_{\Sigma_c[{3\over2}^-]\to \Sigma_c^{*}[{3\over2}^+] \rho \to \Sigma_c^{*}[{3\over2}^+]\pi\pi} = 1\times10^{-4}~{\rm MeV} \, ,
\\
\nonumber &(a3)& \Gamma^D_{\Sigma_c[{1\over2}^-]\to \Sigma_c^{*}[{3\over2}^+] \pi} = 0.10^{+0.45}_{-0.10}~{\rm MeV} \, ,
\\
\nonumber &(b2)& \Gamma^D_{\Sigma_c[{3\over2}^-]\to \Sigma_c[{1\over2}^+] \pi} = 1.2^{+2.4}_{-1.0}~{\rm MeV} \, ,
\\
\nonumber  &(b3)& \Gamma^D_{\Sigma_c[{3\over2}^-]\to \Sigma_c^{*}[{3\over2}^+] \pi} =0.09~{\rm MeV} \, ,
\\
          &(d2)& \Gamma^S_{\Xi_c^{\prime}[{1\over2}^-]\to\Xi_c^{\prime}[{1\over2}^+]\pi} = 12^{+15}_{-~8}~{\rm MeV} \, ,
\\
\nonumber &(e5)& \Gamma^S_{\Xi_c^{\prime}[{3\over2}^-]\to\Xi_c^{*}[{3\over2}^+]\pi} = 3.3^{+4.3}_{-2.3}~{\rm MeV} \, ,
\\
\nonumber &(d7)& \Gamma^S_{\Xi_c^{\prime}[{1\over2}^-]\to\Xi_c[{1\over2}^+]\rho\to\Xi_c[{1\over2}^+]\pi\pi} = 1.7^{+7.6}_{-1.7}~{\rm MeV} \, ,
\\
\nonumber &(d8)& \Gamma^S_{\Xi_c^{\prime}[{1\over2}^-]\to\Lambda_c[{1\over2}^+]\bar K^*\to\Lambda_c[{1\over2}^+]\pi \bar K} = 4\times 10^{-8}~{\rm MeV} \, ,
\\
\nonumber &(d9)& \Gamma^S_{\Xi_c^{\prime}[{1\over2}^-]\to\Xi_c^{\prime}[{1\over2}^+]\rho\to\Xi_c^{\prime}[{1\over2}^+]\pi\pi} = 0.38^{+0.54}_{-0.30}~{\rm MeV} \, ,
\\
\nonumber &(d11)& \Gamma^S_{\Xi_c^{\prime}[{1\over2}^-]\to\Xi_c^{*}[{3\over2}^+]\rho\to\Xi_c^{*}[{3\over2}^+]\pi\pi} = 2\times 10^{-7}~{\rm MeV} \, ,
\\
\nonumber &(e7)& \Gamma^S_{\Xi_c^{\prime}[{3\over2}^-]\to\Xi_c[{1\over2}^+]\rho\to\Xi_c[{1\over2}^+]\pi\pi} = 0.21^{+0.60}_{-0.20}~{\rm MeV} \, ,
\\
\nonumber &(e8)& \Gamma^S_{\Xi_c^{\prime}[{3\over2}^-]\to\Lambda_c[{1\over2}^+]\bar K^*\to\Lambda_c[{1\over2}^+]\pi \bar K} = 2\times 10^{-4}~{\rm MeV} \, ,
\\
\nonumber &(e9)& \Gamma^S_{\Xi_c^{\prime}[{3\over2}^-]\to\Xi_c^{\prime}[{1\over2}^+]\rho\to\Xi_c^{\prime}[{1\over2}^+]\pi\pi} = 0.12^{+0.19}_{-0.10}~{\rm MeV} \, ,
\\
\nonumber &(e11)& \Gamma^S_{\Xi_c^{\prime}[{3\over2}^-]\to\Xi_c^{*}[{3\over2}^+]\rho\to\Xi_c^{*}[{3\over2}^+]\pi\pi} = 0.001~{\rm MeV} \, ,
\\
\nonumber &(d5)& \Gamma^D_{\Xi_c^{\prime}[{1\over2}^-]\to\Xi_c^{*}[{3\over2}^+]\pi} = 0.12^{+0.22}_{-0.10}~{\rm MeV} \, ,
\\
\nonumber &(e3)& \Gamma^D_{\Xi_c^{\prime}[{3\over2}^-]\to\Xi_c^{\prime}[{1\over2}^+]\pi} = 0.67^{+1.06}_{-0.52}~{\rm MeV} \, ,
\\
\nonumber &(e5)& \Gamma^D_{\Xi_c^{\prime}[{3\over2}^-]\to\Xi_c^{*}[{3\over2}^+]\pi} = 0.05~{\rm MeV} \, .
\end{eqnarray}
We summarize the above results in Table~\ref{tab:decay611lambda}.

\subsection{The $[\mathbf{6}_F,2,1,\lambda]$ doublet}

\begin{figure*}[htb]
\begin{center}
\subfigure[]{
\scalebox{0.42}{\includegraphics{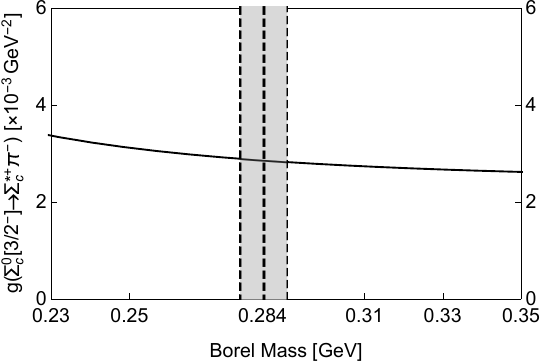}}}~~~~~
\subfigure[]{
\scalebox{0.42}{\includegraphics{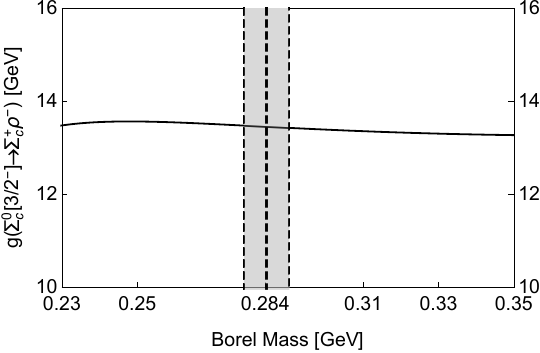}}}~~~~~
\subfigure[]{
\scalebox{0.42}{\includegraphics{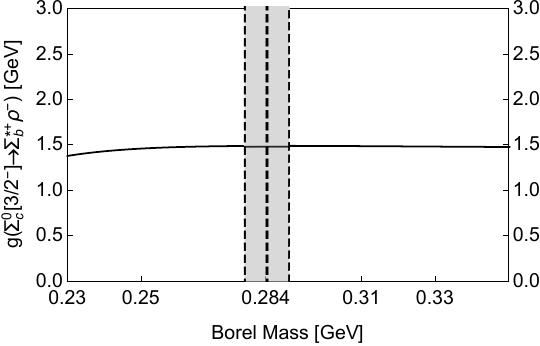}}}~~~~~
\subfigure[]{
\scalebox{0.35}{\includegraphics{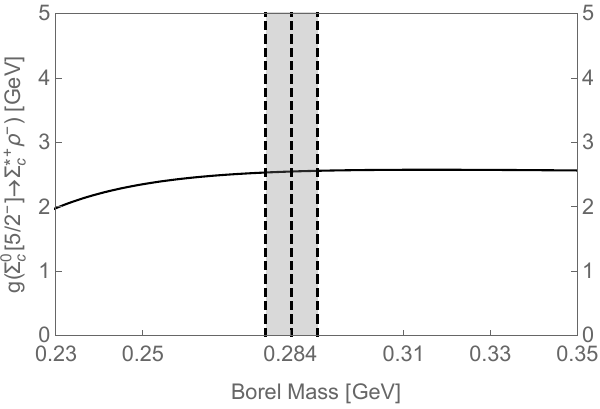}}}
\\
\subfigure[]{
\scalebox{0.33}{\includegraphics{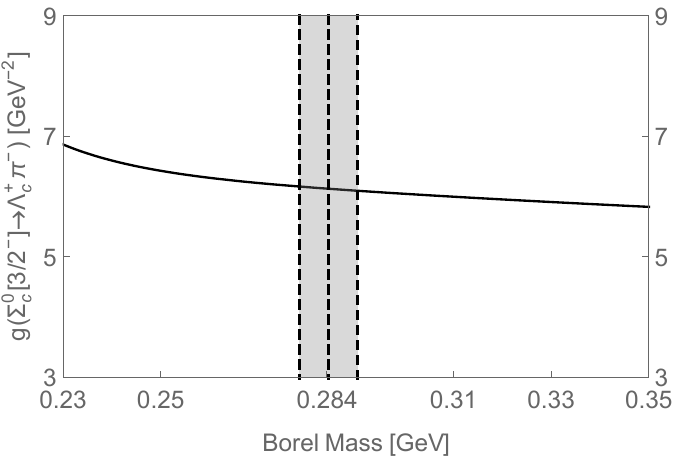}}}~~~~~
\subfigure[]{
\scalebox{0.42}{\includegraphics{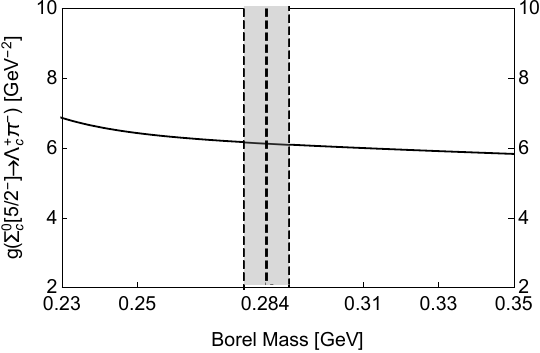}}}~~~~~
\subfigure[]{
\scalebox{0.42}{\includegraphics{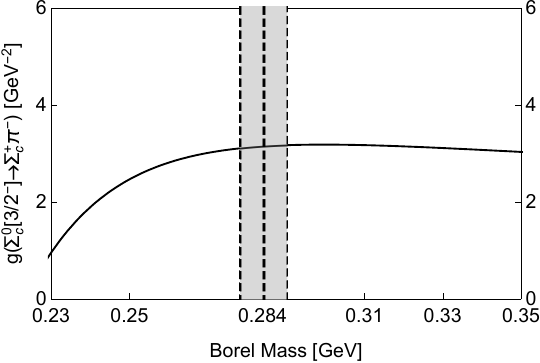}}}~~~~~
\subfigure[]{
\scalebox{0.33}{\includegraphics{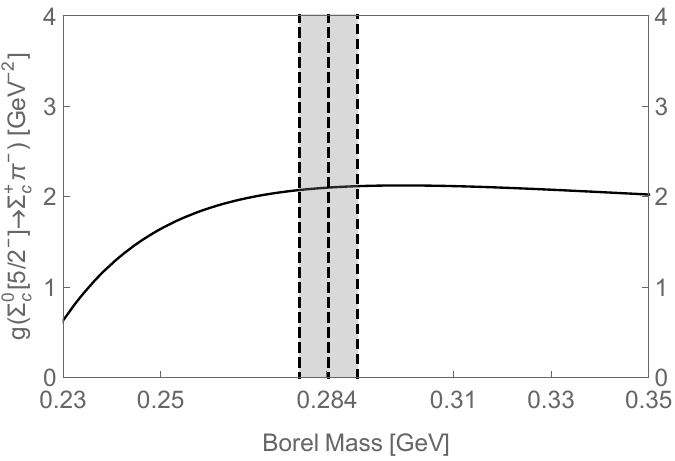}}}
\\
\subfigure[]{
\scalebox{0.42}{\includegraphics{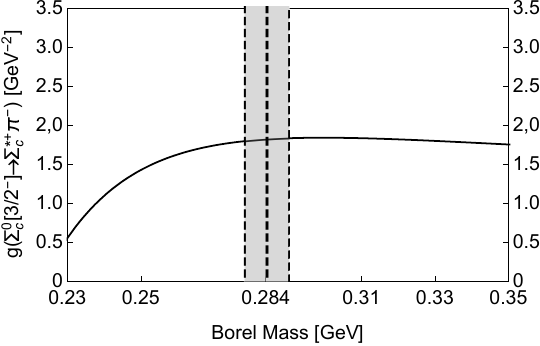}}}~~~~~
\subfigure[]{
\scalebox{0.42}{\includegraphics{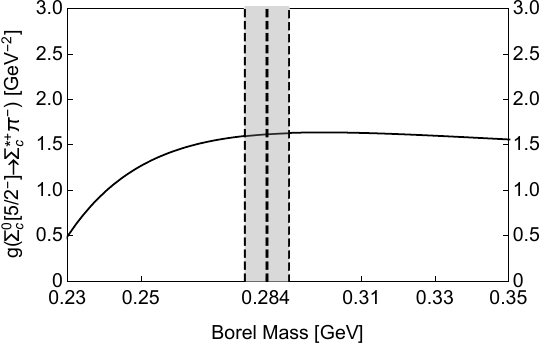}}}~~~~~
\subfigure[]{
\scalebox{0.42}{\includegraphics{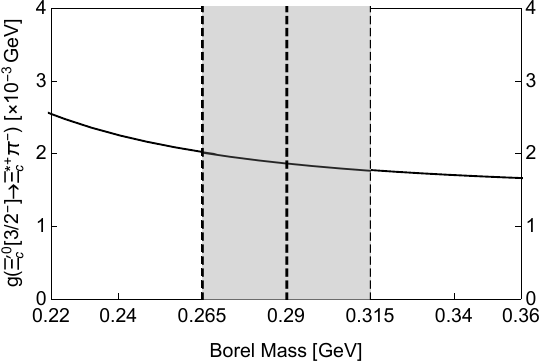}}}~~~~~
\subfigure[]{
\scalebox{0.42}{\includegraphics{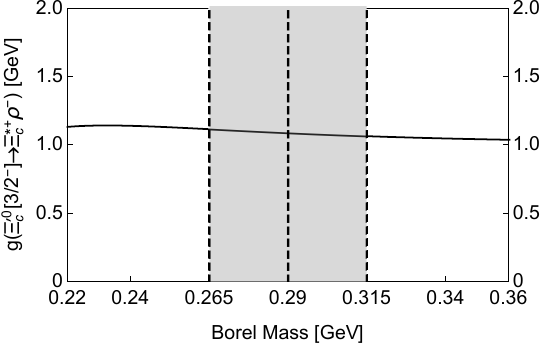}}}
\\
\subfigure[]{
\scalebox{0.3}{\includegraphics{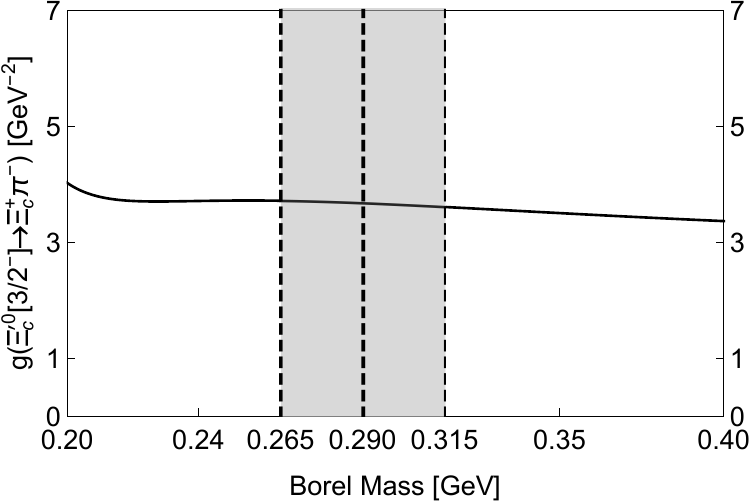}}}~~~~~
\subfigure[]{
\scalebox{0.3}{\includegraphics{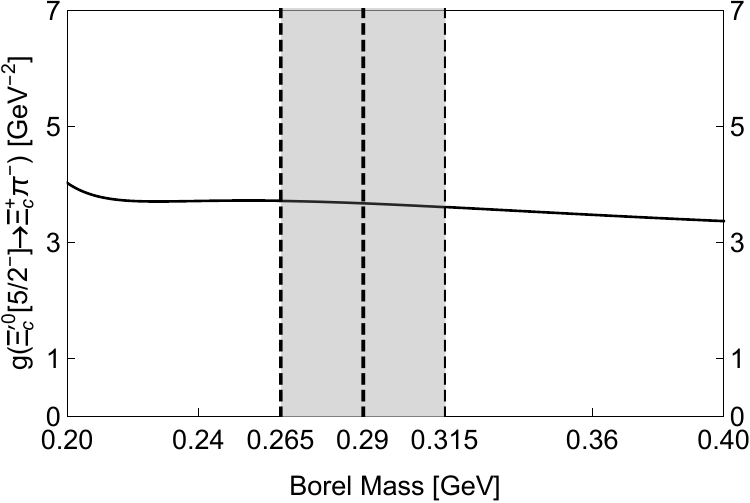}}}~~~~~
\subfigure[]{
\scalebox{0.3}{\includegraphics{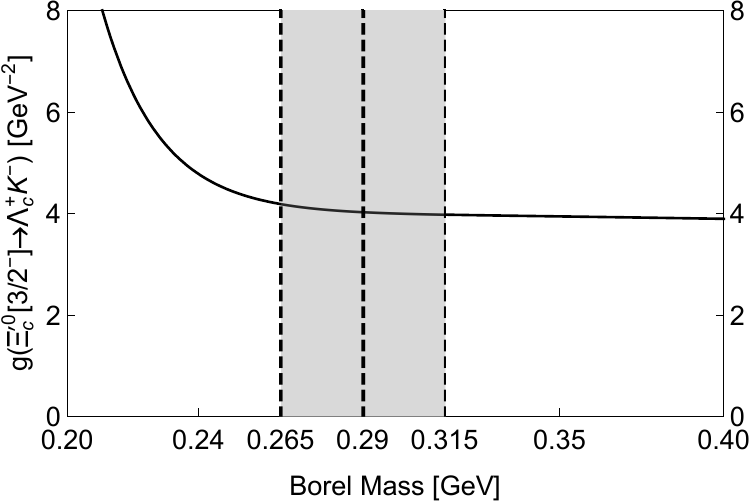}}}~~~~~
\subfigure[]{
\scalebox{0.3}{\includegraphics{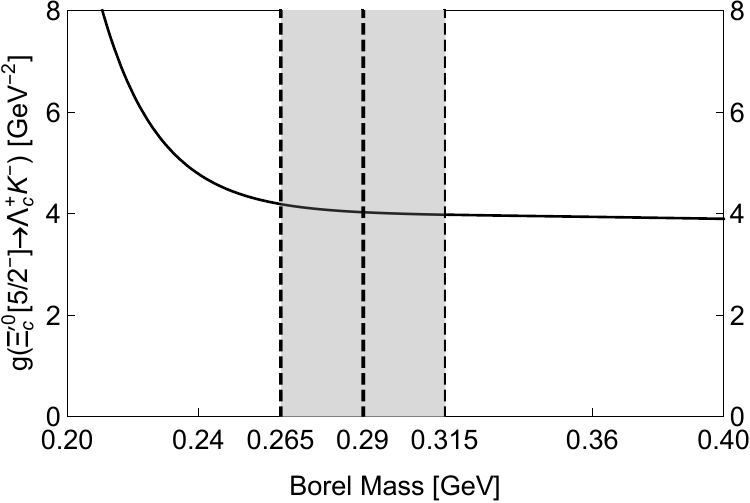}}}
\\
\subfigure[]{
\scalebox{0.3}{\includegraphics{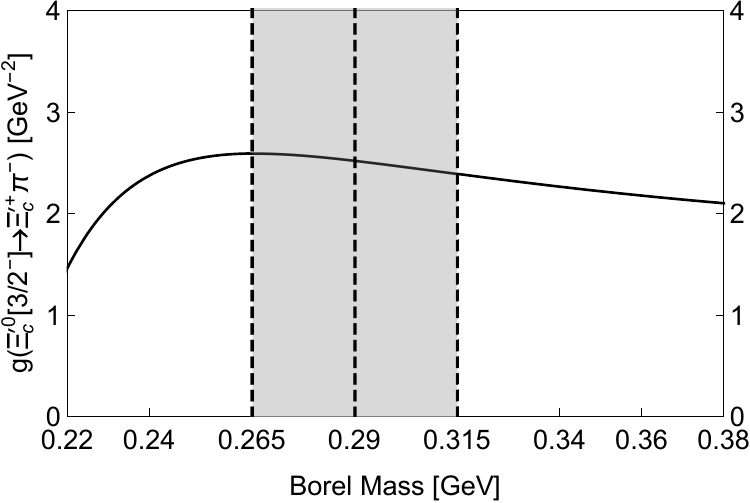}}}~~~~~
\subfigure[]{
\scalebox{0.3}{\includegraphics{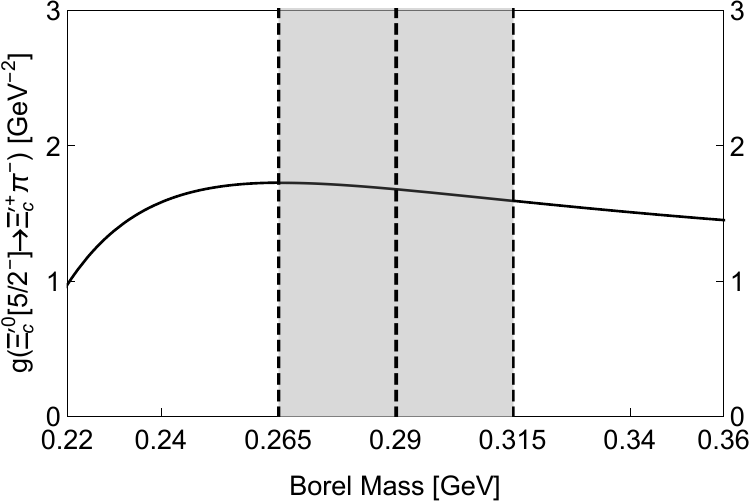}}}~~~~~
\subfigure[]{
\scalebox{0.3}{\includegraphics{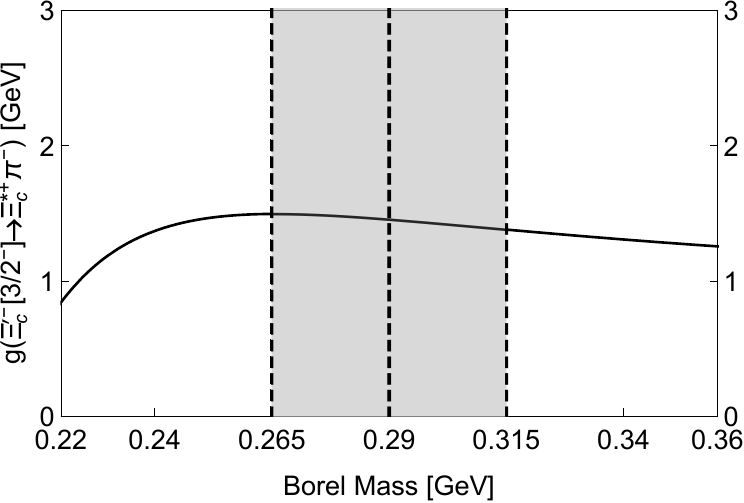}}}~~~~~
\subfigure[]{
\scalebox{0.3}{\includegraphics{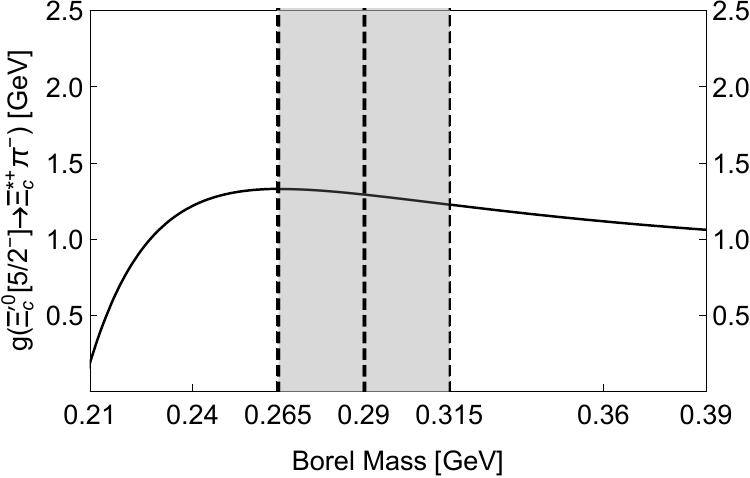}}}
\\
\subfigure[]{
\scalebox{0.3}{\includegraphics{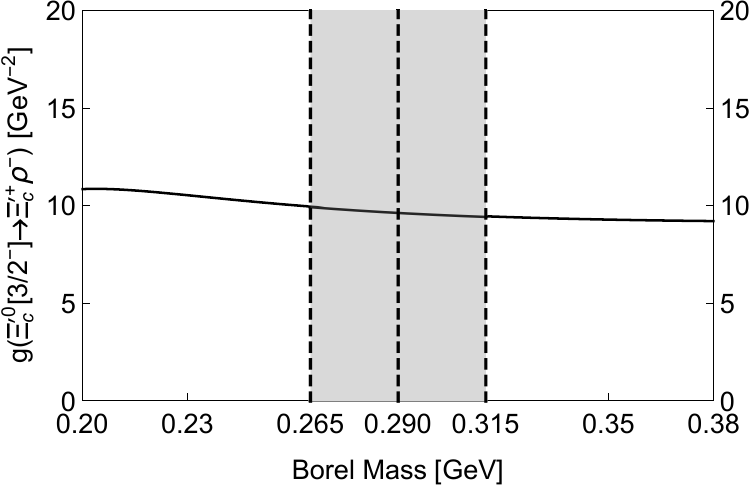}}}~~~~~
\subfigure[]{
\scalebox{0.3}{\includegraphics{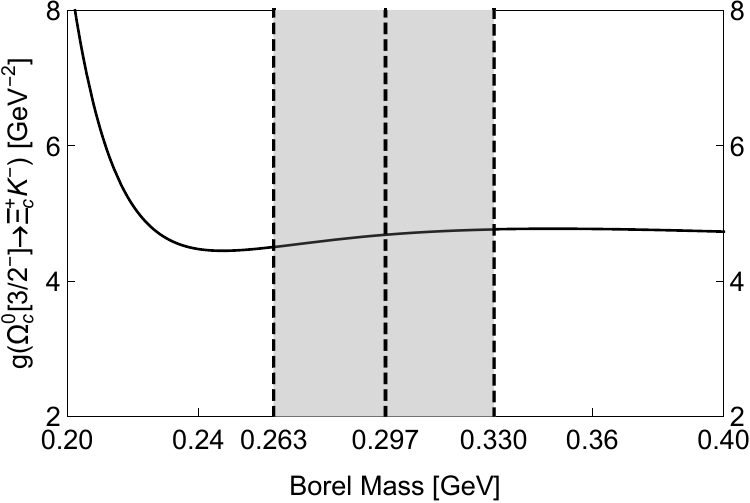}}}~~~~~
\subfigure[]{
\scalebox{0.42}{\includegraphics{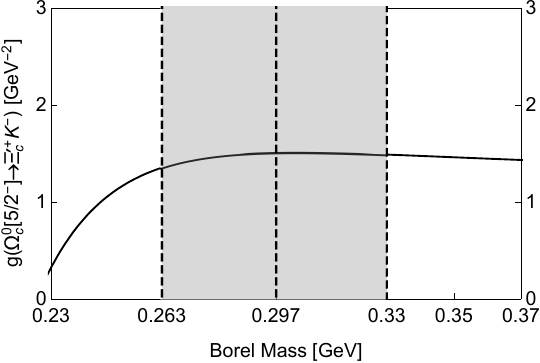}}}~~~~~
\subfigure[]{
\scalebox{0.42}{\includegraphics{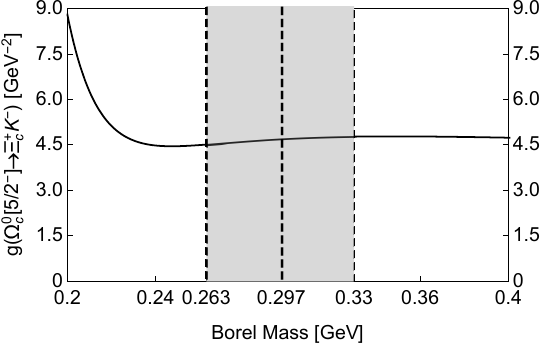}}}
\end{center}
\caption{Coupling constants as functions of the Borel mass $T$:
(a) $g^S_{\Sigma_c^0[{3\over2}^-] \rightarrow \Sigma_c^{*+}\pi^-}$,
(b) $g^S_{\Sigma_c^0[{3\over2}^-] \rightarrow \Sigma_c^{+}\rho^-}$,
(c) $g^S_{\Sigma_c^0[{3\over2}^-] \rightarrow \Sigma_c^{*+}\rho^-}$,
(d) $g^S_{\Sigma_c^0[{5\over2}^-] \rightarrow \Sigma_c^{*+}\rho^-}$,
(e) $g^D_{\Sigma_c^0[{3\over2}^-] \rightarrow \Lambda_c^{+}\pi^-}$,
(f) $g^D_{\Sigma_c^0[{5\over2}^-] \rightarrow \Lambda_c^{+}\pi^-}$,
(g) $g^D_{\Sigma_c^0[{3\over2}^-] \rightarrow \Sigma_c^{+}\pi^-}$,
(h) $g^D_{\Sigma_c^0[{5\over2}^-] \rightarrow \Sigma_c^{+}\pi^-}$,
(i) $g^D_{\Sigma_c^0[{3\over2}^-] \rightarrow \Sigma_c^{*+}\pi^-}$,
(j) $g^D_{\Sigma_c^0[{5\over2}^-] \rightarrow \Sigma_c^{*+}\pi^-}$,
(k) $g^S_{\Xi_c^{\prime0}[{3\over2}^-] \rightarrow \Xi_c^{*+} \pi^-}$,
(l) $g^S_{\Xi_c^{\prime0}[{3\over2}^-] \rightarrow \Xi_c^{*+} \rho^-}$,
(m) $g^D_{\Xi_c^{\prime0}[{3\over2}^-] \rightarrow \Xi_c^{+} \pi^-}$,
(n) $g^D_{\Xi_c^{\prime0}[{5\over2}^-] \rightarrow \Xi_c^{+} \pi^-}$,
(o) $g^D_{\Xi_c^{\prime0}[{3\over2}^-] \rightarrow \Lambda_c^{+} K^-}$,
(p) $g^D_{\Xi_c^{\prime0}[{5\over2}^-] \rightarrow \Lambda_c^{+} K^-}$,
(q) $g^D_{\Xi_c^{\prime0}[{3\over2}^-] \rightarrow \Xi_c^{\prime+} \pi^-}$,
(r) $g^D_{\Xi_c^{\prime0}[{5\over2}^-] \rightarrow \Xi_c^{\prime+} \pi^-}$,
(s) $g^D_{\Xi_c^{\prime0}[{3\over2}^-] \rightarrow \Xi_c^{*+} \pi^-}$,
(t) $g^D_{\Xi_c^{\prime0}[{5\over2}^-] \rightarrow \Xi_c^{*+} \pi^-}$,
(u) $g^S_{\Xi_c^{\prime0}[{3\over2}^-] \rightarrow \Xi_c^{\prime+} \rho^-}$,
(v) $g^D_{\Omega_c^{0}[{3\over2}^-] \rightarrow \Xi_c^{+} K^-}$,
(w) $g^D_{\Omega_c^{0}[{5\over2}^-] \rightarrow \Xi_c^{\prime+} K^-}$,
and
(x) $g^D_{\Omega_c^{0}[{5\over2}^-] \rightarrow \Xi_c^{+} K^-}$.
The charmed baryon doublet $[\mathbf{6}_F, 2, 1, \lambda]$ is investigated here.
\label{fig:621lambda}}
\end{figure*}

The $[\mathbf{6}_F, 2, 1, \lambda]$ doublet contains altogether six charmed baryons: $\Sigma_c({3\over2}^-/{5\over2}^-)$, $\Xi_c({3\over2}^-/{5\over2}^-)$, and $\Omega_c({3\over2}^-/{5\over2}^-)$. We study their $S$- and $D$-wave decays into ground-state charmed baryons together with light pseudoscalar and vector mesons. We derive the following non-zero coupling constants:
\begin{eqnarray}
\nonumber &(b3)& g^S_{\Sigma_c[{3\over2}^-]\to \Sigma_c^{*}[{3\over2}^+]\pi} = 0.003^{+0.002}_{-0.001}~{\rm GeV}^{-2} \, ,
\\
\nonumber &(b5)& g^S_{\Sigma_c[{3\over2}^-]\to \Sigma_c[{1\over2}^+] \rho} = 13.46{^{+7.71}_{-6.18}}~{\rm GeV}^{-2} \, ,
\\
\nonumber &(b6)& g^S_{\Sigma_c[{3\over2}^-]\to \Sigma_c^{*}[{3\over2}^+] \rho } = 1.49~{\rm GeV}^{-2} \, ,
\\
\nonumber &(c4)& g^S_{\Sigma_c[{5\over2}^-]\to \Sigma_c^{*}[{3\over2}^+] \rho} = 2.55^{+1.41}_{-1.11}~{\rm GeV}^{-2} \, ,
\\
\nonumber &(b1)& g^D_{\Sigma_c[{3\over2}^-]\to \Lambda_c[{1\over2}^+]\pi} = 6.13^{+3.54}_{-2.65}~{\rm GeV}^{-2}\, ,
\\
\nonumber &(b2)& g^D_{\Sigma_c[{3\over2}^-]\to \Sigma_c[{1\over2}^+]\pi} = 3.15^{+2.20}_{-1.86}~{\rm GeV}^{-2}\, ,
\\
\nonumber &(b3)& g^D_{\Sigma_c[{3\over2}^-]\to \Sigma_c^{*}[{3\over2}^+]\pi} = 1.82^{+1.27}_{-1.07}~{\rm GeV}^{-2}\, ,
\\
\nonumber &(c1)& g^D_{\Sigma_c[{5\over2}^-]\to \Lambda_c[{1\over2}^+]\pi} = 6.13^{+3.54}_{-2.65}~{\rm GeV}^{-2}\, ,
\\
\nonumber &(c2)& g^D_{\Sigma_c[{5\over2}^-]\to \Sigma_c[{1\over2}^+] \pi} = 2.10^{+1.46}_{-1.24}~{\rm GeV}^{-2} \, ,
\\
\nonumber &(c3)& g^D_{\Sigma_c[{5\over2}^-]\to \Sigma_c^{*}[{3\over2}^+] \pi}= 1.62^{+1.13}_{-0.95}~{\rm GeV}^{-2} \, ,
\\
\nonumber &(e5)& g^S_{\Xi_c^{\prime}[{3\over2}^-]\to \Xi_c^{*}[{3\over2}^+]\pi} = 0.002{^{+0.029}_{-0.001}}~{\rm GeV}^{-2} \, ,
\\
\nonumber &(e6)& g^S_{\Xi_c^{\prime}[{3\over2}^-]\to \Sigma_c^{*}[{3\over2}^+]\bar K} = 0.005~{\rm GeV}^{-2} \, ,
\\
\nonumber &(e9)& g^S_{\Xi_c^{\prime}[{3\over2}^-]\to \Xi_c^{\prime}[{1\over2}^+]\rho} = 9.65{^{+5.94}_{-4.53}}~{\rm GeV}^{-2} \, ,
\\
\nonumber &(e10)& g^S_{\Xi_c^{\prime}[{3\over2}^-]\to \Sigma_c[{1\over2}^+]\bar K^*} = 7.65~{\rm GeV}^{-2} \, ,
\\
\nonumber &(e11)& g^S_{\Xi_c^{\prime}[{3\over2}^-]\to \Xi_c^{*}[{3\over2}^+]\rho} = 1.08~{\rm GeV}^{-2} \, ,
\\
\nonumber &(e12)& g^S_{\Xi_c^{\prime}[{3\over2}^-]\to \Sigma_c^{*}[{3\over2}^+]\bar K^*} = 0.95~{\rm GeV}^{-2} \, ,
\\
\nonumber &(f8)& g^S_{\Xi_c^{\prime}[{5\over2}^-]\to \Xi_c^{*}[{3\over2}^+]\rho} = 0.69~{\rm GeV}^{-2} \, ,
\\
\nonumber &(f7)& g^S_{\Xi_c^{\prime}[{5\over2}^-]\to \Sigma_c^{*}[{3\over2}^+]\bar K^*} = 1.44~{\rm GeV}^{-2} \, ,
\\
\nonumber &(e1)& g^D_{\Xi_c^{\prime}[{3\over2}^-]\to \Xi_c[{1\over2}^+]\pi} = 3.68{^{+2.42}_{-1.69}}~{\rm GeV}^{-2} \, ,
\\
\nonumber &(e2)& g^D_{\Xi_c^{\prime}[{3\over2}^-]\to \Lambda_c[{1\over2}^+]\bar K} = 4.02{^{+2.73}_{-1.95}}~{\rm GeV}^{-2} \, ,
\\
          &(e3)& g^D_{\Xi_c^{\prime}[{3\over2}^-]\to \Xi_c^{\prime}[{1\over2}^+]\pi} = 2.52{^{+1.68}_{-1.31}}~{\rm GeV}^{-2} \, ,
\\
\nonumber &(e4)& g^D_{\Xi_c^{\prime}[{3\over2}^-]\to \Sigma_c[{1\over2}^+]\bar K} = 0.91{^{+1.44}_{-1.22}}~{\rm GeV}^{-2} \, ,
\\
\nonumber &(e5)& g^D_{\Xi_c^{\prime}[{3\over2}^-]\to \Xi_c^{*}[{3\over2}^+]\pi} = 1.45{^{+0.97}_{-0.75}}~{\rm GeV}^{-2} \, ,
\\
\nonumber &(e6)& g^D_{\Xi_c^{\prime}[{3\over2}^-]\to \Sigma_c^{*}[{3\over2}^+]\bar K} = 0.52~{\rm GeV}^{-2} \, ,
\\
\nonumber &(f1)& g^D_{\Xi_c^{\prime}[{5\over2}^-]\to \Xi_c[{1\over2}^+]\pi} = 3.68{^{+2.32}_{-1.69}}~{\rm GeV}^{-2} \, ,
\\
\nonumber &(f2)& g^D_{\Xi_c^{\prime}[{5\over2}^-]\to \Lambda_c[{1\over2}^+]\bar K} = 4.02{^{+2.73}_{-1.95}}~{\rm GeV}^{-2} \, ,
\\
\nonumber &(f3)& g^D_{\Xi_c^{\prime}[{5\over2}^-]\to \Xi_c^{\prime}[{1\over2}^+]\pi} = 1.68{^{+1.12}_{-0.87}}~{\rm GeV}^{-2} \, ,
\\
\nonumber &(f4)& g^D_{\Xi_c^{\prime}[{5\over2}^-]\to \Sigma_c[{1\over2}^+]\bar K} = 0.60{^{+0.96}_{-0.60}}~{\rm GeV}^{-2} \, ,
\\
\nonumber &(f5)& g^D_{\Xi_c^{\prime}[{5\over2}^-]\to \Xi_c^{*}[{3\over2}^+]\pi} = 1.29{^{+0.86}_{-0.67}}~{\rm GeV}^{-2} \, ,
\\
\nonumber &(f6)& g^D_{\Xi_c^{\prime}[{5\over2}^-]\to \Sigma_c^{*}[{3\over2}^+]\bar K} = 0.46{^{+0.74}_{-0.63}}~{\rm GeV}^{-2} \, ,
\\
\nonumber &(h3)& g^S_{\Omega_c[{3\over2}^-]\to \Xi_c^{*}[{3\over2}^+]\bar K} =0.007~{\rm GeV}^{-2}\, ,
\\
\nonumber &(f5)& g^S_{\Omega_c[{3\over2}^-]\to \Xi_c^{\prime}[{1\over2}^+]\bar K^*}= 11.41~{\rm GeV}^{-2} \, ,
\\
\nonumber &(f6)& g^S_{\Omega_c[{3\over2}^-]\to \Xi_c^{*}[{1\over2}^+]\bar K^*}= 1.48~{\rm GeV}^{-2} \, ,
\\
\nonumber &(i4)& g^S_{\Omega_c[{5\over2}^-]\to \Xi_c^{*}[{1\over2}^+]\bar K^*}= 3.01~{\rm GeV}^{-2} \, ,
\\
\nonumber &(g1)& g^D_{\Omega_c[{3\over2}^-]\to \Xi_c[{1\over2}^+]\bar K}= 4.68{^{+3.09}_{-2.35}}~{\rm GeV}^{-2} \, ,
\\
\nonumber &(g2)& g^D_{\Omega_c[{3\over2}^-]\to \Xi_c^{\prime}[{1\over2}^+]\bar K}= 2.26{^{+2.07}_{-1.80}}~{\rm GeV}^{-2} \, ,
\\
\nonumber &(g3)& g^D_{\Omega_c[{3\over2}^-]\to \Xi_c^{*}[{3\over2}^+]\bar K}= 1.31~{\rm GeV}^{-2} \, ,
\\
\nonumber &(i1)& g^D_{\Omega_c[{5\over2}^-]\to \Xi_c[{1\over2}^+]\bar K}= 4.68^{+3.09}_{-2.35}~{\rm GeV}^{-2} \, ,
\\
\nonumber &(i2)& g^D_{\Omega_c[{5\over2}^-]\to \Xi_c^{\prime}[{1\over2}^+]\bar K}= 1.51^{+1.38}_{-1.19}~{\rm GeV}^{-2} \, ,
\\
\nonumber &(h3)& g^D_{\Omega_c[{5\over2}^-]\to \Xi_c^{*}[{3\over2}^+]\bar K}= 1.16~{\rm GeV}^{-2} \, .
\end{eqnarray}
Some of these coupling constants are shown in Fig.~\ref{fig:621lambda} as functions of the Borel mass $T$. We further use these coupling constants to derive the following decay channels that are kinematically allowed:
\begin{eqnarray}
\nonumber &(b3)& \Gamma^S_{\Sigma_c[{3\over2}^-]\to \Sigma_c^{*}[{3\over2}^+]\pi} =6\times10^{-4}~{\rm MeV} \, ,
\\
\nonumber &(b5)& \Gamma^S_{\Sigma_c[{3\over2}^-]\to \Sigma_c[{1\over2}^+] \rho\to\Sigma_c[{1\over2}^+]\pi\pi} =0.72{^{+1.06}_{-0.51}}~{\rm MeV} \, ,
\\
\nonumber &(b6)& \Gamma^S_{\Sigma_c[{3\over2}^-]\to \Sigma_c^{*}[{3\over2}^+] \rho \to \Sigma_c^{*}[{3\over2}^+]\pi\pi} = 5\times10^{-6}~{\rm MeV} \, ,
\\
\nonumber &(c4)& \Gamma^S_{\Sigma_c[{5\over2}^-]\to \Sigma_c^{*}[{3\over2}^+] \rho\to \Sigma_c^{*}[{3\over2}^+]\pi\pi}= 4\times10^{-5}~{\rm MeV} \, ,
\\
\nonumber &(b1)& \Gamma^D_{\Sigma_c[{3\over2}^-]\to \Lambda_c[{1\over2}^+]\pi}= 36{^{+54}_{-25}}~{\rm MeV} \, ,
\\
\nonumber &(b2)& \Gamma^D_{\Sigma_c[{3\over2}^-]\to \Sigma_c[{1\over2}^+]\pi} = 2.5{^{+4.6}_{-2.1}}~{\rm MeV} \, ,
\\
\nonumber &(b3)& \Gamma^D_{\Sigma_c[{3\over2}^-]\to \Sigma_c^{*}[{3\over2}^+]\pi}= 0.18{^{+0.34}_{-0.15}}~{\rm MeV } \, ,
\\
\nonumber &(c1)& \Gamma^D_{\Sigma_c[{5\over2}^-]\to \Lambda_c[{1\over2}^+]\pi}= 12{^{+18}_{-~8}}~{\rm MeV} \, ,
\\
\nonumber &(c2)& \Gamma^D_{\Sigma_c[{5\over2}^-]\to \Sigma_c[{1\over2}^+] \pi} = 0.39{^{+0.72}_{-0.32}}~{\rm MeV} \, ,
\\
\nonumber &(c3)& \Gamma^D_{\Sigma_c[{5\over2}^-]\to \Sigma_c^{*}[{3\over2}^+] \pi}= 0.61{^{+1.14}_{-0.50}}~{\rm MeV} \, ,
\\
\nonumber &(e5)& \Gamma^S_{\Xi_c^{\prime}[{3\over2}^-]\to \Xi_c^{*}[{3\over2}^+]\pi} = 2\times10^{-4}~{\rm MeV} \, ,
\\
\nonumber &(e9)& \Gamma^S_{\Xi_c^{\prime}[{3\over2}^-]\to \Xi_c^{\prime}[{1\over2}^+]\rho\to\Xi_c^{\prime}[{1\over2}^+]\pi\pi} = 1.4{^{+2.2}_{-1.0}}~{\rm MeV} \, ,
\\
\nonumber &(e11)& \Gamma^S_{\Xi_c^{\prime}[{3\over2}^-]\to \Xi_c^{*}[{3\over2}^+]\rho \to \Xi_c^{*}[{3\over2}^+]\pi\pi} = 0.001~{\rm MeV} \, ,
\\
\nonumber &(f8)& \Gamma^S_{\Xi_c^{\prime}[{5\over2}^-]\to \Xi_c^{*}[{3\over2}^+]\rho\to\Xi_c^{*}[{3\over2}^+]\pi\pi} = 0.02~{\rm MeV} \, ,
\\
          &(e1)& \Gamma^D_{\Xi_c^{\prime}[{3\over2}^-]\to \Xi_c[{1\over2}^+]\pi} = 17{^{+30}_{-12}}~{\rm GeV} \, ,
\\
\nonumber &(e2)& \Gamma^D_{\Xi_c^{\prime}[{3\over2}^-]\to \Lambda_c[{1\over2}^+]\bar K} = 9.8{^{+17.9}_{-~7.2}}~{\rm MeV} \, ,
\\
\nonumber &(e3)& \Gamma^D_{\Xi_c^{\prime}[{3\over2}^-]\to \Xi_c^{\prime}[{1\over2}^+]\pi} = 2.3{^{+4.0}_{-1.7}}~{\rm MeV} \, ,
\\
\nonumber &(e4)& \Gamma^D_{\Xi_c^{\prime}[{3\over2}^-]\to \Sigma_c[{1\over2}^+]\pi} = 0.003~{\rm MeV} \, ,
\\
\nonumber &(e5)& \Gamma^D_{\Xi_c^{\prime}[{3\over2}^-]\to \Xi_c^{*}[{3\over2}^+]\pi} = 0.19{^{+0.33}_{-0.14}}~{\rm MeV} \, ,
\\
\nonumber &(f1)& \Gamma^D_{\Xi_c^{\prime}[{5\over2}^-]\to \Xi_c[{1\over2}^+]\pi} = 9.6{^{+15.8}_{-~6.8}}~{\rm MeV} \, ,
\\
\nonumber &(f2)& \Gamma^D_{\Xi_c^{\prime}[{5\over2}^-]\to \Lambda_c[{1\over2}^+]\bar K} = 6.3{^{+11.4}_{-~4.6}}~{\rm MeV} \, ,
\\
\nonumber &(f3)& \Gamma^D_{\Xi_c^{\prime}[{5\over2}^-]\to \Xi_c^{\prime}[{1\over2}^+]\pi} = 0.70{^{+1.25}_{-0.54}}~{\rm MeV} \, ,
\\
\nonumber &(f4)& \Gamma^D_{\Xi_c^{\prime}[{5\over2}^-]\to \Sigma_c[{1\over2}^+]\pi} = 0.02~{\rm MeV} \, ,
\\
\nonumber &(f5)& \Gamma^D_{\Xi_c^{\prime}[{5\over2}^-]\to \Xi_c^{*}[{3\over2}^+]\pi} = 1.5{^{+2.7}_{-1.1}}~{\rm MeV} \, ,
\\
\nonumber &(f6)& \Gamma^D_{\Xi_c^{\prime}[{5\over2}^-]\to \Sigma_c^{*}[{3\over2}^+]\pi} = 0.004~{\rm MeV} \, ,
\\
\nonumber &(g1)& \Gamma^D_{\Omega_c[{3\over2}^-]\to \Xi_c[{1\over2}^+]\bar K}= 9.9{^{+17.4}_{-~7.4}}~{\rm MeV} \, ,
\\
\nonumber &(g2)& \Gamma^D_{\Omega_c[{3\over2}^-]\to \Xi_c^{\prime}[{1\over2}^+]\bar K}= 0.10{^{+0.26}_{-0.09}}~{\rm MeV} \, ,
\\
\nonumber &(i1)& \Gamma^D_{\Omega_c[{5\over2}^-]\to \Xi_c[{1\over2}^+]\bar K}= 5.5{^{+9.6}_{-4.1}}~{\rm MeV} \, ,
\\
\nonumber &(i2)& \Gamma^D_{\Omega_c[{5\over2}^-]\to \Xi_c^{\prime}[{1\over2}^+]\bar K}= 0.03~{\rm MeV} \, .
\end{eqnarray}
We summarize the above results in Table~\ref{tab:decay621lambda}.

%
\section{Mixing between $[\mathbf{6}_F, 1, 1, \lambda]$ and $[\mathbf{6}_F, 2, 1, \lambda]$}
\label{sec:mixing}
%

In this section we investigate the mixing effect between two different HQET multiplets, especially, between the $[\mathbf{6}_F, 1, 1, \lambda]$ and $[\mathbf{6}_F, 2, 1, \lambda]$ multiplets. The mixing between $[\Xi_c^\prime(3/2^-), 1, 1, \lambda]$ and $[\Xi_c^\prime(3/2^-), 2, 1, \lambda]$ has been carefully examined in Ref.~\cite{Yang:2020zjl}, and the same procedures will be applied here.

From Tables~\ref{tab:decay611lambda} and \ref{tab:decay621lambda} we find it possible to interpret the $\Xi_c(2923)^0$, $\Xi_c(2939)^0$, and $\Xi_c(2965)^0$ as the $P$-wave $\Xi_c^\prime$ baryons $[\Xi_c^{\prime}(1/2^-), 1, 1, \lambda]$, $[\Xi_c^{\prime}(3/2^-), 1, 1, \lambda]$, and $[\Xi_c^{\prime}(3/2^-), 2, 1, \lambda]$, respectively. However, there are three discrepancies between these theoretical results and the LHCb experiment~\cite{Aaij:2020yyt}: a) the missing of the $\Lambda^+_c K^-$ decay channel for the $\Xi_c(2923)^0$ and $\Xi_c(2939)^0$, b) the mass splitting between the $\Xi_c(2923)^0$ and $\Xi_c(2939)^0$, and c) total widths of the $\Xi_c(2939)^0$ and $\Xi_c(2965)^0$.

To explain these discrepancies, we recall that the HQET is an effective theory, which works quite well for bottom baryons~\cite{Yang:2020zrh}, but not so perfect for charmed baryons~\cite{Yang:2020zjl}. Therefore, the three $J^P=1/2^-$ charmed baryons can mix, and the three $J^P=3/2^-$ charmed baryons can also mix. Accordingly, we just need a tiny mixing angle $\theta_1 \approx 0^\circ$ to make it possible to observe all the $P$-wave $\Xi_c^\prime$ baryons in the $\Lambda_c K$ decay channel.

In the present study we explicitly consider the mixing between the $[{\bf 6}_F, 1, 1, \lambda]$ and $[{\bf 6}_F, 2, 1, \lambda]$ doublets:
\begin{eqnarray}
\left(\begin{array}{c}
|\Sigma_c(3/2^-)\rangle_1\\
|\Sigma_c(3/2^-)\rangle_2
\end{array}\right)
&=&
\left(\begin{array}{cc}
\cos\theta_2 & \sin\theta_2 \\
-\sin\theta_2 & \cos\theta_2
\end{array}\right)
\\ \nonumber && ~~~~~~~ \times
\left(\begin{array}{c}
|\Sigma_c(3/2^-),1,1,\lambda\rangle\\
|\Sigma_c(3/2^-),2,1,\lambda\rangle
\end{array}\right) \, ,
\\
\left(\begin{array}{c}
|\Xi_c^\prime(3/2^-)\rangle_1\\
|\Xi_c^\prime(3/2^-)\rangle_2
\end{array}\right)
&=&
\left(\begin{array}{cc}
\cos\theta_2 & \sin\theta_2 \\
-\sin\theta_2 & \cos\theta_2
\end{array}\right)
\\ \nonumber && ~~~~~~~ \times
\left(\begin{array}{c}
|\Xi_c^\prime(3/2^-),1,1,\lambda\rangle\\
|\Xi_c^\prime(3/2^-),2,1,\lambda\rangle
\end{array}\right) \, ,
\\
\left(\begin{array}{c}
|\Omega_c(3/2^-)\rangle_1\\
|\Omega_c(3/2^-)\rangle_2
\end{array}\right)
&=&
\left(\begin{array}{cc}
\cos\theta_2 & \sin\theta_2 \\
-\sin\theta_2 & \cos\theta_2
\end{array}\right)
\\ \nonumber && ~~~~~~~ \times
\left(\begin{array}{c}
|\Omega_c(3/2^-),1,1,\lambda\rangle\\
|\Omega_c(3/2^-),2,1,\lambda\rangle
\end{array}\right) \, ,
\end{eqnarray}
where $\theta_2$ is an overall mixing angle.

We fine-tune it to be $\theta_2 = 37\pm5^\circ$, and the obtained results are shown in Table~\ref{tab:result}. This mixing mediates widths of the $[\Xi_c^\prime(3/2^-), 1, 1, \lambda]$ and $[\Xi_c^\prime(3/2^-), 2, 1, \lambda]$, and decreases the mass splitting within the $[\Xi_c^\prime, 1, 1, \lambda]$ doublet:
\begin{eqnarray*}
\nonumber M_{[\Xi_c^{\prime}(3/2^-), 1, 1, \lambda]} &:& 2.95^{+0.12}_{-0.11}~{\rm GeV} \longrightarrow 2.94^{+0.12}_{-0.11}~{\rm GeV} \, ,
\\ \nonumber \Gamma_{[\Xi_c^{\prime}(3/2^-), 1, 1, \lambda]} &:& 4.4^{+4.5}_{-2.3}~{\rm MeV} \longrightarrow 12^{+10}_{-~4}~{\rm MeV} \, ,
\\ M_{[\Xi_c^{\prime}(3/2^-), 2, 1, \lambda]} &:& 2.96^{+0.24}_{-0.15}~{\rm GeV} \longrightarrow 2.97^{+0.24}_{-0.15}~{\rm GeV} \, ,
\\ \nonumber \Gamma_{[\Xi_c^{\prime}(3/2^-), 2, 1, \lambda]} &:& 31^{+35}_{-14}~{\rm MeV} \longrightarrow 19^{+22}_{-~9}~{\rm MeV} \, ,
\\ \nonumber \Delta M_{[\Xi_c^{\prime}, 1, 1, \lambda]} &:& 38^{+15}_{-13}~{\rm MeV} \longrightarrow 27^{+16}_{-27}~{\rm MeV} \, ,
\\ \nonumber \Delta M_{[\Xi_c^{\prime}, 2, 1, \lambda]} &:& 66^{+29}_{-25}~{\rm MeV} \longrightarrow 56^{+30}_{-35}~{\rm MeV} \, .
\end{eqnarray*}
Now the $\Xi_c(2939)^0$ and $\Xi_c(2965)^0$ can be well interpreted as the two $J^P = 3/2^-$ baryons $|\Xi_c^\prime(3/2^-)\rangle_1$ and $|\Xi_c^\prime(3/2^-)\rangle_2$, respectively.

Similarly, we study the mixing between $[\Sigma_c(3/2^-), 1, 1, \lambda]$ and $[\Sigma_c(3/2^-), 2, 1, \lambda]$ as well as the mixing between $[\Omega_c(3/2^-), 1, 1, \lambda]$ and $[\Omega_c(3/2^-), 2, 1, \lambda]$. The obtained results are summarized in Table~\ref{tab:result}, supporting to explain the $\Omega_c(3066)^0$ and $\Omega_c(3090)^0$ as the two $J^P = 3/2^-$ baryons $|\Omega_c(3/2^-)\rangle_1$ and $|\Omega_c(3/2^-)\rangle_2$, respectively.

Note that only those non-negligible decay channels are listed in Table~\ref{tab:result} due to limited spaces, while we have calculated all the possible decay channels, as follows:
\begin{itemize}

\item Partial decay widths of the $|\Sigma_c({3/2}^-)\rangle_1$ are:
\begin{eqnarray}
\nonumber &(b1)& \Gamma^D_{|\Sigma_c({3/2}^-)\rangle_1\to\Lambda_c\pi}=13^{+20}_{-~9}~{\rm MeV} \ ,
\\ \nonumber &(b2)& \Gamma^D_{|\Sigma_c({3/2}^-)\rangle_1\to\Sigma_c\pi}=3.3^{+4.2}_{-2.2}~{\rm MeV} \ ,
\\ \nonumber &(b3)& \Gamma^D_{|\Sigma_c({3/2}^-)\rangle_1\to\Sigma_c^{*}\pi}=0.24^{+0.31}_{-0.16}~{\rm MeV} \ ,
\\ &(b3)& \Gamma^S_{|\Sigma_c({3/2}^-)\rangle_1\to\Sigma_c^{*}\pi}=6.4^{+10.3}_{-~4.7}~{\rm MeV} \ ,
\label{result:mixing1}
\\ \nonumber &(b4)& \Gamma^S_{|\Sigma_c({3/2}^-)\rangle_1\to\Lambda_c\rho\to\Lambda_c\pi\pi}=0.59^{+1.67}_{-0.58}~{\rm MeV} \ ,
\\ \nonumber &(b5)& \Gamma^S_{|\Sigma_c({3/2}^-)\rangle_1\to\Sigma_c\rho\to\Sigma_c\pi\pi}=0.75^{+0.80}_{-0.44}~{\rm MeV} \ ,
\\ \nonumber &(b6)& \Gamma^S_{|\Sigma_c({3/2}^-)\rangle_1\to\Sigma_c^{*}\rho\to\Sigma_c^{*}\pi\pi}=1\times10^{-4}~{\rm MeV} \ .
\label{eq:widths}
\end{eqnarray}

\item Partial decay widths of the $|\Sigma_c({3/2}^-)\rangle_2$ are:
\begin{eqnarray}
\nonumber &(b1)& \Gamma^D_{|\Sigma_c({3/2}^-)\rangle_2\to\Lambda_c\pi}=23^{+35}_{-16}~{\rm MeV} \ ,
\\ \nonumber &(b2)& \Gamma^D_{|\Sigma_c({3/2}^-)\rangle_2\to\Sigma_c\pi}=0.37^{+2.27}_{-0.37}~{\rm MeV} \ ,
\\ \nonumber &(b3)& \Gamma^D_{|\Sigma_c({3/2}^-)\rangle_2\to\Sigma_c^{*}\pi}=0.03~{\rm MeV} \ ,
\\ &(b3)& \Gamma^S_{|\Sigma_c({3/2}^-)\rangle_2\to\Sigma_c^{*}\pi}=3.5^{+6.1}_{-2.7}~{\rm MeV} \ ,
\label{result:mixing2}
\\ \nonumber &(b4)& \Gamma^S_{|\Sigma_c({3/2}^-)\rangle_2\to\Lambda_c\rho\to\Lambda_c\pi\pi}=0.33^{+0.93}_{-0.33}~{\rm MeV} \ ,
\\ \nonumber &(b5)& \Gamma^S_{|\Sigma_c({3/2}^-)\rangle_2\to\Sigma_c\rho\to\Sigma_c\pi\pi}=0.17^{+0.72}_{-0.17}~{\rm MeV} \ ,
\\ \nonumber &(b6)& \Gamma^S_{|\Sigma_c({3/2}^-)\rangle_2\to\Sigma_c^{*}\rho\to\Sigma_c^{*}\pi\pi}=3\times10^{-5}~{\rm MeV} \ .
\end{eqnarray}

\item Partial decay widths of the $|\Xi_c^{\prime}({3/2}^-)\rangle_1$ are:
\begin{eqnarray}
\nonumber &(e2)& \Gamma^D_{|\Xi_c^{\prime}({3/2}^-)\rangle_1\to \Lambda_c \bar K}=2.3^{+4.3}_{-1.7}~{\rm MeV} \ ,
\\ \nonumber &(e1)& \Gamma^D_{|\Xi_c^{\prime}({3/2}^-)\rangle_1\to \Xi_c\pi}=4.6^{+8.1}_{-3.3}~{\rm MeV} \ ,
\\ \nonumber &(e3)& \Gamma^D_{|\Xi_c^{\prime}({3/2}^-)\rangle_1\to\Xi_ c^{\prime}\pi}=2.0^{+2.2}_{-1.2}~{\rm MeV} \ ,
\\ \nonumber &(e5)& \Gamma^S_{|\Xi_c^{\prime}({3/2}^-)\rangle_1\to \Xi_c^{*}\pi}=2.1^{+2.6}_{-1.5}~{\rm MeV} \ ,
\\ &(e5)& \Gamma^D_{|\Xi_c^{\prime}({3/2}^-)\rangle_1\to\Xi_c^{*}\pi}=0.14^{+0.16}_{-0.08}~{\rm MeV} \ ,
\label{result:mixing3}
\\ \nonumber &(e8)& \Gamma^S_{|\Xi_c^{\prime}({3/2}^-)\rangle_1\to\Lambda_c \bar K^{*}\to\Lambda_c \bar K\pi}=2 \times 10^{-6}~{\rm MeV} \ ,
\\ \nonumber &(e7)& \Gamma^S_{|\Xi_c^{\prime}({3/2}^-)\rangle_1\to\Xi_c\rho\to\Xi_c\pi\pi}=0.13^{+0.39}_{-0.13}~{\rm MeV} \ ,
\\ \nonumber &(e9)& \Gamma^S_{|\Xi_c^{\prime}({3/2}^-)\rangle_1\to\Xi_c^{\prime}\rho\to\Xi_c^{\prime}\pi\pi}=0.53^{+0.57}_{-0.31}~{\rm MeV} \ ,
\\ \nonumber &(e11)& \Gamma^S_{|\Xi_c^{\prime}({3/2}^-)\rangle_1\to\Xi_c^{*}\rho\to\Xi_c^{*}\pi\pi}=1 \times 10^{-3}~{\rm MeV} \ .
\end{eqnarray}

\item Partial decay widths of the $|\Xi_c^{\prime}({3/2}^-)\rangle_2$ are:
\begin{eqnarray}
\nonumber &(e2)& \Gamma^D_{|\Xi_c^{\prime}({3/2}^-)\rangle_2\to \Lambda_c \bar K}=6.3^{+11.6}_{-~4.7}~{\rm MeV} \ ,
\\ \nonumber &(e1)& \Gamma^D_{|\Xi_c^{\prime}({3/2}^-)\rangle_2\to \Xi_c\pi}=11^{+19}_{-~8}~{\rm MeV} \ ,
\\ \nonumber &(e4)& \Gamma^D_{|\Xi_c^{\prime}({3/2}^-)\rangle_2\to \Sigma_c \bar K}=4 \times10^{-4}~{\rm MeV} \ ,
\\ \nonumber &(e3)& \Gamma^D_{|\Xi_c^{\prime}({3/2}^-)\rangle_2\to \Xi_c^{\prime}\pi}=0.37^{+1.89}_{-0.37}~{\rm MeV} \ ,
\\ \nonumber &(e5)& \Gamma^S_{|\Xi_c^{\prime}({3/2}^-)\rangle_2\to \Xi_c^{*}\pi}= 1.3^{+1.8}_{-0.9}~{\rm MeV} \ ,
\\ &(e5)& \Gamma^D_{|\Xi_c^{\prime}({3/2}^-)\rangle_2\to \Xi_c^{*}\pi}=0.03~{\rm MeV} \ ,
\label{result:mixing4}
\\ \nonumber &(e8)& \Gamma^S_{|\Xi_c^{\prime}({3/2}^-)\rangle_2\to\Lambda_c \bar K^{*}\to\Lambda_c \bar K\pi}=2 \times 10^{-5}~{\rm MeV} \ ,
\\ \nonumber &(e7)& \Gamma^S_{|\Xi_c^{\prime}({3/2}^-)\rangle_2\to\Xi_c\rho\to\Xi_c\pi\pi}=0.12^{+0.36}_{-0.12}~{\rm MeV} \ ,
\\ \nonumber &(e9)& \Gamma^S_{|\Xi_c^{\prime}({3/2}^-)\rangle_2\to\Xi_c^{\prime}\rho\to\Xi_c^{\prime}\pi\pi}=0.37^{+1.15}_{-0.36}~{\rm MeV} \ ,
\\ \nonumber &(e11)& \Gamma^S_{|\Xi_c^{\prime}({3/2}^-)\rangle_2\to\Xi_c^{*}\rho\to\Xi_c^{*}\pi\pi}=5 \times 10^{-3}~{\rm MeV} \ .
\end{eqnarray}

\item Partial decay widths of the $|\Omega_c({3/2}^-)\rangle_1$ are:
\begin{eqnarray}
&(h1)& \Gamma^D_{|\Omega_c({3/2}^-)\rangle_1\to \Xi_c \bar K}=2.0^{+3.5}_{-1.5}~{\rm MeV} \ .
\label{result:mixing5}
\end{eqnarray}

\item Partial decay widths of the $|\Omega_c({3/2}^-)\rangle_2$ are:
\begin{eqnarray}
\nonumber &(h1)& \Gamma^D_{|\Omega_c({3/2}^-)\rangle_2\to \Xi_c \bar K}=6.3^{+11.2}_{-~4.8}~{\rm MeV} \ ,
\\ &(h2)& \Gamma^D_{|\Omega_c({3/2}^-)\rangle_2\to \Xi_c^{\prime} \bar K}=0.02~{\rm MeV} \ .
\label{result:mixing6}
\end{eqnarray}
\end{itemize}

%
\section{Summary and Discussions}\label{sec:summary}
%

\begin{table*}[hbt]
\begin{center}
\renewcommand{\arraystretch}{1.5}
\caption{Decay properties of $P$-wave charmed baryons belonging to the $[\mathbf{6}_F, 1, 0, \rho]$ doublet, with possible experimental candidates given in the last column.}
\begin{tabular}{ c | c | c | c | c | c | c | c}
\hline\hline
Baryon & ~~~~Mass~~~~ & Difference & \multirow{2}{*}{~~~~~~~~~Decay channels~~~~~~~~~}  & ~$S$-wave width~  & ~$D$-wave width~ & ~Total width~ & \multirow{2}{*}{~Candidate~}
\\ ($j^P$) & ({GeV})& ({MeV}) & & ({MeV}) & ({MeV}) & ({MeV})
\\ \hline\hline
\multirow{4}{*}{$\Sigma_c({1\over2}^-)$}&\multirow{4}{*}{$2.77^{+0.16}_{-0.12}$}&\multirow{8}{*}{$15^{+6}_{-5}$}&$\Sigma_c({1\over2}^-)\to \Sigma_c\pi$&$380^{+630}_{-270}$&--&\multirow{4}{*}{$390^{+630}_{-270}$}&\multirow{4}{*}{--}
\\ \cline{4-6}
&&&$\Sigma_c({1\over2}^-)\to \Sigma_c^{*} \pi$  & --  & $ 0.82^{+1.40}_{-0.59}$ &
\\ \cline{4-6}
&&&$\Sigma_c({1\over2}^-)\to \Lambda_c\rho\to \Lambda_c\pi\pi$& \multicolumn{2}{c|}{$0.06$ }&
\\ \cline{4-6}
&&&$\Sigma_c({1\over2}^-)\to \Sigma_c\rho\to \Sigma_c\pi\pi$& \multicolumn{2}{c|}{$3\times 10^{-5}$}&
\\ \cline{1-2} \cline{4-8}
\multirow{4}{*}{$\Sigma_c({3\over2}^-)$}&\multirow{4}{*}{$2.79^{+0.16}_{-0.12}$}&&$\Sigma_c({3\over2}^-)\to \Sigma_c \pi$ & -- & $3.1^{+4.6}_{-2.3}$&\multirow{4}{*}{$220^{+360}_{-150}$}&\multirow{4}{*}{--}
\\ \cline{4-6}
&&&$\Sigma_c({3\over2}^-)\to \Sigma_c^{*} \pi$ & $220^{+360}_{-150}$& $ 0.21^{+0.34}_{-0.15}$&
\\ \cline{4-6}
&&&$\Sigma_c({3\over2}^-)\to \Lambda_c\rho\to \Lambda_c\pi\pi$&\multicolumn{2}{c|}{$0.08$}&
\\ \cline{4-6}
&&&$\Sigma_c({3\over2}^-)\to \Sigma_c\rho\to \Sigma_c\pi\pi$&\multicolumn{2}{c|}{$4\times10^{-5}$}&
\\ \hline
\multirow{4}{*}{$\Xi^\prime_c({1\over2}^-)$}&\multirow{4}{*}{$2.88^{+0.15}_{-0.13}$}&\multirow{8}{*}{$13^{+6}_{-5}$}&$\Xi_c^{\prime}({1\over2}^-)\to \Xi_c^{\prime}\pi$&$110^{+170}_{-~80}$&--&\multirow{4}{*}{$110^{+170}_{-~80}$}&\multirow{4}{*}{--}
\\ \cline{4-6}
&&&$\Xi_c^{\prime}({1\over2}^-)\to\Xi_c^{*}\pi$&--&$0.15^{+0.23}_{-0.11}$&
\\ \cline{4-6}
&&&$\Xi_c^{\prime}({1\over2}^-)\to\Xi_c\rho\to\Xi_c\pi\pi$&\multicolumn{2}{c|}{$2\times10^{-4}$}&
\\ \cline{4-6}
&&&$\Xi_c^{\prime}({1\over2}^-)\to\Xi_c^{\prime}\rho\to\Xi_c^{\prime}\pi\pi$&\multicolumn{2}{c|}{$5\times10^{-9}$}&
\\ \cline{1-2} \cline{4-8}
\multirow{4}{*}{$\Xi_c^{\prime}({3\over2}^-)$}&\multirow{4}{*}{$2.89^{+0.15}_{-0.13}$}&&$\Xi_c^{\prime}({3\over2}^-)\to\Xi_ c^{\prime}\pi$&--&$0.63^{+0.99}_{-0.45}$&\multirow{4}{*}{$59^{+88}_{-39}$}&\multirow{4}{*}{--}
\\ \cline{4-6}
&&&$\Xi_c^{\prime}({3\over2}^-)\to\Xi_c^{*}\pi$&$58^{+88}_{-39}$&$0.03^{+0.05}_{-0.02}$&
\\ \cline{4-6}
&&&$\Xi_c^{\prime}({3\over2}^-)\to \Xi_c\rho\to\Xi_c\pi\pi$&\multicolumn{2}{c|}{$5\times10^{-4}$}&
\\ \cline{4-6}
&&&$\Xi_c^{\prime}({3\over2}^-)\to \Xi_c^{\prime}\rho\to\Xi_c^{\prime}\pi\pi$&\multicolumn{2}{c|}{$2\times10^{-9}$}&
\\ \hline
$\Omega_c({1\over2}^-)$&$2.99^{+0.15}_{-0.15} $&\multirow{2}{*}{$12^{+5}_{-5}$}&\multicolumn{3}{c|}{--}&$\sim~0$&\multirow{2}{*}{$\Omega_c(3000)^0$}
\\ \cline{1-2} \cline{4-7}
$\Omega_c({3\over2}^-)$&$3.00^{+0.15}_{-0.15}$&&\multicolumn{3}{c|}{--}&$\sim~0$
\\ \hline\hline
\end{tabular}
\label{tab:decay610rho}
\end{center}
\end{table*}

\begin{table*}[hbt]
\begin{center}
\renewcommand{\arraystretch}{1.5}
\caption{Decay properties of $P$-wave charmed baryons belonging to the $[\mathbf{6}_F,0,1,\lambda]$ singlet, with possible experimental candidates given in the last column.}
\begin{tabular}{c|c|c|c|c|c|c|c}
\hline\hline
Baryon & ~~~~Mass~~~~ & Difference & \multirow{2}{*}{~~~~~~~~~Decay channels~~~~~~~~~}  & ~$S$-wave width~  & ~$D$-wave width~ & ~Total width~ & \multirow{2}{*}{~Candidate~}
\\ ($j^P$) & ({GeV})& ({MeV}) & & ({MeV}) & ({MeV}) & ({MeV})
\\ \hline\hline
\multirow{3}{*}{$\Sigma_c({1\over2}^-)$}&\multirow{3}{*}{$2.83^{+0.06}_{-0.04}$}&\multirow{3}{*}{--}&$\Sigma_c({1\over2}^-)\to\Lambda_c\pi$&$610^{+860}_{-410}$&--&\multirow{3}{*}{$610^{+860}_{-410}$}&\multirow{3}{*}{--}
\\ \cline{4-6}
&&&$\Sigma_c({1\over2}^-)\to\Sigma_c\rho\to\Sigma_c\pi\pi$&\multicolumn{2}{c|}{$1.1^{+1.4}_{-0.7}$}&
\\ \cline{4-6}
&&&$\Sigma_c({1\over2}^-)\to\Sigma_c^{*}\rho\to\Sigma_c^{*}\pi\pi$&\multicolumn{2}{c|}{$0.03$}&
\\ \hline
\multirow{3}{*}{$\Xi_c^{\prime}({1\over2}^-)$}&\multirow{3}{*}{$2.90^{+0.13}_{-0.12}$}&\multirow{3}{*}{--}&$\Xi_c^{\prime}({1\over2}^-)\to\Xi_c\pi$&$360^{+550}_{-250}$&--&\multirow{3}{*}{$760^{+820}_{-370}$}&\multirow{3}{*}{--}
\\ \cline{4-6}
&&&$\Xi_c^{\prime}({1\over2}^-)\to\Lambda_c \bar K$&$400^{+610}_{-270}$&--&
\\ \cline{4-6}
&&&$\Xi_c^{\prime}({1\over2}^-)\to\Xi_c^{\prime}\rho\to\Xi_c^{\prime}\pi\pi$&\multicolumn{2}{c|}{$0.03$}&&
\\ \hline
$\Omega_c({1\over2}^-)$&$3.03^{+0.18}_{-0.19}$&--&$\Omega_c({1\over2}^-)\to\Xi_c \bar K$&$980^{+1530}_{-~670}$&--&$980^{+1530}_{-~670}$&--
\\ \hline\hline
\end{tabular}
\label{tab:decay601lambda}
\end{center}
\end{table*}

\begin{table*}[hbt]
\begin{center}
\renewcommand{\arraystretch}{1.5}
\caption{Decay properties of $P$-wave charmed baryons belonging to the $[\mathbf{6}_F, 1, 1, \lambda]$ doublet, with possible experimental candidates given in the last column.}
\begin{tabular}{ c | c | c | c | c | c | c | c}
\hline\hline
Baryon & ~~~~Mass~~~~ & Difference & \multirow{2}{*}{~~~~~~~~~Decay channels~~~~~~~~~}  & ~$S$-wave width~  & ~$D$-wave width~ & ~Total width~ & \multirow{2}{*}{~Candidate~}
\\ ($j^P$) & ({GeV})& ({MeV}) & & ({MeV}) & ({MeV}) & ({MeV})
\\ \hline\hline
\multirow{5}{*}{$\Sigma_c({1\over2}^-)$}&\multirow{5}{*}{$2.73^{+0.17}_{-0.18}$}&\multirow{10}{*}{$41^{+17}_{-15}$}&$\Sigma_c({1\over2}^-)\to \Sigma_c\pi$&$37^{+60}_{-28}$&--&\multirow{5}{*}{$48^{+70}_{-29}$}&\multirow{10}{*}{$\Sigma_c(2800)^0$}
\\ \cline{4-6}
&&&$\Sigma_c({1\over2}^-)\to \Sigma_c^{*} \pi$  & --  & $ 0.10^{+0.45}_{-0.10}$ &
\\ \cline{4-6}
&&&$\Sigma_c({1\over2}^-)\to \Lambda_c\rho\to \Lambda_c\pi\pi$& \multicolumn{2}{c|}{$9.2^{+37.0}_{-~9.2}$ }&
\\ \cline{4-6}
&&&$\Sigma_c({1\over2}^-)\to \Sigma_c\rho\to \Sigma_c\pi\pi$& \multicolumn{2}{c|}{$1.2^{+2.1}_{-1.0}$}&
\\ \cline{4-6}
&&&$\Sigma_c({1\over2}^-)\to \Sigma_c^{*}\rho\to \Sigma_c^{*}\pi\pi$& \multicolumn{2}{c|}{$1\times10^{-4}$}&
\\ \cline{1-2} \cline{4-7}
\multirow{5}{*}{$\Sigma_c({3\over2}^-)$}&\multirow{5}{*}{$2.77^{+0.17}_{-0.17}$}&&$\Sigma_c({3\over2}^-)\to \Sigma_c \pi$ & -- & $1.2^{+2.4}_{-1.0}$&\multirow{5}{*}{$13^{+17}_{-~8}$}&
\\ \cline{4-6}
&&&$\Sigma_c({3\over2}^-)\to \Sigma_c^{*} \pi$ & $10^{+16}_{-~8}$& $ 0.09$&
\\ \cline{4-6}
&&&$\Sigma_c({3\over2}^-)\to \Lambda_c\rho\to \Lambda_c\pi\pi$&\multicolumn{2}{c|}{$0.92^{+2.58}_{-0.91}$}&
\\ \cline{4-6}
&&&$\Sigma_c({3\over2}^-)\to \Sigma_c\rho\to \Sigma_c\pi\pi$&\multicolumn{2}{c|}{$0.20^{+0.36}_{-0.16}$}&
\\ \cline{4-6}
&&&$\Sigma_c({3\over2}^-)\to \Sigma_c^{*}\rho\to \Sigma_c^{*}\pi\pi$& \multicolumn{2}{c|}{$2\times10^{-4}$}&
\\ \hline
\multirow{6}{*}{$\Xi^\prime_c({1\over2}^-)$}&\multirow{6}{*}{$2.91^{+0.13}_{-0.12}$}&\multirow{12}{*}{$38^{+15}_{-13}$}&$\Xi_c^{\prime}({1\over2}^-)\to \Xi_c^{\prime}\pi$&$12^{+15}_{-~8}$&--&\multirow{6}{*}{$14^{+17}_{-~8}$}&\multirow{6}{*}{$\Xi_c(2923)^0$}
\\ \cline{4-6}
&&&$\Xi_c^{\prime}({1\over2}^-)\to\Xi_c^{*}\pi$&--&$0.12^{+0.22}_{-0.10}$&
\\ \cline{4-6}
&&&$\Xi_c^{\prime}({1\over2}^-)\to\Lambda_c \bar K^*\to\Lambda_c \bar K \pi$&\multicolumn{2}{c|}{$4\times10^{-8}$}&
\\ \cline{4-6}
&&&$\Xi_c^{\prime}({1\over2}^-)\to\Xi_c\rho\to\Xi_c\pi\pi$&\multicolumn{2}{c|}{$1.7^{+7.6}_{-1.7}$}&
\\ \cline{4-6}
&&&$\Xi_c^{\prime}({1\over2}^-)\to\Xi_c^{\prime}\rho\to\Xi_c^{\prime}\pi\pi$&\multicolumn{2}{c|}{$0.38^{+0.54}_{-0.30}$}&
\\ \cline{4-6}
&&&$\Xi_c^{\prime}({1\over2}^-)\to\Xi_c^{*}\rho\to\Xi_c^{*}\pi\pi$&\multicolumn{2}{c|}{$2\times10^{-7}$}&
\\ \cline{1-2} \cline{4-8}
\multirow{6}{*}{$\Xi_c^{\prime}({3\over2}^-)$}&\multirow{6}{*}{$2.95^{+0.12}_{-0.11}$}&&$\Xi_c^{\prime}({3\over2}^-)\to\Xi_ c^{\prime}\pi$&--&$0.67^{+1.06}_{-0.52}$&\multirow{6}{*}{$4.4^{+4.5}_{-2.3}$}&\multirow{6}{*}{$\Xi_c(2939)^0$}
\\ \cline{4-6}
&&&$\Xi_c^{\prime}({3\over2}^-)\to\Xi_c^{*}\pi$&$3.3^{+4.3}_{-2.3}$&$0.05$&
\\ \cline{4-6}
&&&$\Xi_c^{\prime}({3\over2}^-)\to\Lambda_c \bar K^*\to\Lambda_c \bar K \pi$&\multicolumn{2}{c|}{$2\times10^{-4}$}&
\\ \cline{4-6}
&&&$\Xi_c^{\prime}({3\over2}^-)\to \Xi_c\rho\to\Xi_c\pi\pi$&\multicolumn{2}{c|}{$0.21^{+0.60}_{-0.20}$}&
\\ \cline{4-6}
&&&$\Xi_c^{\prime}({3\over2}^-)\to \Xi_c^{\prime}\rho\to\Xi_c^{\prime}\pi\pi$&\multicolumn{2}{c|}{$0.12^{+0.19}_{-0.10}$}&
\\ \cline{4-6}
&&&$\Xi_c^{\prime}({3\over2}^-)\to\Xi_c^{*}\rho\to\Xi_c^{*}\pi\pi$&\multicolumn{2}{c|}{$1\times10^{-3}$}&
\\ \hline
$\Omega_c({1\over2}^-)$&$3.04^{+0.11}_{-0.09} $&\multirow{2}{*}{$36^{+14}_{-13}$}&\multicolumn{3}{c|}{--}&$\sim~0$&$\Omega_c(3050)^0$
\\ \cline{1-2} \cline{4-8}
$\Omega_c({3\over2}^-)$&$3.07^{+0.10}_{-0.09}$&&\multicolumn{3}{c|}{--}&$\sim~0$&$\Omega_c(3066)^0$
\\ \hline\hline
\end{tabular}
\label{tab:decay611lambda}
\end{center}
\end{table*}

\begin{table*}[hbt]
\begin{center}
\renewcommand{\arraystretch}{1.5}
\caption{Decay properties of $P$-wave charmed baryons belonging to the $[\mathbf{6}_F, 2, 1, \lambda]$ doublet, with possible experimental candidates given in the last column.}
\begin{tabular}{ c | c | c | c | c | c | c | c}
\hline\hline
Baryon & ~~~~Mass~~~~ & Difference & \multirow{2}{*}{~~~~~~~Decay channels~~~~~~~}  & ~$S$-wave width~  & ~$D$-wave width~ & ~Total width~ & \multirow{2}{*}{~Candidate~}
\\ ($j^P$) & ({GeV})& ({MeV}) & & ({MeV}) & ({MeV}) & ({MeV})
\\ \hline\hline
\multirow{5}{*}{$\Sigma_c({3\over2}^-)$}&\multirow{5}{*}{$2.78^{+0.13}_{-0.12}$}&\multirow{9}{*}{$86^{+38}_{-33}$}&$\Sigma_c({3\over2}^-)\to\Lambda_c\pi$&--&$36^{+54}_{-25}$&\multirow{5}{*}{$40^{+54}_{-25}$}&\multirow{9}{*}{$\Sigma_c(2800)^0$}
\\ \cline{4-6}
&&&$\Sigma_c({3\over2}^-)\to \Sigma_c \pi$ &--& $2.5{^{+4.6}_{-2.1}}$&
\\ \cline{4-6}
&&&$\Sigma_c({3\over2}^-)\to \Sigma_c^{*}\pi$ & $6\times10^{-4}$ & $ 0.18^{+0.34}_{-0.15}$&
\\ \cline{4-6}
&&&$\Sigma_c({3\over2}^-)\to\Sigma_c\rho\to\Sigma_c\pi\pi$&\multicolumn{2}{c|}{$0.72^{+1.06}_{-0.51}$}&
\\ \cline{4-6}
&&&$\Sigma_c({3\over2}^-)\to\Sigma_c^{*}\rho\to\Sigma_c^{*}\pi\pi$&\multicolumn{2}{c|}{$5\times10^{-6}$}&
\\ \cline{1-2} \cline{4-7}
\multirow{4}{*}{$\Sigma_c({5\over2}^-)$}&\multirow{4}{*}{$2.87^{+0.12}_{-0.11}$}&&$\Sigma_c({5\over2}^-)\to\Lambda_c \pi$&--&$12^{+18}_{-~8}$&\multirow{4}{*}{$13^{+18}_{-~8}$}&
\\ \cline{4-6}
&&&$\Sigma_c({5\over2}^-)\to\Sigma_c\pi$&--&$0.39^{+0.72}_{-0.32}$&
\\ \cline{4-6}
&&&$\Sigma_c({5\over2}^-)\to\Sigma_c^{*}\pi$&--&$0.61^{+1.14}_{-0.50}$&
\\ \cline{4-6}
&&&$\Sigma_c({5\over2}^-)\to\Sigma_c^{*}\rho\to\Sigma_c^{*}\pi\pi$&\multicolumn{2}{c|}{$4\times10^{-5}$}&
\\ \hline
\multirow{7}{*}{$\Xi_c^{\prime}({3\over2}^-)$}&\multirow{7}{*}{$2.96^{+0.24}_{-0.15}$}&\multirow{13}{*}{$66^{+29}_{-25}$}&$\Xi_c^{\prime}({3\over2}^-)\to \Lambda_c \bar K$ &--& $9.8^{+17.9}_{-~7.2}$ & \multirow{7}{*}{$31{^{+35}_{-14}}$}&\multirow{7}{*}{$\Xi_c(2965)^0$}
\\ \cline{4-6}
&&&$\Xi_c^{\prime}({3\over2}^-)\to \Xi_c \pi$ &--& $17^{+30}_{-12}$&
\\ \cline{4-6}
&&&$\Xi_c^{\prime}({3\over2}^-)\to \Sigma_c \bar K$ &--& $0.003$&
\\ \cline{4-6}
&&&$\Xi_c^{\prime}({3\over2}^-)\to \Xi_c^{\prime}\pi$ &--&$2.3^{+4.0}_{-1.7}$&
\\ \cline{4-6}
&&&$\Xi_c^{\prime}({3\over2}^-)\to \Xi_c^{*}\pi$ &$2\times10^{-4}$&$0.19^{+0.33}_{-0.14}$&
\\ \cline{4-6}
&&&$\Xi_c^{\prime}({3\over2}^-)\to\Xi_c^{\prime}\rho\to\Xi_c^{\prime}\pi\pi$&\multicolumn{2}{c|}{$ 1.4^{+2.2}_{-1.0}$}&
\\ \cline{4-6}
&&&$\Xi_c^{\prime}({3\over2}^-)\to\Xi_c^{*}\rho\to\Xi_c^{*}\pi\pi$&\multicolumn{2}{c|}{$ 1\times10^{-3}$}&
\\ \cline{1-2} \cline{4-8}
\multirow{6}{*}{$\Xi^\prime_c({5\over2}^-)$}&\multirow{6}{*}{$3.02^{+0.23}_{-0.14}$}&&$\Xi_c^{\prime}({5\over2}^-)\to\Lambda_c \bar K$&--&$6.3^{+11.4}_{-~4.6}$&\multirow{7}{*}{$18^{+20}_{-~8}$}&\multirow{7}{*}{--}
\\ \cline{4-6}
&&&$\Xi_c^{\prime}({5\over2}^-)\to \Xi_c \pi$ &--& $9.6^{+15.8}_{-~6.8}$&
\\ \cline{4-6}
&&&$\Xi_c^{\prime}({5\over2}^-)\to \Sigma_c \bar K$ &--& $0.02^{+0.09}_{-0.02}$&
\\ \cline{4-6}
&&&$\Xi_c^{\prime}({5\over2}^-)\to \Xi_c^{\prime}\pi$ &--&$0.70^{+1.25}_{-0.54}$&
\\ \cline{4-6}
&&&$\Xi_c^{\prime}({5\over2}^-)\to \Sigma_c^{*} \bar K$ &--&$4\times10^{-3}$&
\\ \cline{4-6}
&&&$\Xi_c^{\prime}({5\over2}^-)\to \Xi_c^{*} \pi$ &--&$1.5^{+2.6}_{-1.1}$&
\\ \cline{4-6}
&&&$\Xi_c^{\prime}({3\over2}^-)\to\Xi_c^{*}\rho\to\Xi_c^{*}\pi\pi$&\multicolumn{2}{c|}{$ 0.02$}&
\\ \hline
\multirow{2}{*}{$\Omega_c({3\over2}^-)$}&\multirow{2}{*}{$3.08^{+0.22}_{-0.17}$}&\multirow{4}{*}{$59^{+26}_{-22}$}&$\Omega_c({3\over2}^-)\to \Xi_c \bar K$&--& $9.9^{+17.4}_{-~7.4}$&\multirow{2}{*}{$10^{+17}_{-~7}$}&\multirow{2}{*}{$\Omega_c(3090)^0$}
\\ \cline{4-6}
&&&$\Omega_c({3\over2}^-)\to\Xi_c^{\prime} \bar K$&--&$0.10^{+0.26}_{-0.09}$&
\\ \cline{1-2} \cline{4-8}
\multirow{2}{*}{$\Omega_c({5\over2}^-)$}&\multirow{2}{*}{$3.14^{+0.21}_{-0.15}$}&&$\Omega_c({5\over2}^-)\to \Xi_c \bar K$ &--& $5.5^{+9.6}_{-4.1}$&\multirow{2}{*}{$5.5^{+9.6}_{-4.1}$}&\multirow{2}{*}{$\Omega_c(3119)^0$}
\\ \cline{4-6}
&&&$\Omega_c({5\over2}^-)\to\Xi_c^{\prime} \bar K$&--&$0.03^{+0.07}_{-0.02}$&
\\ \hline \hline
\end{tabular}
\label{tab:decay621lambda}
\end{center}
\end{table*}

In this paper we perform a rather complete study on $P$-wave charmed baryons of to the $SU(3)$ flavor $\mathbf{6}_F$. We use the method of QCD sum rules to study their mass spectra. We also use the method of light-cone sum rules to study their decay properties, including their $S$- and $D$-wave decays into ground-state charmed baryons together with pseudoscalar mesons $\pi/K$ and vector mesons $\rho/K^*$. We work within the framework of heavy quark effective theory, and we also consider the mixing effect between different HQET multiplets.

Accordingly to the heavy quark effective theory, we categorize $P$-wave charmed baryons of the $SU(3)$ flavor $\mathbf{6}_F$ into four multiplets: $[\mathbf{6}_F, 1, 0, \rho]$, $[\mathbf{6}_F, 0, 1, \lambda]$, $[\mathbf{6}_F, 1, 1, \lambda]$, and $[\mathbf{6}_F, 2, 1, \lambda]$. Their results are separately summarized in Tables~\ref{tab:decay610rho}/\ref{tab:decay601lambda}/\ref{tab:decay611lambda}/\ref{tab:decay621lambda}. Besides, we explicitly consider the mixing between the $[{\bf 6}_F, 1, 1, \lambda]$ and $[{\bf 6}_F, 2, 1, \lambda]$ doublets, and the obtained results are summarized in Table~\ref{tab:result}. Based on these results, we can well understand many excited charmed baryons as a whole:
\begin{itemize}

\item The $[\mathbf{6}_F,0,1,\lambda]$ singlet contains three charmed baryons: $\Sigma_c({1\over2}^-)$, $\Xi^\prime_c({1\over2}^-)$, and $\Omega_c({1\over2}^-)$. Their widths are all calculated to be very large, so they are difficult to be observed in experiments.

\item The $[\mathbf{6}_F, 1, 0, \rho]$ doublet contains six charmed baryons: $\Sigma_c({1\over2}^-/{3\over2}^-)$, $\Xi^\prime_c({1\over2}^-/{3\over2}^-)$, and $\Omega_c({1\over2}^-/{3\over2}^-)$. We find it possible to interpret the $\Omega_c(3000)^0$ as the $P$-wave $\Omega_c$ baryon of either $J^P=1/2^-$ or $3/2^-$, belonging to this doublet. However, total widths of $\Sigma_c({1\over2}^-/{3\over2}^-)$ and $\Xi^\prime_c({1\over2}^-/{3\over2}^-)$ are calculated to be quite large.

\item The $[\mathbf{6}_F,1,1,\lambda]$ doublet contains six charmed baryons: $\Sigma_c({1\over2}^-/{3\over2}^-)$, $\Xi^\prime_c({1\over2}^-/{3\over2}^-)$, and $\Omega_c({1\over2}^-/{3\over2}^-)$. The $[\mathbf{6}_F,2,1,\lambda]$ doublet also contains six charmed baryons: $\Sigma_c({3\over2}^-/{5\over2}^-)$, $\Xi^\prime_c({3\over2}^-/{5\over2}^-)$, and $\Omega_c({3\over2}^-/{5\over2}^-)$. We explicitly consider the mixing between the two charmed baryons of $J^P = 3/2^-$:
    \begin{eqnarray}
    \nonumber |\Sigma_c(3/2^-),1,1,\lambda\rangle \longleftrightarrow |\Sigma_c(3/2^-),2,1,\lambda\rangle \, ,
    \\ \nonumber |\Xi_c^\prime(3/2^-),1,1,\lambda\rangle \longleftrightarrow |\Xi_c^\prime(3/2^-),2,1,\lambda\rangle \, ,
    \\ \nonumber |\Omega_c(3/2^-),1,1,\lambda\rangle \longleftrightarrow |\Omega_c(3/2^-),2,1,\lambda\rangle \, ,
    \end{eqnarray}
     with the mixing angle fine-tuned to be $\theta_2 = 37\pm5^\circ$.

    Our results suggest: a) the $\Xi_c(2923)^0$ and $\Omega_c(3050)^0$ can be interpreted as the $P$-wave $\Xi_c^\prime$ and $\Omega_c$ baryons of $J^P=1/2^-$, belonging to $[\mathbf{6}_F,1,1,\lambda]$; b) the $\Omega_c(3119)^0$ can be interpreted the $P$-wave $\Omega_c$ baryon of $J^P=5/2^-$, belonging to $[\mathbf{6}_F,2,1,\lambda]$; c) the $\Xi_c(2939)^0$, $\Xi_c(2965)^0$, $\Omega_c(3066)^0$, and $\Omega_c(3090)^0$ can be interpreted the $P$-wave $\Xi_c^\prime$ and $\Omega_c$ baryons of $J^P=3/2^-$, belonging to the mixing of the $[\mathbf{6}_F,1,1,\lambda]$ and $[\mathbf{6}_F,2,1,\lambda]$ doublets.

\end{itemize}
Besides, our results suggest that the $\Sigma_c(2800)^0$ can be explained as the combination of the $P$-wave $\Sigma_c$ baryons belonging to the $[\mathbf{6}_F,1,1,\lambda]$ and $[\mathbf{6}_F,2,1,\lambda]$ doublets. We summarize all the above interpretations in Table~\ref{tab:result}, but note that there exist many other possible explanations.

To arrive at our interpretations, we need to pay attention to: there exist considerable uncertainties in our results for absolute values of charmed baryon masses due to their dependence on the charm quark mass~\cite{Chen:2015kpa,Mao:2015gya}; however, mass splittings within the same doublets do not depend much on this, and are calculated with much less uncertainties; moreover, we can extract more useful information from decay properties of charmed baryons.

\begin{figure*}[htb]
\begin{center}
\scalebox{1.0}{\includegraphics{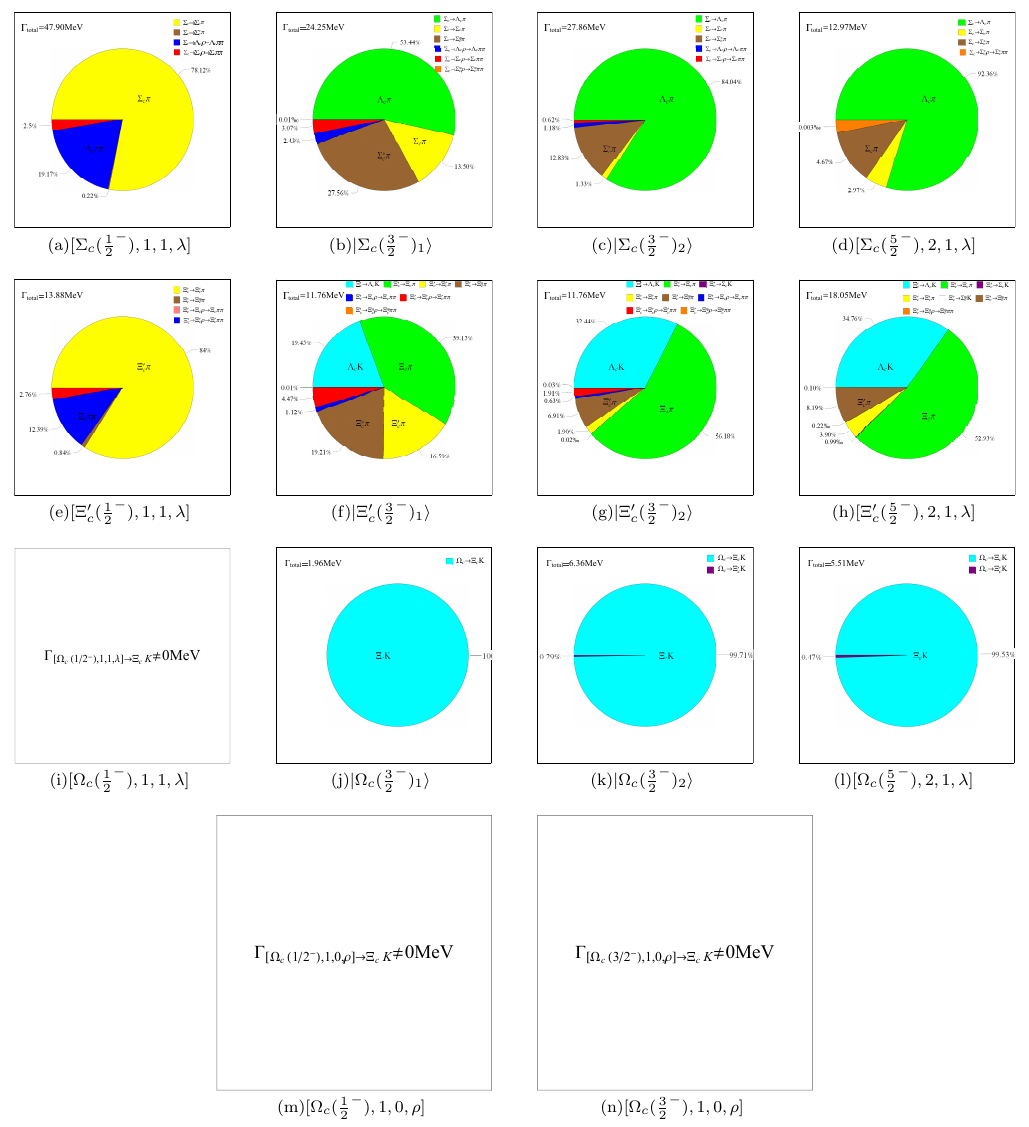}}
\end{center}
\caption{Branching ratios of fourteen $P$-wave charmed baryons of the $SU(3)$ flavor $\mathbf{6}_F$, with limited and non-zero widths.
\label{fig:pie}}
\end{figure*}

Summarizing the above results, the present sum rule study within the heavy quark effective theory can explain many excited singly charmed baryons as a whole, including the $\Sigma_c(2800)^0$, $\Xi_c(2923)^0$, $\Xi_c(2939)^0$, $\Xi_{c}(2965)^{0}$, $\Omega_c(3000)^0$, $\Omega_c(3050)^0$, $\Omega_c(3066)^0$,  $\Omega_c(3090)^0$, and $\Omega_c(3119)^0$. Their masses, mass splittings within the same multiplets, and decay properties are all calculated and summarized in Table~\ref{tab:result}. We suggest the Belle-II, CMS, and LHCb Collaborations to further study them to verify our interpretations. Especially, we propose to further study the $\Sigma_c(2800)^0$ to examine whether it can be separated into several excited charmed baryons. For convenience, we show their total widths and branching ratios in Fig.~\ref{fig:pie} using pie charts.

To end this paper, we note that the $\rho$-mode multiplet $[\mathbf{6}_F, 1, 0, \rho]$ is found in our QCD sum rule approach to be lower than the $\lambda$-mode multiplets $[{\bf 6}_F, 1, 1, \lambda]$ and $[{\bf 6}_F, 2, 1, \lambda]$. This behaviour is consistent with our previous QCD sum rule results~\cite{Chen:2015kpa,Mao:2015gya}, but in contrast to the quark model expectation~\cite{Copley:1979wj,Yoshida:2015tia}. However, this can be possible because mass differences between different multiplets have considerable (theoretical) uncertainties in our framework. We propose to verify whether the $\rho$-mode heavy baryon exists or not by carefully examining the $\Omega_c(3000)^0$ as a possible $\rho$-mode candidate.

\begin{table*}[hbtp]
\begin{center}
\caption{Decay properties of $P$-wave charmed baryons belonging to the $SU(3)$ flavor $\mathbf{6}_F$ representation. The first column lists charmed baryons categorized according to the heavy quark effective theory (HQET), and the third column lists the results after considering the mixing effect between different HQET multiplets. Possible experimental candidates are given in the last column for comparisons. Note that in this table we only list those non-negligible decay channels due to limited spaces, while we have calculated all the possible decay channels as given in Tables~\ref{tab:decay610rho}-\ref{tab:decay611lambda} and Eqs.~(\ref{result:mixing1}-\ref{result:mixing6}).}
\renewcommand{\arraystretch}{1.36}
\scalebox{0.93}{\begin{tabular}{   c|c | c | c | c | c | c | c}
\hline\hline
  \multirow{2}{*}{HQET state}&\multirow{2}{*}{Mixing}&\multirow{2}{*}{Mixed state} & Mass & Difference & ~~~~~~~~~~~Main Decay channel~~~~~~~~~~~ & Width  & \multirow{2}{*}{Candidate}
\\  &&&   ({GeV}) & ({MeV}) & ({MeV})& ({MeV}) &
\\ \hline\hline
$[\Sigma_c({1\over2}^-),0,1,\lambda]$&\multirow{5}{*}{$\theta_1\approx 0^\circ$}&$[\Sigma_c({1\over2}^-),0,1,\lambda]$&$2.83^{+0.06}_{-0.04}$& -- &
$\begin{array}{l}
\Gamma_S\left(\Sigma_c({1/2}^-)\to\Lambda_c \pi\right)=610^{+860}_{-410}
\end{array}$&$610^{+860}_{-410}$&--
\\ \cline{1-1}\cline{3-8}
$[\Sigma_c({1\over2}^-),1,1,\lambda]$&&$[\Sigma_c({1\over2}^-),1,1,\lambda]$&$2.73^{+0.17}_{-0.18}$&\multirow{4}{*}{$23^{+19}_{-43}$}&
$\begin{array}{l}
\Gamma\left(\Sigma_c({1/2}^-)\to\Lambda_c \pi\right) \neq 0 \\
\Gamma_S\left(\Sigma_c({1/2}^-)\to\Sigma_c\pi\right)=37^{+60}_{-28}\\
\Gamma_S\left(\Sigma_c({1/2}^-)\to \Lambda_c\rho\to\Lambda_c\pi\pi\right)=9.2^{+37.0}_{-~9.2}\\
\Gamma_S\left(\Sigma_c({1/2}^-)\to \Sigma_c\rho\to\Sigma_c\pi\pi\right)=1.2^{+2.1}_{-1.0}
\end{array}$&$48^{+70}_{-29}$ & \multirow{10}{*}{$\Sigma_c(2800)^0$}
\\ \cline{1-4} \cline{6-7}
$[\Sigma_c({3\over2}^-),1,1,\lambda]$&\multirow{3}{*}{$\theta_2={37\pm5^\circ}$}&$|\Sigma_c({3\over2}^-)\rangle_1$&$2.75^{+0.17}_{-0.17}$&&
$\begin{array}{l}
\Gamma_D\left(\Sigma_c({3/2}^-)\to\Lambda_c\pi\right)=13^{+20}_{-~9}\\
\Gamma_D\left(\Sigma_c({3/2}^-)\to\Sigma_c\pi\right)=3.3^{+4.2}_{-2.2}\\
\Gamma_S\left(\Sigma_c({3/2}^-)\to\Sigma_c^{*}\pi\right)=6.4^{+10.3}_{-~4.7}
\end{array}$&$24^{+23}_{-10}$&
\\ \cline{1-1} \cline{3-7}
$[\Sigma_c({3\over2}^-),2,1,\lambda]$&&$|\Sigma_c({3\over2}^-)\rangle_2$&$2.80^{+0.14}_{-0.12}$&\multirow{4}{*}{$68^{+51}_{-51}$}&
$\begin{array}{l}
\Gamma_D\left(\Sigma_c({3/2}^-)\to\Lambda_c\pi\right)=23^{+35}_{-16}\\
\Gamma_S\left(\Sigma_c({3/2}^-)\to\Sigma_c^{*}\pi\right)=3.5^{+6.1}_{-2.7}\\
\end{array}$&$28^{+36}_{-16}$&
\\ \cline{1-4} \cline{6-7}
$[\Sigma_c({5\over2}^-),2,1,\lambda]$&--&$[\Sigma_c({5\over2}^-),2,1,\lambda]$&$2.87^{+0.12}_{-0.11}$&&
$\begin{array}{l}
\Gamma_D\left(\Sigma_c({5/2}^-)\to\Lambda_c\pi\right)=12^{+18}_{-~8}\\
\Gamma_D\left(\Sigma_c({5/2}^-)\to \Sigma_c\pi\right)=0.39^{+0.72}_{-0.32}\\
\Gamma_D\left(\Sigma_c({5/2}^-)\to\Sigma_c^{*}\pi\right)=0.61^{+1.14}_{-0.50}
\end{array}$&$13^{+18}_{-~8}$&
\\ \hline
$[\Xi_c^\prime({1\over2}^-),0,1,\lambda]$&\multirow{6}{*}{$\theta_1 \approx 0^\circ$}&$[\Xi_c^{\prime}({1\over2}^-),0,1,\lambda]$&$2.90^{+0.13}_{-0.12}$& -- &
$\begin{array}{l}
\Gamma_S\left(\Xi_c^{\prime}({1/2}^-)\to\Lambda_c \bar K\right)=400^{+610}_{-270}\\
\Gamma_S\left(\Xi_c^{\prime}({1/2}^-)\to \Xi_c \pi\right)=360^{+550}_{-250}
\end{array}$
&$760^{+820}_{-370}$&--
\\ \cline{1-1}\cline{3-8}
$[\Xi_c^\prime({1\over2}^-),1,1,\lambda]$&&$[\Xi_c^\prime({1\over2}^-),1,1,\lambda]$&$2.91^{+0.13}_{-0.12}$&\multirow{5}{*}{$27^{+16}_{-27}$}&
$\begin{array}{l}
\Gamma\left(\Xi_c^{\prime}({1/2}^-)\to\Lambda_c \bar K\right) \neq 0 \\
\Gamma\left(\Xi_c^{\prime}({1/2}^-)\to \Xi_c \pi\right)  \neq 0 \\
\Gamma_S\left(\Xi_c^{\prime}({1/2}^-)\to \Xi_c^{\prime}\pi\right)=12^{+15}_{-~8}\\
\Gamma_S\left(\Xi_c^{\prime}({1/2}^-)\to\Xi_c\rho\to\Xi_c\pi\pi\right)=1.7^{+7.6}_{-1.7}
\end{array}$&$14^{+17}_{-~8}$&$\Xi_c(2923)^0$
\\ \cline{1-4}\cline{6-8}
$[\Xi_c^{\prime}({3\over2}^-),1,1,\lambda]$&\multirow{4}{*}{$\theta_2={37\pm5^\circ}$}&$|\Xi_c^\prime({3\over2}^-)\rangle_1$&$2.94^{+0.12}_{-0.11}$&&
$\begin{array}{l}
\Gamma_D\left(\Xi_c^{\prime}({3/2}^-)\to \Lambda_c \bar K\right)=2.3^{+4.3}_{-1.7}\\
\Gamma_D\left(\Xi_c^{\prime}({3/2}^-)\to \Xi_c\pi\right)=4.6^{+8.1}_{-3.3}\\
\Gamma_D\left(\Xi_c^{\prime}({3/2}^-)\to\Xi_ c^{\prime}\pi\right)=2.0^{+2.2}_{-1.2}\\
\Gamma_S\left(\Xi_c^{\prime}({3/2}^-)\to \Xi_c^{*}\pi\right)=2.1^{+2.6}_{-1.5}
\end{array}$&$12^{+10}_{-~4}$
& $\Xi_c(2939)^0 $
\\ \cline{1-1}\cline{3-8}
$[\Xi_c^\prime({3\over2}^-),2,1,\lambda]$&&$|\Xi_c^\prime({3\over2}^-)\rangle_2$&$2.97^{+0.24}_{-0.15}$&\multirow{4}{*}{$56^{+30}_{-35}$}&
$\begin{array}{l}
\Gamma_D\left(\Xi_c^{\prime}({3/2}^-)\to \Lambda_c \bar K\right)=6.3^{+11.6}_{-~4.7}\\
\Gamma_D\left(\Xi_c^{\prime}({3/2}^-)\to \Xi_c\pi\right)=11^{+19}_{-~8}\\
\Gamma_S\left(\Xi_c^{\prime}({3/2}^-)\to \Xi_c^{*}\pi\right)= 1.3^{+1.80}_{-0.94}
\end{array}$&$19^{+22}_{-~9}$&$\Xi_c(2965)^0$
\\ \cline{1-4} \cline{6-8}
$[\Xi^\prime_c({5\over2}^-),2,1,\lambda]$&--&$[\Xi^\prime_c({5\over2}^-),2,1,\lambda]$&$3.02^{+0.23}_{-0.14}$&&
$\begin{array}{l}
\Gamma_D\left(\Xi_c^{\prime}({5/2}^-)\to \Lambda_c \bar K\right)=6.3^{+11.4}_{-~4.6}\\
\Gamma_D\left(\Xi_c^{\prime}({5/2}^-)\to \Xi_c \pi\right)=9.6^{+15.8}_{-~6.8}\\
\Gamma_D\left(\Xi_c^{\prime}({5/2}^-)\to \Xi_c^{*} \pi\right)=1.5^{+2.6}_{-1.1}
\end{array}$&$18^{+20}_{-~8}$
&--
\\ \hline
$[\Omega_c({1\over2}^-),1,0,\rho]$&$\theta_1^\prime \approx0^\circ$&$[\Omega_c({1\over2}^-),1,0,\rho]$&$2.99^{+0.15}_{-0.15}$& \multirow{2}{*}{$12^{+5}_{-5}$} & $\Gamma\left(\Omega_c({1/2}^-)\to\Xi_c \bar K\right) \neq 0$ &$\sim~0$&\multirow{2}{*}{$\Omega_c(3000)^0$}
\\ \cline{1-4} \cline{6-7}
$[\Omega_c({3\over2}^-),1,0,\rho]$&$\theta_2^\prime \approx0^\circ$&$[\Omega_c({3\over2}^-),1,0,\rho]$&$3.00^{+0.15}_{-0.15}$&& $\Gamma\left(\Omega_c({3/2}^-)\to\Xi_c \bar K\right) \neq 0$ &$\sim~0$&
\\ \hline
$[\Omega_c({1\over2}^-),0,1,\lambda]$&\multirow{2}{*}{$\theta_1 \approx0^\circ$}&$[\Omega_c({1\over2}^-),0,1,\lambda]$&$3.03^{+0.18}_{-0.19}$& -- &$\begin{array}{l}
\Gamma_S\left(\Omega_c({1/2}^-)\to\Xi_c \bar K\right)=980^{+1530}_{-~670}
\end{array}$&$980^{+1530}_{-~670}$&--
\\ \cline{1-1} \cline{3-8}
$[\Omega_c({1\over2}^-),1,1,\lambda]$&&$[\Omega_c({1\over2}^-),1,1,\lambda]$&$3.04^{+0.11}_{-0.09}$& \multirow{2}{*}{$27^{+15}_{-23}$} & $\Gamma\left(\Omega_c({1/2}^-)\to\Xi_c \bar K\right) \neq 0$ &$\sim~0$&$\Omega_c(3050)^0$
\\ \cline{1-4} \cline{6-8}
$[\Omega_c({3\over2}^-),1,1,\lambda]$&\multirow{2}{*}{$\theta_2 \approx37\pm5^\circ$}&$|\Omega_c({3\over2}^-)\rangle_1$&$3.06^{+0.10}_{-0.09}$&  &
$\begin{array}{l}
\Gamma_D\left(\Omega_c({3/2}^-)\to \Xi_c \bar K\right)=2.0^{+3.5}_{-1.5}
\end{array}$&$2.0^{+3.5}_{-1.5}$&$\Omega_c(3066)^0$
\\ \cline{1-1} \cline{3-8}
$[\Omega_c({3\over2}^-),2,1,\lambda]$&&$|\Omega_c({3\over2}^-)\rangle_2$&$3.09^{+0.22}_{-0.17}$&\multirow{3}{*}{$51^{+26}_{-29}$}&$\begin{array}{l}
\Gamma_D\left(\Omega_c({3/2}^-)\to \Xi_c \bar K\right)=6.3^{+11.2}_{-~4.8}
\end{array}$&$6.4^{+11.2}_{-~4.8}$&$\Omega_c(3090)^0$
\\ \cline{1-4} \cline{6-8}
$[\Omega_c({5\over2}^-),2,1,\lambda]$&--&$[\Omega_c({5\over2}^-),2,1,\lambda]$&$3.14^{+0.21}_{-0.15}$&&
$\begin{array}{l}
\Gamma_D\left(\Omega_c({3/2}^-)\to \Xi_c \bar K\right)=5.5^{+9.6}_{-4.1}
\end{array}$&$5.5^{+9.6}_{-4.1}$&$\Omega_c(3119)^0$
\\ \hline\hline
\end{tabular}}
\label{tab:result}
\end{center}
\end{table*}

%
\section*{Acknowledgments}

We thank Er-Liang Cui and Qiang Mao for useful discussions.
This project is supported by the National Natural Science Foundation of China under Grants No.~11722540 and No.~12075019,
and
the Fundamental Research Funds for the Central Universities.

\appendix

\section{Some sum rule equations}
\label{sec:othersumrule}

In this appendix we give several examples of sum rule equations, which are used to study $D$-wave decays of $P$-wave charmed baryons into ground-state charmed baryons and light pseudoscalar mesons.

\begin{widetext}
The sum rule equation for the $\Sigma_c^0[{1\over2}^-]$ belonging to $[\mathbf{6}_F, 1 , 0, \rho]$ is
\begin{eqnarray}
&&G_{\Sigma_c^0[{1\over2}^-]\to\Sigma_c^{*+}\pi^-}^D(\omega,\omega^{\prime})= \frac{g_{\Sigma_c^0[{1\over2}^-]\to\Sigma_c^{*+}\pi^- } f_{\Sigma_c^0[{1\over2}^-]}f_{\Sigma_c^{*+}}}{(\bar{\Lambda}_{\Sigma_c^0[{1\over2}^-]}-\omega^{\prime})(\bar{\Lambda}_{\Sigma_c^{*+}}-\omega)}
\\ \nonumber&=& \int_0^\infty dt \int_0^1 du e^{i(1-u)\omega^\prime t} e^{iu\omega t}\times 8\times \Big(\frac{f_{\pi} u}{24}\langle\bar q q\rangle \phi_{2;\pi}(u)
+\frac{f_{\pi} m_{\pi}^2 u}{24 (m_u+m_d)\pi^2 t^2}\phi_{3;\pi}^{\sigma}(u)+\frac{f_{\pi} u t^2}{384}\langle\bar q q\rangle \phi_{4;\pi}(u)
\\ \nonumber &&+\frac{f_{\pi} u t^2}{384}\langle g_s \bar q \sigma G q\rangle \phi_{2;\pi}(u)+\frac{f_{\pi} u t^4}{6144}\langle g_s \bar q \sigma G q\rangle \phi_{4;\pi}(u)\Big)
\\ \nonumber &-&\int_0^\infty dt \int_0^1 du \int \mathcal{D}\underline{\alpha}e^{i\omega^\prime t(\alpha_2+u\alpha_3)}e^{i\omega t(1-\alpha_2-u\alpha_3)}\times\Big(-\frac{i f_{3\pi} u v\cdot q}{4\pi^2 t}\Phi_{3;\pi}(\underline{\alpha})+\frac{i f_{3\pi} \alpha_2 u v\cdot q}{4 \pi^2 t}\Phi_{3;\pi}(\underline{\alpha})
\\ \nonumber &&+\frac{i f_{3\pi}\alpha_3 u v\cdot q}{4\pi^2 t}\Phi_{3;\pi}(\underline{\alpha})-\frac{i f_{3\pi} u v\cdot q}{4\pi^2 t}\Phi_{3;\pi}(\underline{\alpha})+\frac{i f_{3\pi}\alpha_2 v\cdot q}{4\pi^2 t}\Phi_{3\pi}(\underline{\alpha})-\frac{i f_{3\pi} v\cdot q}{4\pi^2 t}\Phi_{3;\pi}(\underline{\alpha})+\frac{f_{3\pi} u}{4\pi^2 t^2}\Phi_{3;\pi}(\underline{\alpha})\Big) \, .
\end{eqnarray}

The sum rule equation for the $\Xi_c^{\prime0}[{3\over2}^-]$ belonging to $[\mathbf{6}_F, 1 , 0, \rho]$ is
\begin{eqnarray}
&&G_{\Xi_c^{\prime0}[{3\over2}^-]\to\Xi_c^{\prime+}\pi^-}^D(\omega,\omega^{\prime})= \frac{g_{\Xi_c^{\prime0}[{3\over2}^-]\to\Xi_c^{\prime+}\pi^- } f_{\Xi_c^{\prime0}[{3\over2}^-]}f_{\Xi_c^{\prime+}}}{(\bar{\Lambda}_{\Xi_c^{\prime0}[{3\over2}^-]}-\omega^{\prime})(\bar{\Lambda}_{\Xi_c^{\prime+}}-\omega)}
\\ \nonumber &=& \int_0^\infty dt \int_0^1 du e^{i(1-u)\omega^\prime t} e^{iu\omega t}\times 4\times \Big(\frac{f_{\pi} m_s u}{8\pi^2 t^2}\phi_{2;\pi}(u)+\frac{f_{\pi} m_{\pi}^2 u}{24 (m_u+m_d)\pi^2 t^2}\phi_{3;\pi}^{\sigma}(u)
\\ \nonumber &&+\frac{f_{\pi} m_s u}{128 \pi^2}\phi_{4;\pi}(u)+\frac{f_{\pi} u}{24}\langle\bar s s\rangle \phi_{2;\pi}(u)+\frac{f_{\pi} m_s  m_{\pi}^2 u t^2}{576 (m_u+m_d)}\langle\bar s s\rangle\phi_{3;\phi}^{\sigma}(u)+\frac{f_{\pi} u t^2}{384}\langle\bar s s\rangle\phi_{4;\pi}(u)
\\ \nonumber &&+\frac{f_{\pi} u t^2}{384}\langle g_s \bar s\sigma G s\rangle\phi_{2;\pi}(u)+\frac{f_{\pi} u t^4}{6144}\langle g_s \bar s\sigma G s\rangle\phi_{4;\pi}(u)\Big)
\\ \nonumber &-&\int_0^\infty dt \int_0^1 du \int \mathcal{D}\underline{\alpha}e^{i\omega^\prime t(\alpha_2+u\alpha_3)}e^{i\omega t(1-\alpha_2-u\alpha_3)}\times{1\over2}\times\Big(-\frac{f_{3\pi} u}{4\pi^2 t^2}\Phi_{3;\pi}(\underline{\alpha})+\frac{if_{3\pi}\alpha_3 u^2 v\cdot q}{4\pi^2 t}\Phi_{3;\pi}(\underline{\alpha})
\\ \nonumber &&+\frac{if_{3\pi}\alpha_2 u v\cdot q}{4\pi^2 t}\Phi_{3;\pi}(\underline{\alpha})+frac{i f_{3\pi}\alpha_3 u v\cdot q}{4\pi^2 t}\Phi_{3;\pi}(\underline{\alpha})-\frac{if_{3\pi} u v\cdot q}{4\pi^2 t}\Phi_{3;\pi}(\underline{\alpha})+\frac{i f_{3\pi}\alpha_3 v\cdot q}{4\pi^2 t}\Phi_{3;\pi}(\underline{\alpha})
\\ \nonumber &&-\frac{i f_{3\pi} v\cdot q}{4\pi^2 t}\Phi_{3;\pi}(\underline{\alpha})\Big) \, .
\end{eqnarray}

The sum rule equation for the $\Omega_c^0[{3\over2}^-]$ belonging to $[\mathbf{6}_F, 1 , 0, \rho]$ is
\begin{eqnarray}
&&G_{\Omega_c^0[{3\over2}^-]\to\Xi_c^{*+}K^-}^D(\omega,\omega^{\prime})= \frac{g_{\Omega_c^0[{3\over2}^-]\to\Xi_c^{*+}K^- } f_{\Omega_c^0[{3\over2}^-]}f_{\Xi_c^{*+}}}{(\bar{\Lambda}_{\Omega_c^0[{3\over2}^-]}-\omega^{\prime})(\bar{\Lambda}_{\Xi_c^{*+}}-\omega)}
\\ \nonumber&=& \int_0^\infty dt \int_0^1 du e^{i(1-u)\omega^\prime t} e^{iu\omega t}\times 8\times \Big(\frac{f_{K} m_s u}{24 \pi^2 t^2}\phi_{2;K}(u)+\frac{f_{K} m_{K}^2 u}{72(m_u+m_d)\pi^2 t^2}\phi_{3;K}^{\sigma}(u)
\\ \nonumber &&+\frac{f_{\pi} m_s u}{384 \pi^2}\phi_{4;K}(u)+\frac{f_{K} u}{72}\langle\bar s s\rangle\phi_{2;K}(u)+\frac{f_{K} m_s m_{K}^2 u t^2}{1728(m_u+m_d)}\langle\bar s s\rangle\phi_{3;K}^{\sigma}(u)+\frac{f_{K} u t^2}{1152}\langle\bar s s \rangle\phi_{4;K}(u)
\\ \nonumber &&+\frac{f_{K} u t^2}{1152}\langle g_s \bar s \sigma G s\rangle\phi_{2;K}(u)+\frac{f_{K} u t^4}{18432}\langle g_s \bar s \sigma G s\rangle\phi_{4;K}(u)\Big)
\\ \nonumber &-&\int_0^\infty dt \int_0^1 du \int \mathcal{D}\underline{\alpha}e^{i\omega^\prime t(\alpha_2+u\alpha_3)}e^{i\omega t(1-\alpha_2-u\alpha_3)}\times\Big(\frac{i f_{3K}\alpha_3 u^2 v\cdot q}{12\pi^2 t}\Phi_{3;K}(\underline{\alpha})+\frac{f_{3K} u v\cdot q}{12\pi^2 t}\Phi_{3;K} (\underline{\alpha})
\\ \nonumber &&+\frac{i f_{3K}\alpha_2 u v \cdot q}{12\pi^2 t}\Phi_{3;K}(\underline{\alpha})+\frac{i f_{3K}\alpha_3 u v \cdot q}{12 \pi^2}\Phi_{3;K}(\underline{\alpha})-\frac{i f_{3K} u v \cdot q}{12\pi^2 t}\Phi_{3;K}(\underline{\alpha})+\frac{i f_{3K} \alpha_2 v \cdot q}{12 \pi^2 t}\Phi_{3;K}(\underline{\alpha})
\\ \nonumber &&-\frac{i f_{3K} v \cdot q}{12\pi^2 t}\Phi_{3;K}(\underline{\alpha})\Big) \, .
\end{eqnarray}

The sum rule equation for the $\Sigma_c^0[{1\over2}^-]$ belonging to $[\mathbf{6}_F, 0 , 1, \lambda]$ is
\begin{eqnarray}
&&G_{\Sigma_c^0[{1\over2}^-]\to\Lambda_c^{+}\pi^-}^S(\omega,\omega^{\prime})= \frac{g_{\Sigma_c^0[{1\over2}^-]\to\Lambda_c^{+}\pi^- } f_{\Sigma_c^0[{1\over2}^-]}f_{\Lambda_c^{+}}}{(\bar{\Lambda}_{\Sigma_c^0[{1\over2}^-]}-\omega^{\prime})(\bar{\Lambda}_{\Xi_c^{+}}-\omega)}
\\ \nonumber &=& \int_0^\infty dt \int_0^1 du e^{i(1-u)\omega^\prime t} e^{iu\omega t}\times 8\times \Big(-\frac{3 f_{\pi} m_{\pi}^2}{4\pi^2(m_u+m_d) t^4}\phi_{3;\pi}^p(u)-\frac{i f_{\pi} m_{\pi}^2 v \cdot q}{8\pi^2(m_u+m_d) t^3}\phi_{3;\pi}^{\sigma}(u)
\\ \nonumber &&+\frac{i f_{\pi}}{16 t v \cdot q}\langle \bar q q \rangle\psi_{4;\pi}(u)+\frac{i f_{\pi} t}{256 v \cdot q}\langle g_s \bar q \sigma G q\rangle\psi_{4;\pi}(u)\Big) \, .
\end{eqnarray}

The sum rule equation for the $\Xi_c^{\prime0}[{1\over2}^-]$ belonging to $[\mathbf{6}_F, 0 , 1, \lambda]$ is
\begin{eqnarray}
&&G_{\Xi_c^{\prime0}[{1\over2}^-]\to\Lambda_c^{+}K^-}^S(\omega,\omega^{\prime})= \frac{g_{\Xi_c^{\prime0}[{1\over2}^-]\to\Lambda_c^{+}K^- } f_{\Xi_c^{\prime0}[{1\over2}^-]}f_{\Lambda_c^{+}}}{(\bar{\Lambda}_{\Xi_c^{\prime0}[{1\over2}^-]}-\omega^{\prime})(\bar{\Lambda}_{\Lambda_c^{+}}-\omega)}
\\ \nonumber &=& \int_0^\infty dt \int_0^1 du e^{i(1-u)\omega^\prime t} e^{iu\omega t}\times 4\times \Big(-\frac{3 f_{K} m_{K}^2}{4\pi^2(m_u+m_s) t^4}\phi_{3;K}^p(u)-\frac{i f_{K} m_{K}^2 v \cdot q}{8\pi^2(m_u+m_s) t^3}\phi_{3;K}^{\sigma}(u)
\\ \nonumber &&+\frac{i f_{K}}{16 t v \cdot q}\langle \bar q q \rangle\psi_{4;K}(u)+\frac{i f_{K} t}{256 v \cdot q}\langle g_s \bar q \sigma G q\rangle\psi_{4;K}(u)\Big) \, .
\end{eqnarray}

The sum rule equation for the $\Omega_c^{0}[{1\over2}^-]$ belonging to $[\mathbf{6}_F, 0 , 1, \lambda]$ is
\begin{eqnarray}
&&G_{\Omega_c^{0}[{1\over2}^-]\to\Xi_c^{+}K^-}^S(\omega,\omega^{\prime})= \frac{g_{\Omega_c^{0}[{1\over2}^-]\to\Xi_c^{+}K^- } f_{\Omega_c^{0}[{1\over2}^-]}f_{\Xi_c^{+}}}{(\bar{\Lambda}_{\Omega_c^{0}[{1\over2}^-]}-\omega^{\prime})(\bar{\Lambda}_{\Xi_c^{+}}-\omega)}
\\ \nonumber &=& \int_0^\infty dt \int_0^1 du e^{i(1-u)\omega^\prime t} e^{iu\omega t}\times 8\times \Big(-\frac{3 f_{K} m_{K}^2}{4\pi^2(m_u+m_d) t^4}\phi_{3;K}^p(u)-\frac{i f_{K} m_{\pi}^2 v \cdot q}{8\pi^2(m_u+m_d) t^3}\phi_{3;K}^{\sigma}(u)
\\ \nonumber &&+\frac{i f_{K}}{16 t v \cdot q}\langle \bar s s \rangle\psi_{4;K}(u)+\frac{i f_{K} t}{256 v \cdot q}\langle g_s \bar s \sigma G s\rangle\psi_{4;K}(u)+\frac{3 if_{K} m_s}{16\pi^2 t^3 v \cdot q}\psi_{4;K}(u)-\frac{f_{K} m_s m_{K}^2}{32 (m_u+m_d)}\langle \bar s s\rangle\phi_{3;K}^p(u)
\\ \nonumber &&-\frac{if_{K} m_s m_{K}^2 t v \cdot q}{192(m_u+m_d)}\langle \bar s s\rangle\phi_{3;K}^{\sigma}(u)\Big) \, .
\end{eqnarray}

The sum rule equation for the $\Xi_c^{\prime0}[{1\over2}^-]$ belonging to $[\mathbf{6}_F, 1 , 1, \lambda]$ is
\begin{eqnarray}
&&G_{\Xi_c^{\prime0}[{1\over2}^-]\to\Xi_c^{*+}\pi^-}^D(\omega,\omega^{\prime})= \frac{g_{\Xi_c^{\prime0}[{1\over2}^-]\to\Xi_c^{*+}\pi^- } f_{\Xi_c^{\prime0}[{1\over2}^-]}f_{\Xi_c^{*+}}}{(\bar{\Lambda}_{\Xi_c^{\prime0}[{1\over2}^-]}-\omega^{\prime})(\bar{\Lambda}_{\Xi_c^{*+}}-\omega)}
\\ \nonumber&=& \int_0^\infty dt \int_0^1 du e^{i(1-u)\omega^\prime t} e^{iu\omega t}\times 4\times \Big(-\frac{f_{\pi} u}{4\pi^2 t^3}\phi_{2;\pi}(u)+\frac{f_{\pi} m_s m_{\pi}^2 u}{48(m_u+m_d) \pi^2 t}\phi_{3;\pi}^{\sigma}(u)
\\ \nonumber &&-\frac{f_{\pi} u}{64 \pi^2 t}\phi_{4;\pi}(u)-\frac{f_{\pi} m_s u t}{96}\langle\bar s s\rangle \phi_{2;\pi}(u)+\frac{f_{\pi} m_{\pi}^2 u t}{144(m_u+m_d)}\langle \bar s s \rangle \phi_{3;\pi}^{\sigma}(u)-\frac{f_{\pi} m_s t^3}{1536}\langle\bar s s \rangle \phi_{4;\pi}(u)
\\ \nonumber &&+\frac{f_{\pi} m_{\pi}^2 t^3}{2304(m_u+m_d)}\langle g_s \bar s \sigma G s \rangle\phi_{3;\pi}^{\sigma}(u)\Big)
\\ \nonumber &-&\int_0^\infty dt \int_0^1 du \int \mathcal{D}\underline{\alpha}e^{i\omega^\prime t(\alpha_2+u\alpha_3)}e^{i\omega t(1-\alpha_2-u\alpha_3)}\times{1\over2}\times\Big(-\frac{f_{\pi}\alpha_3 u^2}{8\pi^2 t}\Phi_{4;\pi}(\underline{\alpha})-\frac{f_{\pi}\alpha_2 u}{8\pi^2 t}\Phi_{4;\pi}(\underline{\alpha})
\\ \nonumber &&-\frac{f_{\pi}\alpha_3 u}{16\pi^2 t}\Phi_{4;\pi}(\underline{\alpha})-\frac{f_{\pi}\alpha_3 u}{16\pi^2 t}\widetilde\Phi_{4;\pi}(\underline{\alpha})+\frac{f_{\pi} u}{8\pi^2 t}\Phi_{4;\pi}(\underline{\alpha})-\frac{f_{\pi}\alpha_2}{16\pi^2 t}\Phi_{4;\pi}(\underline{\alpha})
-\frac{f_{\pi}\alpha_2}{16\pi^2 t}\widetilde\Phi{4;\pi}(\underline{\alpha})
\\ \nonumber &&+\frac{f_{\pi}}{16\pi^2 t}\Phi_{4;\pi}(\underline{\alpha})+\frac{f_{\pi}}{16\pi^2 t}\widetilde\Phi_{4;\pi}(\underline{\alpha})+\frac{if_{\pi} u}{8\pi^2 t^2 v\cdot q}\Psi_{4;\pi}(\underline{\alpha})
+\frac{3 i f_{\pi}}{8\pi^2 t^2 v\cdot q}\widetilde\Psi_{4;\pi}(\underline{\alpha})
-\frac{i f_{\pi}}{8\pi^2  t^2 v \cdot q}\Phi_{4;\pi}(\underline{\alpha})
\\ \nonumber &&-\frac{3 i f_{\pi}}{8\pi^2 t^2 v\cdot q}\widetilde\Phi_{4;\pi}(\underline{\alpha})+\frac{i f_{\pi}}{8\pi^2 t^2 v\cdot q}\Psi_{4;\pi}(\underline{\alpha})
-\frac{i f_{\pi}}{8\pi^2 t^2 v\cdot q}\widetilde\Psi_{4;\pi}(\underline{\alpha})\Big) \, .
\end{eqnarray}

The sum rule equation for the $\Sigma_c^{0}[{3\over2}^-]$ belonging to $[\mathbf{6}_F, 1 , 1, \lambda]$ is
\begin{eqnarray}
&&G_{\Sigma_c^{0}[{3\over2}^-]\to\Sigma_c^{+}\pi^-}^D(\omega,\omega^{\prime})= \frac{g_{\Sigma_c^{0}[{3\over2}^-]\to\Sigma_c^{+}\pi^- } f_{\Sigma_c^{0}[{3\over2}^-]}f_{\Sigma_c^{+}}}{(\bar{\Lambda}_{\Sigma_c^{0}[{3\over2}^-]}-\omega^{\prime})(\bar{\Lambda}_{\Sigma_c^{+}}-\omega)}
\\ \nonumber &=& \int_0^\infty dt \int_0^1 du e^{i(1-u)\omega^\prime t} e^{iu\omega t}\times 8\times \Big(\frac{i f_{\pi} u}{4\pi^2 t^3}\phi_{2;\pi}(u)+\frac{i f_{\pi} u}{64\pi^2 t}\phi_{4;\pi}(u)
\\ \nonumber&&-\frac{i f_{\pi} m_{\pi}^2 u t}{144(m_u+m_d)}\langle\bar q q\rangle\phi_{3;\pi}^{\sigma}(u)-\frac{i f_{\pi} m_{\pi}^2 u t^3}{2304(m_u+m_d)}\langle g_s\bar q \sigma G q\rangle\phi_{3;\pi}^{\sigma}(u)\Big)
\\ \nonumber &-&\int_0^\infty dt \int_0^1 du \int \mathcal{D}\underline{\alpha}e^{i\omega^\prime t(\alpha_2+u\alpha_3)}e^{i\omega t(1-\alpha_2-u\alpha_3)}\times\Big(\frac{i f_{\pi}\alpha_3 u^2}{8\pi^2 t}\Phi_{4;\pi}(\underline{\alpha})+\frac{i f_{\pi}\alpha_2 u}{8\pi^2 t}\Phi_{4;\pi}(\underline{\alpha})
\\ \nonumber &&+\frac{i f_{\pi}\alpha_3 u}{16\pi^2 t}\Phi_{4;\pi}(\underline{\alpha})+\frac{i f_{\pi}\alpha_3 u}{16\pi^2 t}\widetilde\Phi_{4;\pi}(\underline{\alpha})-\frac{i f_{\pi} u}{8\pi^2 t}\Phi_{4;\pi}(\underline{\alpha})+\frac{i f_{\pi}\alpha_2}{16\pi^2 t}\Phi_{4;\pi}(\underline{\alpha})+\frac{i f_{\pi}\alpha_2}{16\pi^2 t}\widetilde\Phi_{4;\pi}(\underline{\alpha})
\\ \nonumber &&-\frac{i f_{\pi}}{16\pi^2 t}\Phi_{4;\pi}(\underline{\alpha})-\frac{i f_{\pi}}{16\pi^2 t}\widetilde\Phi_{4;\pi}(\underline{\alpha})+\frac{f_{\pi} u}{8\pi^2 t^2 v\cdot q}\Psi_{4;\pi}(\underline{\alpha})+\frac{3 f_{\pi} u}{8\pi^2 t^2 v\cdot q}\widetilde\Psi_{4;\pi}(\underline{\alpha})-\frac{f_{\pi}}{8\pi^2 t^2 v\cdot q}\Phi_{4;\pi}(\underline{\alpha})
\\ \nonumber &&-\frac{3 f_{\pi}}{8\pi^2 t^2 v\cdot q}\widetilde\Phi_{4;\pi}(\underline{\alpha})+\frac{f_{\pi}}{8\pi^2 t^2 v\cdot q}\Psi_{4;\pi}(\underline{\alpha})-\frac{f_{\pi}}{8\pi^2 t^2 v\cdot q}\widetilde\Psi_{4;\pi}(\underline{\alpha})\Big) \, .
\end{eqnarray}

The sum rule equation for the $\Omega_c^{0}[{3\over2}^-]$ belonging to $[\mathbf{6}_F, 1 , 1, \lambda]$ is
\begin{eqnarray}
&&G_{\Omega_c^{0}[{3\over2}^-]\to\Xi_c^{\prime+}K^-}^D(\omega,\omega^{\prime})= \frac{g_{\Omega_c^{0}[{3\over2}^-]\to\Xi_c^{\prime+}K^- } f_{\Omega_c^{0}[{3\over2}^-]}f_{\Xi_c^{\prime+}}}{(\bar{\Lambda}_{\Omega_c^{0}[{3\over2}^-]}-\omega^{\prime})(\bar{\Lambda}_{\Xi_c^{\prime+}}-\omega)}
\\ \nonumber &=& \int_0^\infty dt \int_0^1 du e^{i(1-u)\omega^\prime t} e^{iu\omega t}\times 8\times \Big(\frac{i f_{K} u}{4\pi^2 t^3}\phi_{2;K}(u)+\frac{i f_{K} u}{64\pi^2 t}\phi_{4;K}(u)-\frac{i f_{K} m_{K}^2 u t}{144(m_u+m_s)}\langle\bar s s\rangle\phi_{3;K}^{\sigma}(u)
\\ \nonumber &&-\frac{i f_{K} m_{K}^2 u t^3}{2304(m_u+m_s)}\langle g_s\bar s \sigma G s\rangle\phi_{3;K}^{\sigma}(u)\Big)
\\ \nonumber &-&\int_0^\infty dt \int_0^1 du \int \mathcal{D}\underline{\alpha}e^{i\omega^\prime t(\alpha_2+u\alpha_3)}e^{i\omega t(1-\alpha_2-u\alpha_3)}\times\Big(\frac{i f_{K}\alpha_3 u^2}{8\pi^2 t}\Phi_{4;K}(\underline{\alpha})+\frac{i f_{K}\alpha_2 u}{8\pi^2 t}\Phi_{4;K}(\underline{\alpha})
\\ \nonumber &&+\frac{i f_{K}\alpha_3 u}{16\pi^2 t}\Phi_{4;K}(\underline{\alpha})+\frac{i f_{K}\alpha_3 u}{16\pi^2 t}\widetilde\Phi_{4;K}(\underline{\alpha})-\frac{i f_{K} u}{8\pi^2 t}\Phi_{4;K}(\underline{\alpha})+\frac{i f_{K}\alpha_2}{16\pi^2 t}\Phi_{4;K}(\underline{\alpha})+\frac{i f_{K}\alpha_2}{16\pi^2 t}\widetilde\Phi_{4;K}(\underline{\alpha})
\\ \nonumber &&-\frac{i f_{K}}{16\pi^2 t}\Phi_{4;K}(\underline{\alpha})-\frac{i f_{K}}{16\pi^2 t}\widetilde\Phi_{4;\pi}(\underline{\alpha})+\frac{f_{\pi} u}{8\pi^2 t^2 v\cdot q}\Psi_{4;K}(\underline{\alpha})+\frac{3 f_{K} u}{8\pi^2 t^2 v\cdot q}\widetilde\Psi_{4;K}(\underline{\alpha})-\frac{f_{K}}{8\pi^2 t^2 v\cdot q}\Phi_{4;K}(\underline{\alpha})
\\ \nonumber &&-\frac{3 f_{K}}{8\pi^2 t^2 v\cdot q}\widetilde\Phi_{4;K}(\underline{\alpha})+\frac{f_{K}}{8\pi^2 t^2 v\cdot q}\Psi_{4;K}(\underline{\alpha})-\frac{f_{K}}{8\pi^2 t^2 v\cdot q}\widetilde\Psi_{4;K}(\underline{\alpha})\Big) \, .
\end{eqnarray}

The sum rule equation for the $\Sigma_c^{0}[{3\over2}^-]$ belonging to $[\mathbf{6}_F, 2 , 1, \lambda]$ is
\begin{eqnarray}
&&G_{\Sigma_c^{0}[{3\over2}^-]\to\Lambda_c^{+}\pi^-}^D(\omega,\omega^{\prime})= \frac{g_{\Sigma_c^{0}[{3\over2}^-]\to\Lambda_c^{+}\pi^- } f_{\Sigma_c^{0}[{3\over2}^-]}f_{\Lambda_c^{+}}}{(\bar{\Lambda}_{\Sigma_c^{0}[{3\over2}^-]}-\omega^{\prime})(\bar{\Lambda}_{\Lambda_c^{+}}-\omega)}
\\ \nonumber &=& \int_0^\infty dt \int_0^1 du e^{i(1-u)\omega^\prime t} e^{iu\omega t}\times 8\times \Big(\frac{f_{\pi} m_{\pi}^2 u}{12(m_u+m_d)\pi^2 t^2}\phi_{3;\pi}^{\sigma}(u)+\frac{f_{\pi} u}{12}\langle\bar q q\rangle\phi_{2;\pi}(u)
\\ \nonumber &&+\frac{f_{\pi} u t^2}{192}\langle\bar q q\rangle\phi_{4;\pi}(u)+\frac{f_{\pi} u t^2}{192}\langle g_s \bar q \sigma G q\rangle\phi_{2;\pi}(u)+\frac{f_{\pi} u t^4}{3072}\langle g_s \bar q \sigma G q\rangle\phi_{4;\pi}(u)\Big)
\\ \nonumber  &-&\int_0^\infty dt \int_0^1 du \int \mathcal{D}\underline{\alpha}e^{i\omega^\prime t(\alpha_2+u\alpha_3)}e^{i\omega t(1-\alpha_2-u\alpha_3)}\times\Big(\frac{f_{3\pi} u}{2\pi t^2}\Phi_{3;\pi}(\underline{\alpha})-\frac{f_{3\pi}}{2\pi^2 t^2}\Phi_{3;\pi}(\underline{\alpha})
\\ \nonumber &&+\frac{i f_{3\pi} u^2 \alpha_3 v\cdot q}{2\pi^2 t}\Phi_{3;\pi}(\underline{\alpha})+\frac{i f_{3\pi} u \alpha_2 v\cdot q}{2\pi^2 t}\Phi_{3;\pi}(\underline{\alpha})-\frac{i f_{3\pi} u v\cdot q}{2\pi^2 t}\Phi_{3;\pi}(\underline{\alpha})\Big) \, .
\end{eqnarray}

The sum rule equation for the $\Xi_c^{\prime0}[{3\over2}^-]$ belonging to $[\mathbf{6}_F, 2 , 1, \lambda]$ is
\begin{eqnarray}
&&G_{\Xi_c^{\prime0}[{3\over2}^-]\to\Xi_c^{\prime+}\pi^-}^D(\omega,\omega^{\prime})= \frac{g_{\Xi_c^{\prime0}[{3\over2}^-]\to\Xi_c^{\prime+}\pi^- } f_{\Xi_c^{\prime0}[{3\over2}^-]}f_{\Xi_c^{\prime+}}}{(\bar{\Lambda}_{\Xi_c^{\prime0}[{3\over2}^-]}-\omega^{\prime})(\bar{\Lambda}_{\Xi_c^{\prime+}}-\omega)}
\\ \nonumber &=& \int_0^\infty dt \int_0^1 du e^{i(1-u)\omega^\prime t} e^{iu\omega t}\times 4\times \Big(\frac{3i f_{\pi} u}{4\pi^2 t^3}\phi_{2;\pi}(u)-\frac{i f_{\pi} m_{\pi}^2 m_s u}{16 (m_u+m_d)\pi^2 t}\phi_{3;\pi}^{\sigma}(u)
\\ \nonumber &&+\frac{3 i f_{\pi} u}{64\pi^2 t}\phi_{4;\pi}(u)+\frac{i f_{\pi} m_s u t}{32}\langle\bar s s\rangle\phi_{2;\pi}(u)-\frac{i f_{\pi} m_{\pi}^2 u t}{48(m_u+m_d)}\langle\bar s s\rangle\phi_{3;\pi}^{\sigma}(u)+\frac{i f_{\pi} m_s u t^3}{512}\langle\bar s s\rangle\phi_{4;\pi}(u)
\\ \nonumber &&-\frac{i f_{\pi} m_{\pi}^2 u t^3}{768(m_u+m_d)}\langle g_s \bar s \sigma G s\rangle\phi_{3;\pi}^{sigma}(u)\Big)
\\ \nonumber &-&\int_0^\infty dt \int_0^1 du \int \mathcal{D}\underline{\alpha}e^{i\omega^\prime t(\alpha_2+u\alpha_3)}e^{i\omega t(1-\alpha_2-u\alpha_3)}\times{1\over2}\times\Big(\frac{3f_{\pi} u}{8\pi^2 t^2 v\cdot q}\Psi_{4;\pi}(\underline{\alpha})-\frac{3 f_{\pi} u}{8\pi^2 t62 v\cdot q}\widetilde\Psi_{4;\pi}(\underline{\alpha})
\\ \nonumber &&-\frac{3 f_{\pi}}{8\pi^2 t^2 v\cdot q}\Phi_{4;\pi}(\underline{\alpha})+\frac{3 f_{\pi}}{8\pi^2 t^2 v\cdot q}\widetilde\Phi_{4;\pi}(\underline{\alpha})-\frac{3 f_{\pi} }{8\pi^2 t^2 v\cdot q}\Psi_{4;\pi}(\underline{\alpha})+\frac{3 f_{\pi}}{8\pi^2 t^2 v\cdot q}\widetilde\Psi_{4;\pi}(\underline{\alpha})
\\ \nonumber &&+\frac{3 i f_{\pi}\alpha_3 u^2}{8\pi^2 t}\Phi_{4;\pi}(\underline{\alpha})+\frac{3 i f_{\pi}\alpha_2 u}{8\pi^2 t}\Phi_{4;\pi}(\underline{\alpha})+\frac{3 i f_{\pi}\alpha_3 u}{16\pi^2 t}\Phi_{4;\pi}(\underline{\alpha})+\frac{3 i f_{\pi}\alpha_3 u}{16\pi^2 t}\widetilde\Phi_{4;\pi}(\underline{\alpha})
\\ \nonumber &&-\frac{3 i f_{\pi} u}{8\pi^2 t}\Phi_{4;\pi}(\underline{\alpha})+\frac{3 i f_{\pi}\alpha_2}{16\pi^2 t}\Phi_{4;\pi}(\underline{\alpha})+\frac{3 if_{\pi}\alpha_2}{16\pi^2 t}\widetilde\Phi_{4;\pi}(\underline{\alpha})-\frac{3 i f_{\pi}}{16\pi^2 t}\Phi_{4;\pi}(\underline{\alpha})-\frac{3 i f_{\pi}}{16\pi^2 t}\widetilde\Phi_{4;\pi}(\underline{\alpha})\Big) \, .
\end{eqnarray}

The sum rule equation for the $\Omega_c^{0}[{5\over2}^-]$ belonging to $[\mathbf{6}_F, 2 , 1, \lambda]$ is
\begin{eqnarray}
&&G_{\Omega_c^{0}[{5\over2}^-]\to\Xi_c^{+}K^-}^D(\omega,\omega^{\prime})= \frac{g_{\Omega_c^{0}[{5\over2}^-]\to\Xi_c^{+}K^- } f_{\Omega_c^{0}[{5\over2}^-]}f_{\Xi_c^{+}}}{(\bar{\Lambda}_{\Omega_c^{0}[{5\over2}^-]}-\omega^{\prime})(\bar{\Lambda}_{\Xi_c^{+}}-\omega)}
\\ \nonumber &=& \int_0^\infty dt \int_0^1 du e^{i(1-u)\omega^\prime t} e^{iu\omega t}\times 8\times \Big(\frac{3 f_{K} m_s u}{20\pi^2 t^2}\phi_{2;K}(u)+\frac{f_{K} m_{K}^2 u}{20(m_u+m_d)\pi^2}\phi_{3;K}^{\sigma}(u)
\\ \nonumber &&+\frac{3 f_{K} m_s u}{320\pi^2}\phi_{4;K}(u)+\frac{f_{K} u}{20}\langle\bar s s\rangle\phi_{2;K}(u)+\frac{f_{K} m_s m_{K}^2 u t^2}{480(m_u+m_d)}\langle\bar s s\rangle\phi_{3;\pi}^{\sigma}(u)+\frac{f_{K} u t^2}{320}\langle\bar s s\rangle\phi_{4;\pi}(u)
\\ \nonumber &&+\frac{f_{K} u t^2}{320}\langle g_s\bar s \sigma G s\rangle\phi_{2;K}(u)+\frac{f_{K} u t^4}{5120}\langle g_s\bar s \sigma G s\rangle\phi_{4;K}(u)\Big)
\\ \nonumber &-&\int_0^\infty dt \int_0^1 du \int \mathcal{D}\underline{\alpha}e^{i\omega^\prime t(\alpha_2+u\alpha_3)}e^{i\omega t(1-\alpha_2-u\alpha_3)}\times\Big(\frac{3f_{3K} u}{10\pi^2 t^2}\Phi_{3;K}(\underline{\alpha})-\frac{3f_{3K}}{10\pi^2 t^2}\Phi_{3;K}(\underline{\alpha})
\\ \nonumber &&=\frac{3i f_{3K}\alpha_3 u^2 v\cdot q}{10\pi^2 t}\Phi_{3;K}(\underline{\alpha})+\frac{3if_{3K}\alpha_2 u v\cdot q}{10\pi^2 t}\Phi_{3;K}(\underline{\alpha})-\frac{3i f_{3K} u v\cdot q}{10\pi^2 t}\Phi_{3;K}(\underline{\alpha})\Big) \, .
\end{eqnarray}

\end{widetext}

%

%

\end{document}